\documentclass[9pt,a4paper]{article}
\usepackage[utf8]{inputenc}
\usepackage[english]{babel}
\usepackage{authblk}
\usepackage{physics}
\usepackage{amsmath}
\usepackage{amsfonts}
\usepackage{amssymb}
\usepackage{graphicx}
\usepackage{lmodern}
\usepackage[margin=.75in]{geometry}
\usepackage[none]{hyphenat}
\usepackage[nottoc, notlof, notlot]{tocbibind}
\usepackage[colorlinks=true, breaklinks, linktocpage=true]{hyperref}
\usepackage{enumitem}	
\usepackage{bbm}
\usepackage{bm}
\usepackage{subfig}
\usepackage{cancel}
\usepackage{dsfont}

\usepackage{fancyhdr}
\usepackage{cite}
\usepackage{simpler-wick}
\usepackage{tikz}
\usetikzlibrary{shapes.geometric, arrows, positioning, fit}

\makeatletter
\g@addto@macro\bfseries{\boldmath}
\makeatother

\numberwithin{equation}{section}

\begin{document}

\title{\bfseries Superspace invariants and correlators in $4d$ $\mathcal{N}=1$ superconformal field theories}
\author{{Aditya Jain and Amin A. Nizami}
\thanks{email: adityaj2807@gmail.com, amin.nizami@ashoka.edu.in}}
\affil{\small Department of Physics, Ashoka University, \\
Rajiv Gandhi Education City, Rai, NCR, India, 131029}
\date{} 
\maketitle

{\abstract  Using polarization spinor methods in conjunction with the superspace formalism, we construct 3-point superconformal invariants that are used to determine the form of 3-point correlators of spinning superfield operators in $\mathcal{N}=1$ superconformal field theories (SCFTs) in 4-dimensions.  We enumerate the structural form of various spinning 3-point correlators using these invariants and find additional constraints on their form when the operators are conserved supercurrents. For these purposes, we first construct the invariants and 3-point correlators in non-supersymmetric $4d$ CFTs which are then extended using superspace methods to $4d$ SCFTs.}
\vfill
\pagebreak

\setcounter{tocdepth}{2}
\tableofcontents

\vspace{2em}
\hrule

\setlength{\parskip}{.5em}
\setlength{\parindent}{0em}

\section{Introduction}

The constraints due to symmetries on correlation functions in quantum field theories (QFTs) is a much explored topic. In particular, as is well known, conformal symmetry fixes the structure of two and three point functions in a conformal field theory (CFT). Supersymmetry imposes additional restrictions and superconformal invariance imposes the most limiting constraints as it is the largest (spacetime) symmetry possible in a QFT. Further constraints can arise from Ward-Takahashi identities when the operators are conserved - that is they satisfy a conservation equation: $\partial_{\mu} J^{\mu\mu_2 \mu_3\hdots\mu_s}=0$. Such conserved operators typically saturate the unitarity bound on scaling dimensions.

The primary aim of this work is to study the constraint of $\mathcal{N}=1$ superconformal invariance on 3-point correlation functions of spinning superfield operators in superconformal field theories (SCFTs) in four spacetime dimensions. The spinning operators live in supermultiplets whose lowest component is a superconformal primary. To study correlators of these superfield operators it is thus useful to use superspace methods. We also extensively use the polarization spinor formalism of \cite{GPY2011} for $3d$ CFTs to efficiently encode the spin of the operators.  Moving from three to four dimensions introduces new features which we highlight in this work. We build superconformal invariants in superspace, construct 3-point functions of spinning superfield operators, and study the additional constraints on these correlators due to one or more operators in the 3-point correlator being a conserved supercurrent. Our study also relies on the foundational works \cite{Park4dN1SCFT1997,Park4dSCFT1999,Park6d4dSCFT1998,Osborn4dN1SCFT1998,Osborn4dSCFTchiral2000,KuzenkoN2SCFT1999} on superconformally covariant structures in superspace for $4d$ SCFTs.

Correlators in (super)conformal theories have been the focus of various studies in the literature, with diverse approaches. The systematic computation of spinning $n$-point correlators in $d$-dimensional CFT on grounds of symmetry was first done in \cite{OsbornCFTGenD1993,OsbornCFTGenD1996}, taking inspiration from \cite{Polyakov1970,Schreier1971} and other foundational works. These studies were extended to ${\cal N}$-extended SCFTs in $3d$ and $4d$ using the superspace formalism in \cite{Park4dN1SCFT1997,Park6d4dSCFT1998,Osborn4dN1SCFT1998,Park4dSCFT1999,Park3DSCFT1999,Osborn4dSCFTchiral2000,KuzenkoN2SCFT1999}. Later on, correlators in CFTs with higher-spin symmetry were extensively studied in \cite{GPY2011,CPPR2011,MZ3dCFTHigh2011,MZ3dCFTBroken2012,Stanev2012,Todorov2011,Karateev4d2014,AlbaDiabd42013,AlbaDiab2015}. In \cite{GPY2011}, Giombi \textit{et al.} used auxiliary polarization spinors to study 3-point correlators in $3d$ CFTs by constructing conformal invariants, and showed evidence for the presence of parity-violating structures present in these correlators. Similar constructions using embedding space, auxiliary vectors, and spinors were done in \cite{CPPR2011,Stanev2012,Todorov2011} respectively for the case of $4d$ CFTs. The supersymmetric extension of \cite{GPY2011} was done in \cite{Nizami2013}, where fermionic (or grassmannian) superconformal invariants were built to classify spinning correlators in $3d$ ${\cal N}=1$ SCFT. More recently, in \cite{JainNizami2022}, we used a similar formalism to build invariants and write down correlators in the $3d$ ${\cal N}=2$ theory. An alternative approach has been used by Buchbinder \textit{et al.} for (S)CFTs in many of their works \cite{BuchKuz3dSCFT2015,Buch3dSCFTGen2023,Buch3dSCFTMixed2021,Buch3dSCFTN1spin22021,Buch3dSCFTOdd2023,Buch4DSCFTN12022,Buch4dSCFTN1higherspin2021,Buch4dCFTFerm2022,Buch4dSCFTN1spinor2021,Buch4dCFTGen2023,Buch4dSCFTGen2024,StoneThesis2024}. Their approach builds on the seminal work  \cite{OsbornCFTGenD1993} which enumerates the structures of 3-point correlators of conserved (super)currents. A similar formalism was used in \cite{Manenti4d3pt2018} to study $R$-current correlators in $4d$ SCFT. Momentum space and twistor variables have been used recently \cite{SachinJain2023} for SCFT correlators, but our focus will be on position space exclusively.

Some more comments on the approach of \cite{Buch4dSCFTGen2024} are in order. Like our work, the focus there is on 3-point correlators in $4d$ SCFTs, but the methods and formalism used are different from our approach. In \cite{Buch4dSCFTGen2024}, the authors generalize their earlier works on (super)conformal theories to $4d$ SCFTs, extending some of their recent works \cite{Buch4dSCFTN1spinor2021,Buch4dSCFTN1higherspin2021,Buch4DSCFTN12022}. Essentially, the authors in \cite{Buch4dSCFTGen2024} extend Osborn and Petkou's analysis \cite{OsbornCFTGenD1993} to SCFTs and determine the covariant structures used to construct 3-point correlators, whereas in this paper, we build on the work of Giombi \textit{et al.} \cite{GPY2011} and construct superconformal \textit{invariants} for $4d$ SCFTs, which are then utilized to construct the correlators. Also, while the focus of \cite{Buch4dSCFTGen2024} is exclusively on correlators of conserved higher-spin supercurrents, our analysis gives the general form of correlators containing non-conserved spinning superfield operators (see Section \ref{sec:scftcorrstr}). Furthermore, our formalism facilitates the constraints of the conservation condition on each of the three operators individually, hence we can obtain 3-point correlators containing a mixture of conserved supercurrents and non-conserved superfield operators.

\paragraph{Outline} The structure of the paper is as follows. To keep the paper self-contained we review some essential foundational material in Sections \ref{sec:cftreview} and \ref{sec:scftbb}. In Section \ref{sec:cftreview}, we review the auxiliary polarization spinor formalism for $4d$ CFTs, and construct 2-point and 3-point structures which are conformally invariant, and have a definite parity. These structures are then used to express 3-point correlators of integer-spin symmetric traceless operators, including the special case of operators that are conserved currents. In Section \ref{sec:scftbb}, we extend this formalism to $4d$ ${\cal N}=1$ SCFTs and subsequently build the superspace 2-point and 3-point superconformal building blocks, which are used together with the auxiliary polarization spinors to construct superconformal and superinversion invariants in Section \ref{sec:scftinv}. We also utilize the analysis of Section \ref{sec:cftreview} to construct novel purely grassmanian superconformal invariants. In Section \ref{sec:scftcorrstr}, we determine the structure of spinning 3-point correlators in terms of the derived invariant structures. Finally, in Section \ref{sec:scftconscorr}, the shortening constraints are implemented on various superfield operators to give us 3-point correlators of supercurrents. 

The appendices provide some useful supplementary details: Appendix \ref{appsec:4dconv} establishes our notations and conventions. Appendix \ref{appsec:cftconservation}  enumerates many examples of correlators with (non-)conserved operators in $4d$ CFTs. Appendix \ref{appsec:relations} gives a comprehensive list of the relations between the superconformal invariants constructed in Section \ref{sec:scftbb}. Appendix \ref{appsec:morescft3ptcorr} provides further examples of 3-point SCFT correlators involving supercurrents.










\section{CFT correlators in $4d$}
\label{sec:cftreview}

In this section, we present a detailed review of our formalism for $4d$ CFTs, which will be systematically extended to the supersymmetric case in later sections. This section can be considered as the $4d$ version of the analysis of Giombi \textit{et al.} \cite{GPY2011}, and takes insights from the works of Stanev \cite{Stanev2012}, Todorov \cite{Todorov2011}, Alba and Diab \cite{AlbaDiabd42013}, and Costa \textit{et al.} \cite{CPPR2011}. In this approach, we utilize conformal symmetry and the inversion transformation in conjunction with auxiliary polarization spinors to construct conformally invariant structures. These structures are eventually used to write down the form of general spinning 3-point correlators in $4d$ CFT. Our analysis is general,  and so we also obtain conformal invariants which are odd under parity, and this consequently gives us parity-violating contributions to spinning 3-point correlators in $4d$ CFTs. We also observe that employing the conservation constraint on one (or more) of the operators reduces the number of undetermined coefficients and fixes the form of the 3-point correlator.\footnote{The unknown coefficients are theory-dependent OPE coefficients, and are calculable through 3-point vertices in specific CFTs.} In extending this formalism to $4d$ SCFTs and determining 3-point correlators of spinning superfield operators, we will see that apart from the construction of superconformal invariants by the superspace extension of known conformal invariants, one can also build new grassmanian invariants using  the relations between the conformal invariants derived in this section. Hence, an extensive analysis for the non-supersymmetric CFT case in $4d$ is indispensable.

We work in 4-dimensional Minkowski spacetime $\mathbb{R}^{3,1}$ equipped with a `mostly plus' metric. The conventions and notations are primarily taken from \cite{BuchKuzSUGRA1998}, and a complete list  can be found in Appendix \ref{appssec:4dconv-cft}. The $Spin$ group counterpart for $4d$ Lorentz group is $SL(2,\mathbb{C})$, where we choose the invariant tensors
\begin{align}
(\sigma_{\mu})_{\alpha\dot\alpha}=(\mathds{1},\sigma^{i}),\quad (\tilde\sigma_{\mu})^{\dot\alpha\alpha}=(\mathds{1},-\sigma^{i})\,,
\end{align}
such that $(\tilde\sigma_{\mu})^{\dot\alpha\alpha}=\epsilon^{\dot\alpha\dot\beta}\epsilon^{\alpha\beta}(\sigma_{\mu})_{\beta\dot\beta}$. Here, $\sigma^{i}$ are the standard $2\times 2$ Pauli matrices.

We use $\sigma,\tilde\sigma$ to convert spacetime 4-vectors to spinor matrices,
\begin{gather}
X_{\alpha\dot\alpha}=x^{\mu}(\sigma_{\mu})_{\alpha\dot\alpha}\,,\quad
\tilde X^{\dot\alpha\alpha}=x^{\mu}(\tilde \sigma_{\mu})^{\dot\alpha\alpha}\,,\quad
X\cdot \tilde X=-x^{2}\mathds{1}=\tilde X\cdot X\,,
\label{eq:defcft1pt}
\end{gather}
where $x^{2}=x^{\mu}x_{\mu}$. This defines the inverses
\begin{align}
(X^{-1})^{\dot\alpha\alpha}=-\frac{1}{x^{2}}\tilde X^{\dot\alpha\alpha}\,,\quad (\tilde X^{-1})_{\alpha\dot\alpha}=-\frac{1}{x^{2}}X_{\alpha\dot\alpha}\,.
\label{eq:defcft1ptinv}
\end{align}
For convenience, we will occasionally denote $X_{i},\tilde X_{i},X_{i}^{-1},\tilde X_{i}^{-1}$ as 1-point objects.

We augment the coordinate space with commuting \textit{auxiliary polarization spinors} $\lambda^{\alpha},\bar\lambda^{\dot\alpha}$, which are related to the polarization vectors $\varepsilon^{\mu}$ through
\begin{align}
\varepsilon^{\mu}=\lambda^{\alpha}(\sigma^{\mu})_{\alpha\dot\alpha}\bar\lambda^{\dot\alpha}=\bar\lambda_{\dot\alpha}(\tilde \sigma^{\mu})^{\dot\alpha\alpha}\lambda_{\alpha}\,,
\end{align}
where $(\tilde\sigma_{\mu})^{\dot\alpha\alpha}=\epsilon^{\dot\alpha\dot\beta}\epsilon^{\alpha\beta}(\sigma_{\mu})_{\beta\dot\beta}$, so these polarization spinors are commuting left- and right-handed Weyl spinors (Appendix \ref{appssec:4dconv-cft}). Both the polarizations are null
\begin{align}
\varepsilon^{\mu}\varepsilon_{\mu}=0=\lambda^{\alpha}\lambda_{\alpha}=\bar\lambda_{\dot\alpha}\bar\lambda^{\dot\alpha}\,.	
\end{align}
The polarization spinors carry a scaling weight of $\frac12$. Under dilatation $x^{\mu}\rightarrow \eta x^{\mu}$,
\begin{align}
\lambda^{\alpha}\rightarrow \eta^{\frac12}\lambda^{\alpha}\,,\qquad
\bar\lambda^{\dot\alpha}\rightarrow \eta^{\frac12}\bar\lambda^{\dot\alpha}\,.
\label{eq:lamscaling}
\end{align}
The spinors $\lambda,\bar\lambda$ are among the building blocks in the construction of conformal invariants. They also help us encode the spin of a spinning operator in the following way. Consider a symmetric rank-$s$ (or spin-$s$) tensor $O_{\mu_{1}\hdots \mu_{s}}(x)$. Using $\sigma$'s (or equivalently $\tilde \sigma$'s), one can transform it into a multispinor with $s$ undotted and $s$ dotted indices
\begin{align}
O_{\alpha(s)\dot\alpha(s)}(x)
\equiv
O_{(\alpha_{1}\hdots \alpha_{s})(\dot\alpha_{1}\hdots\dot\alpha_{s})}(x)
=
O_{\mu_{1}\hdots \mu_{s}}(x)\,
\sigma^{\mu_{1}}_{\alpha_{1}\dot\alpha_{1}}\hdots\sigma^{\mu_{s}}_{\alpha_{s}\dot\alpha_{s}}\,,
\end{align}
i.e. it transforms under the irreducible $(s/2,s/2)$ rep of the Lorentz group $SL(2,\mathbb{C})$. Note that we have introduced the shorthand notation for a general symmetric sequence of spinor indices, $\alpha(s)\equiv (\alpha_{1}\hdots\alpha_{s})$, and so on. Contracting the spinor indices with polarization spinors $\lambda,\bar\lambda$, gives us an index-free form
\begin{align}
O_{s}(x,\lambda,\bar\lambda)=\lambda^{\alpha_{1}}\hdots \lambda^{\alpha_{s}}\, O_{\alpha(s)\dot\alpha(s)}(x)\,\bar\lambda^{\dot\alpha_{1}}\hdots \bar\lambda^{\dot\alpha_{s}}\,,
\end{align}
where the spin of the operator is encoded in the number of $\lambda,\bar\lambda$'s in $O_{s}(x,\lambda,\bar\lambda)$. 

Schematically, this looks like
\begin{align}
\underbrace{O_{\mu_{1}\hdots\mu_{s}}(x)}_{\text{$s$ vector indices}}
\xrightarrow{\text{contract with $\sigma$}}\quad
\underbrace{O_{(\alpha_{1}\hdots \alpha_{s})(\dot\alpha_{1}\hdots\dot\alpha_{s})}(x)}_\text{$2s$ spinor indices}\quad
\xrightarrow{\text{contract with $\lambda,\bar\lambda$}}\quad
\underbrace{O_{s}(x,\lambda,\bar\lambda)}_{\text{index-free}}
\label{eq:indexfree}
\end{align}

Thus, the general 3-point correlator containing spin-$s$ symmetric traceless primary operators can be written as \footnote{Throughout this work we will work with well-separated points so that there are no contact terms.}
\begin{align}
\left\langle O_{\mu_{1}\hdots \mu_{s_{1}}}(x_{1})
O_{\nu_{1}\hdots \nu_{s_{2}}}(x_{2})
O_{\tau_{1}\hdots \tau_{s_{3}}}(x_{3})\right\rangle\longrightarrow
F(\{x_{i},\lambda_{i},\bar\lambda_{i},s_{i}\})\equiv
\left\langle
\prod_{i=1}^{3}O_{s_{i}}(x_{i},\lambda_{i},\bar\lambda_{i})
\right\rangle\,.
\label{eq:cft3ptfunction}
\end{align}

\subsection{Inversion}
\label{ssec:cftreview-inver}

Conformal symmetry requires symmetry under Poincar\'e $+$ dilatation $+$ special conformal transformation. As is well known, special conformal transformation $K$ can be written as the composition $K=I\cdot P\cdot I$, where $P, I$ correspond to translation and inversion transformations, respectively. The constraints of special conformal transformations are especially hard to implement, however it was shown in \cite{GPY2011} that augmented coordinate space objects transforming covariantly under inversion can be utilized to construct conformally invariant structures in $3d$. We explore how this formalism works for 4-dimensional CFT.

Under inversion, 
\begin{align}
x^{\mu}\rightarrow x'^{\mu}=\frac{x^{\mu}}{x^{2}}\,, \qquad x^{2}\rightarrow\frac{1}{x^{2}}\,,
\label{eq:defcftinversion}
\end{align}
which gives\footnote{The spinor indices follow the conventions, and hence will be sporadically suppressed for simplicity.}
\begin{align}
X\rightarrow -\tilde X^{-1}\,,\quad \tilde X\rightarrow -X^{-1}\,, \quad X^{-1}\rightarrow -\tilde X\,,\quad \tilde X^{-1}\rightarrow -X\,,
\label{eq:1ptinv}
\end{align}
where we have used Eqs. (\ref{eq:defcft1pt}, \ref{eq:defcft1ptinv}, \ref{eq:defcftinversion}).

Similarly, using the rules of a general coordinate transformation, we have the inversion of the polarization vectors
\begin{align}
\varepsilon^{\mu}\rightarrow\frac{\partial x'^{\mu}}{\partial x^{\nu}}\varepsilon^{\nu}=\frac{\varepsilon^{\mu}}{x^{2}}-\frac{2\varepsilon\cdot x\,x^{\mu}}{x^{4}}\,.
\end{align}
Since $\varepsilon$'s decompose into $\lambda,\bar\lambda$, the above defines for us the inversion of $\lambda,\bar\lambda$ as
\begin{align}
\lambda^{\alpha}\rightarrow i \bar\lambda_{\dot\alpha}(X^{-1})^{\dot\alpha\alpha}\,,\quad
\lambda_{\alpha}\rightarrow -i (\tilde X^{-1})_{\alpha\dot\alpha}\bar\lambda^{\dot\alpha}\,,\qquad 
\bar\lambda^{\dot\alpha}\rightarrow -i (X^{-1})^{\dot\alpha\alpha}\lambda_{\alpha}\,,\quad
\bar\lambda_{\dot\alpha}\rightarrow i \lambda^{\alpha}(\tilde X^{-1})_{\alpha\dot\alpha}\,,
\label{eq:laminversion}
\end{align}
i.e. under inversion the polarization spinors transform into their conjugates dotted with the 1-point coordinate space objects.\footnote{The factor of $\pm i$ is conventional. If $\lambda$ transforms under $U(1)$ with an $\exp(i\alpha)$ factor, then $\bar \lambda$ transforms with $\exp(-i\alpha)$. We thank the referee for pointing this out.}

The above transformation along with Eq. (\ref{eq:lamscaling}) implies that for a conformal primary spin-$s$ operator $O_{\mu_{1}\hdots\mu_{s}}(x)$ with conformal dimension $\Delta$, the index-free operator $O_{s}(x,\lambda,\bar\lambda)$ transforms under inversion as
\begin{align}
O_{s}(x,\lambda,\bar\lambda)\longrightarrow (x^{2})^{\Delta-s}O_{s}(x,\lambda,\bar\lambda)\,,
\label{eq:cftscaling}
\end{align}
i.e. $O_{s}$ transforms like a scalar operator of dimension $\Delta-s$.

Hence, the general 3-point spinning conformal correlator of Eq. (\ref{eq:cft3ptfunction}) transforms under inversion as
\begin{align}
F(\{x_{i},\lambda_{i},\bar\lambda_{i},s_{i}\})\longrightarrow
x_{1}^{2\tau_{1}}x_{2}^{2\tau_{2}}x_{3}^{2\tau_{3}}\,F(\{x_{i},\lambda_{i},\bar\lambda_{i},s_{i}\})\,,\qquad 
\tau_{i}=\Delta_{i}-s_{i}\,,
\label{eq:cftdilatation}
\end{align}
where $\tau_{i}=\Delta_{i}-s_{i}$ is also known as the twist of a spin-$s_{i}$ operator with conformal dimension $\Delta_{i}$.

\subsection{Building blocks}
\label{ssec:cftreview-bb}

The construction of conformal invariants closely follows \cite{GPY2011}: we consider translationally invariant building blocks (and their products) that are made Poincar\'e invariant by contracting with appropriate polarization spinors ($\lambda,\bar\lambda$). These blocks will have a fixed weight under dilatation. To account for special conformal symmetry, their transformation under inversion is analyzed. Poincar\'e invariant objects with a fixed scaling dimension which are invariant under inversion (upto a sign and conjugation, see below), will be used to construct 3-point conformal invariants.

To this end, we consider 2-point building blocks which are invariant under space-time translations and Lorentz covariant,
\begin{align}
X_{ij}=X_{i}-X_{j}\,,\quad
\tilde X_{ij}=\tilde X_{i}-\tilde X_{j}\,,\qquad X_{ij}\cdot\tilde X_{ij}=-x_{ij}^{2}\mathds{1}=\tilde X_{ij}\cdot X_{ij}\,,
\end{align}
where $i,j$ are coordinate point labels (and not space-time component indices). We also obtain the inverses
\begin{align}
X_{ij}^{-1}=-\frac{\tilde X_{ij}}{x_{ij}^{2}}\,,\qquad
\tilde X_{ij}^{-1}=-\frac{X_{ij}}{x_{ij}^{2}}\,.
\end{align}

Using Eq. (\ref{eq:1ptinv}), it is easy to check that under inversion these 2-point objects transform as
\begin{align}
\begin{gathered}
X_{ij}\rightarrow \tilde X_{i}^{-1} \tilde X_{ij} \tilde X_{j}^{-1}\,,\qquad
\tilde X_{ij}\rightarrow X_{i}^{-1}  X_{ij} X_{j}^{-1}\,,\\
X_{ij}^{-1}\rightarrow \tilde X_{i}\tilde X_{ij}^{-1}\tilde X_{j}\,,\qquad
\tilde X_{ij}^{-1}\rightarrow X_{i}X_{ij}^{-1}X_{j}\,,
\end{gathered}\qquad
x_{ij}^{2}\rightarrow \frac{x_{ij}^{2}}{x_{i}^{2}x_{j}^{2}}\,.
\label{eq:2ptinversion}
\end{align}
Here, the 1-point factors can be switched as well for each transformation, i.e. $X_{ij}\rightarrow \tilde X_{j}^{-1} \tilde X_{ij} \tilde X_{i}^{-1}$ works as well. 

\paragraph{Homogeneity upto conjugation}

In Eq. (\ref{eq:2ptinversion}), the inversion transformation of the 2-point building blocks is homogeneous \textit{upto conjugation}. In $4d$, since the Lorentz group is $SL(2,\mathbb{C})$, there are two copies of Lorentz invariant tensors $\sigma,\tilde \sigma$, and the space of covariant $n$-point objects is decomposable into two subspaces characterized by $\sigma,\tilde \sigma$ respectively. Consequently, under inversion the 2-point subspace transforms into its conjugate subspace and vice-versa, which is also observed in Eq. (\ref{eq:2ptinversion}). Since this transformation is homogeneous upto conjugation, we can still expect invariance under special conformal transformations (since they require $I\cdot P\cdot I$, i.e. the application of inversion twice). This implies that invariance under inversion guarantees special conformal invariance.\footnote{Inversion invariance (along with Poincar\'e and scale symmetry) implies conformal invariance, but the converse is not generally true \cite{GilliozCFT2022}. In the case of $3d$ CFT \cite{GPY2011}, the Lorentz group is isomorphic to $SL(2,\mathbb{R})$, and there is only a single $\sigma$, hence the 2-point blocks transform truly homogeneously under inversion, and conformal invariance does imply inversion invariance.} Also note that the polarization spinors transform homogeneously upto conjugation as well in Eq. (\ref{eq:laminversion}). 


\subsection{Conformal invariants}
\label{ssec:cftreview-inv}

Using the polarization spinors and (products of) 2-point blocks defined in the last subsection, we can now construct objects that transform invariantly under conformal transformations. We first define
\begin{gather}
P_{12}=\lambda_{1}^{\alpha}(\tilde X_{12}^{-1})_{\alpha\dot\alpha}\bar\lambda_{2}^{\dot\alpha}\,\  \text{and 5 perm}\,,\\[.5em]
S_{13}=\frac{\lambda_{1}^{\alpha}(X_{12}\tilde X_{23})_{\alpha}^{\ \beta}\lambda_{3\beta}}{|x_{12}||x_{23}||x_{31}|}\,\ \text{and 2 perm}\,,\\[.5em] 
\bar S_{13}=\frac{\bar \lambda_{1\dot\alpha}(\tilde X_{12}X_{23})^{\dot\alpha}_{\ \dot\beta}\bar \lambda_{3}^{\dot\beta}}{|x_{12}||x_{23}||x_{31}|}\,\  \text{and 2 perm}\,,\\[.5em]
Q_{1}= \lambda_{1}^{\alpha}(\tilde X_{12}^{-1}\tilde X_{23}\tilde X_{31}^{-1})_{\alpha\dot\alpha}\bar\lambda_{1}^{\dot\alpha}\,\  \text{and 2 perm}\,.
\end{gather}
Note that $P_{12}$ and $P_{21}$ describe distinct structures (since they contain $\lambda_{1}\bar\lambda_{2}$ and $\lambda_{2}\bar\lambda_{1}$, respectively), which is why we have six $P$'s. Also, it turns out that $S_{31}=-S_{13},\ \bar S_{31}=-\bar S_{13}$, hence the number of permutations.

Under inversion, using Eqs. (\ref{eq:laminversion}, \ref{eq:2ptinversion}) we get the transformation of the objects defined above,
\begin{align}
P_{ij}\rightarrow -P_{ji}\,,\quad S_{ij}\rightarrow -\bar S_{ij}\,,\quad \bar S_{ij}\rightarrow -S_{ij}\,,\quad Q_{i}\rightarrow Q_{i}\,.
\end{align}
These objects are manifestly Poincar\'e invariant, and have a fixed scaling weight (zero). Under inversion, they are invariant upto a sign and conjugation. Since special conformal transformation requires the application of inversion twice, all of them are indeed \textit{conformal invariants}. The analogous $P_{ij},Q_{i},S_{ij}$ are also the invariants constructed for $3d$ CFT by \cite{GPY2011}.\footnote{There are no barred variables in $3d$ since the Lorentz group is $SL(2,\mathbb{R})$.}

Note that conformal invariants such as $P_{12}$, $S_{12}$ can not occur alone in a 3-point correlator. This is because, by construction, such a correlator has to have the same number of $\lambda$'s and $\bar{\lambda}$'s, since we want to describe correlators of bosonic integer spin operators. Hence we are ultimately interested in conformal invariants which are also \textit{inversion invariants}. To this end, we consider combinations of conformal invariants that transform invariantly upto a sign under inversion,\footnote{Henceforth, we refer to structures which transform invariantly upto a sign under inversion as inversion invariants.} 
\begin{align}
\begin{gathered}
Q_{i}\,,\quad \hat P_{ij}\equiv P_{ij}P_{ji}=\hat P_{ji}\,,\quad \hat S_{ij}\equiv S_{ij}\bar S_{ij}=\hat S_{ji}\,,\\
P_{123-}\equiv P_{12}P_{23}P_{31}-P_{21}P_{32}P_{13}\,,\quad P_{123+}\equiv P_{12}P_{23}P_{31}+P_{21}P_{32}P_{13}\,.
\end{gathered}
\end{align}
Their transformation under inversion is straightforward:
\begin{align}
\begin{gathered}
Q_{i}\to Q_{i}\,,\quad 
\hat P_{ij}\to \hat P_{ij}\,,\quad
\hat S_{ij}\to \hat S_{ij}\,,\quad
P_{123-}\to P_{123-}\,,\\
P_{123+}\to -P_{123+}\,.
\end{gathered}
\end{align}
However, some of them are not independent, and we find the following relations
\begin{gather}
\hat S_{ij}-\hat P_{ij}-Q_{i}Q_{j}=0\,,\label{eq:s12cft}
\\
P_{123-}-Q_{1}Q_{2}Q_{3}-\sum_{\rm cyc}Q_{1}\hat P_{23}=0\,.\label{eq:GPY4d}
\end{gather}
The former equation effectively removes $\hat S_{ij}$ from the list of independent invariants, and the latter removes $P_{123-}$. Eq. (\ref{eq:GPY4d}) also looks like the $4d$ equivalent of the relation (2.15) in \cite{GPY2011}. The above two equations will also be important later (in sec. \ref{ssec:scftinv-grass}) when we construct superconformal invariants.

It is easy to check that structures which transform invariantly with a $+/-$ sign under inversion also transform with a $+/-$ sign under parity transformation. This implies that the invariants that we have obtained can be classified as parity-even and parity-odd.\footnote{For a more detailed discussion, refer to \cite{Buch4dCFTGen2023}.}

Thus, we have the following independent conformal (and inversion) invariants which carry a definite parity
\begin{align}
\begin{aligned}
\text{parity-even}:&\quad \hat P_{ij}\,,\ Q_{i}\,\\
\text{parity-odd}:&\quad P_{123+}\,.
\end{aligned}
\label{eq:cftinvariants}
\end{align}
Such invariants were also constructed using auxiliary vectors by Stanev \cite{Stanev2012}, auxiliary spinors by Todorov \cite{Todorov2011}, and embedding space by Costa \textit{et al.} \cite{CPPR2011}.
    
The product of two parity-odd invariants, i.e. $P_{123+}^{2}$ can be expressed as a combination of parity-even invariants, and indeed we find
\begin{align}
P_{123+}^{2}=Q_{1}^{2}Q_{2}^{2}Q_{3}^{2}+2Q_{1}Q_{2}Q_{3}\sum_{\rm cyc}Q_{1}\hat P_{23}+\left(\sum_{\rm cyc}Q_{1}\hat P_{23}\right)^{2}+4\hat P_{12}\hat P_{23}\hat P_{31}\,.
\label{eq:p123square}
\end{align}
Thus, $P_{123+}$ can only occur linearly in a correlator.

\paragraph{Permutation symmetry} If the 3-point correlator contains identical operators, it will be symmetric under exchange of those operators.\footnote{We are only considering correlators of integer spin operators, hence there are no fermionic operators which are anti-symmetric under an exchange.} Consequently, the invariant structures that describe that correlator must preserve this permutation symmetry (also referred to as point-switch symmetry). For example, under a $(x_{2},\lambda_{2})\rightarrow(x_{3},\lambda_{3})$ exchange (henceforth called a $2\leftrightarrow 3$ swap), the correlator $\langle O_{s}O_{1}O_{1}\rangle$ containing identical spin-1 operators must stay unchanged. To implement this point-switch symmetry, we determine the transformation of the conformal invariants under an $i\leftrightarrow j$ swap:
\begin{center}
\begin{tikzpicture}
	\node (p12) {$\hat P_{12}$};
	\node (p1231)[below right = .4em and 2cm of p12] {$\hat P_{23}$};
	\node (p1212)[above right = .4em and 2cm of p12] {$\hat P_{12}$};
	\node (p1223)[right = 2cm of p12] {$\hat P_{31}$};
	\draw [->]  (p12.north east)|- node[midway,above,pos=0.75]{$1\leftrightarrow 2$}(p1212.west);
	\draw [->]  (p12.south east)|-node[midway,above,pos=0.75]{$3\leftrightarrow 1$}(p1231.west);
	\draw [->]  (p12.east)--node[midway,above]{$2\leftrightarrow 3$}(p1223.west);
\end{tikzpicture}$\qquad$
\begin{tikzpicture}
	\node (q1) {$Q_{1}$};
	\node (q131)[below right = .6em and 2cm of q1] {$-Q_{3}$};
	\node (q112)[above right = .6em and 2cm of q1] {$- Q_{2}$};
	\node (q123)[right = 2cm of q1] {$- Q_{1}$};
	\draw [->]  (q1.north east)|- node[midway,above,pos=0.75]{$1\leftrightarrow 2$}(q112.west);
	\draw [->]  (q1.south east)|-node[midway,above,pos=0.75]{$3\leftrightarrow 1$}(q131.west);
	\draw [->]  (q1.east)--node[midway,above]{$2\leftrightarrow 3$}(q123.west);
\end{tikzpicture}$\qquad$
\begin{tikzpicture}
	\node (p12) {$P_{123+}$};
	\node (p1223)[right = 2cm of p12] {$P_{123+}$};
	\draw [->]  (p12.east)--node[midway,above]{any swap}(p1223.west);
	\node (p12p) [below=.5cm of p12]{};
	\node (p1223p)[right = 2cm of p12p] {};
\end{tikzpicture}
\end{center}
The rest of the swaps are easily obtained by permuting the indices.

\paragraph{Conserved currents}

For a CFT exhibiting higher spin symmetry, there exists a conserved spin-$s$ symmetric traceless current $J^{\mu_{1}\hdots\mu_{s}}(x)$, that satisfies the conservation equation
\begin{align}
\frac{\partial }{\partial x^{\mu_{1}}} J^{\mu_{1}\hdots\mu_{s}}(x)=0\,.
\end{align}
For the index-free form $J_{s}(x,\lambda,\bar\lambda)$ that we get after applying (\ref{eq:indexfree}) to the spin-$s$ conserved current, the above conservation equation takes the form
\begin{align}
\left[\frac{\partial}{\partial \bar\lambda^{\dot\alpha}}(\tilde \sigma^{\mu})^{\dot\alpha\alpha}\frac{\partial}{\partial x^{\mu}}\frac{\partial}{\partial \lambda^{\alpha}}
\right]
J_{s}(x,\lambda,\bar\lambda)=0\,.
\label{eq:conslaw}
\end{align}

\subsection{Conformal correlators}
\label{ssec:cftreview-confcorr}

We can now use all the machinery developed in this section to construct 3-point correlators of spinning operators. A general 3-point correlator containing index-free conformal primary operators $O_{s_{i}}(x_{i},\lambda_{i},\bar\lambda_{i})$ of spin $s_{i}$ and conformal dimension $\Delta_{i}$ can be expressed as
\begin{align}
\begin{gathered}
\langle
O_{s_{1}}O_{s_{2}}O_{s_{3}}
\rangle=\frac{1}{|x_{12}|^{\tau_{12,3}}|x_{23}|
^{\tau_{23,1}}|x_{31}|^{\tau_{31,2}}}
\left(\sum_{m}a_{m}{\cal T}^{\rm even}_{m}(\hat P_{ij},Q_{k})
+\sum_{n}b_{n}{\cal T}^{\rm odd}_{n}(\hat P_{ij},Q_{k})P_{123+}\right)\,,\\[.5em]
\text{where}\qquad
\tau_{ij,k}=\tau_{i}+\tau_{j}-\tau_{k}\,,\qquad
\tau_{i}=\Delta_{i}-s_{i}\,.
\label{eq:cftcorrelator}
\end{gathered}
\end{align}
 Here, each ${\cal T}^{\rm even}_{m}$ and ${\cal T}^{\rm odd}_{n}P_{123+}$ is a linearly independent parity-even/odd conformally invariant structure made up of the (products of) invariants listed in (\ref{eq:cftinvariants}) with homogeneity
\begin{align*}
{\cal T}^{\rm even}_{m}&:\ \lambda_{1}^{s_{1}}\lambda_{2}^{s_{2}}\lambda_{3}^{s_{3}}\bar\lambda_{1}^{s_{1}}\bar\lambda_{2}^{s_{2}}\bar\lambda_{3}^{s_{3}}\,,\\
{\cal T}^{\rm odd}_{n}&:\ \lambda_{1}^{s_{1}-1}\lambda_{2}^{s_{2}-1}\lambda_{3}^{s_{3}-1}\bar\lambda_{1}^{s_{1}-1}\bar\lambda_{2}^{s_{2}-1}\bar\lambda_{3}^{s_{3}-1}\,.
\end{align*}
Each parity-odd structure must contain $P_{123+}$ \textit{linearly}, since it is the only possible parity-odd invariant, and due to Eq. (\ref{eq:p123square}). Consequently, the $\lambda,\bar\lambda$ content of ${\cal T}^{\rm odd}_{n}$ has been adjusted above. Furthermore, the point-switch symmetry of the correlator is respected by each of the parity-even/odd structures. Owing to Eq. (\ref{eq:cftdilatation}), the 3-point scalar factor fixes the transformation of the full correlator under inversion, whilst also preserving Poincar\'e symmetry. 

When the 3-point correlator contains conserved current(s), it is subject to more constraints due to the current conservation condition Eq. (\ref{eq:conslaw}). It is well known that the conformal dimension $\Delta_{i}$ of a spin-$s_{i}$ conserved current in a $d$-dimensional CFT saturates the unitarity bound, i.e. $\Delta_{i}=s_{i}+d-2$, which is the canonical dimension of the operator. 
For $d=4$, this gives $\Delta_{i}=s_{i}+2$, and thus, $\tau_{i}=2$. On applying the conservation condition at $x_{i}$, and restricting the value of conformal dimension of $J_{s_{i}}(x_{i},\lambda_{i},\bar\lambda_{i})$\footnote{In our notation, the general spin-$s$ conformal primary operator is denoted as $O_{s}$, while the conserved operator is $J_{s}$.} to be canonical, we find relations between various $a_{m},b_{n}$ coefficients, and the form of the correlator is fixed in terms of fewer unknown coefficients. Since our analysis is completely general, one could impose conservation on any number of operators in the 3-point correlator, and obtain the form of the correlator containing (non-)conserved operators as well.

The formalism developed can also be used to write down the 2-point correlators, since the invariant $\hat P_{12}$ is actually constructed out of two augmented coordinate space points. Hence, the 2-point function for a spin-$s$ primary operator $O_{s}$ with conformal dimension $\Delta$ is
\begin{align}
\langle
O_{s}(x_{1},\lambda_{1},\bar\lambda_{1})O_{s}(x_{2},\lambda_{2},\bar\lambda_{2})
\rangle\propto \frac{(\hat P_{12})^{s}}{(x_{12}^{2})^{\Delta-s}}\,,
\end{align}
that is, it is fixed upto an overall coefficient. The numerator $(\hat P_{12})^{s}$ takes care of conformal invariance, and the denominator fixes the transformation under inversion, considering Eq. (\ref{eq:cftscaling}). 

We now carry out a case-by-case analysis for 3-point correlators
\begin{enumerate}[label=$\bullet$]

	\item containing at least one scalar operator
	
	\item containing all spinning operators

\end{enumerate}
The above distinction is made since the scalar operator would amount to having no $\lambda,\bar\lambda$, and a conservation condition in the form of Eq. (\ref{eq:conslaw}) would not apply. Moreover, there is no parity-odd contribution if a scalar operator is present in the correlator, since the only parity-odd invariant $P_{123+}$ contains $\lambda,\bar\lambda$ at all three points. We present a few examples for illustration, and a comprehensive list of results is relegated to the Appendix \ref{appsec:cftconservation}.

\subsubsection*{$\langle O_{1}O_{1}O_{0}\rangle$}

This correlator contains two identical spin-1 operators along with a scalar operator. It has a homogeneity $\lambda_{1}\lambda_{2}\bar\lambda_{1}\bar\lambda_{2}$, and exhibits point-switch symmetry under $1\leftrightarrow2$ swap. The possible conformally invariant structures with this homogeneity in $\lambda,\bar\lambda$ are
\begin{align*}
Q_{1}Q_{2}\,,\ \hat P_{12}\,.
\end{align*}
Both of these structures are symmetric under $1\leftrightarrow 2$ swap, and are linearly independent. 

Thus, the  correlator takes the form
\begin{align}
\langle O_{1}O_{1}O_{0}\rangle=
\frac{1}{|x_{12}|^{\tau_{12,3}}|x_{23}|
^{\tau_{23,1}}|x_{31}|^{\tau_{31,2}}}\left(a_{1} Q_{1}Q_{2}+a_{2}\hat P_{12}\right)\,.
\end{align}

On imposing conservation on \textit{either} of the two spin-1 operators, we obtain (equivalent) relations between $a_{1},a_{2}$,
\begin{align}
\text{conservation at $x_{1}$:}&\qquad
a_{2}=\tfrac{2(\Delta_{2}-\Delta_{3})}{-3+\Delta_{2}-\Delta_{3}}a_{1}\,,\\[.5em]
\text{conservation at $x_{2}$:}&\qquad
a_{2}=\tfrac{2(\Delta_{1}-\Delta_{3})}{-3+\Delta_{1}-\Delta_{3}}a_{1}\,
\end{align}
where $\Delta_{3}$ is the conformal dimension of the scalar operator at $x_{3}$.

Finally, the form of the correlator when both the spin-1 operators are conserved is completely fixed upto an overall (omitted) coefficient and depends on the dimension of the scalar operator,
\begin{align}
\langle J_{1}J_{1}O_{0}\rangle=\frac{1}{|x_{12}|^{4-\Delta_{3}}|x_{23}|^{\Delta_{3}}|x_{31}|^{\Delta_{3}}}
\left(
Q_{1}Q_{2}+\tfrac{2(-3+\Delta_{3})}{\Delta_{3}}\hat P_{12}
\right)\,.
\end{align}

\subsubsection*{$\langle O_{2}O_{2}O_{0}\rangle$}

The correlator is symmetric under a $1\leftrightarrow2$ swap. The linearly independent structures that preserve this symmetry are
\begin{align*}
Q_{1}^{2}Q_{2}^{2}\,,\ 
Q_{1}Q_{2}\hat P_{12}\,,\ 
\hat P_{12}^{2}\,.
\end{align*}
The non-conserved correlator is thus fixed upto 3 parity-even terms,
\begin{align}
\langle O_{2}O_{2}O_{0}\rangle=
\frac{1}{|x_{12}|^{\tau_{12,3}}|x_{23}|
^{\tau_{23,1}}|x_{31}|^{\tau_{31,2}}}\left(
a_{1}Q_{1}^{2}Q_{2}^{2}+
a_{2}Q_{1}Q_{2}\hat P_{12}+
a_{3}\hat P_{12}^{2}
\right)\,.
\end{align}
The analysis of employing the conservation condition follows that of $\langle O_{1}O_{1}O_{0}\rangle$. Imposing conservation on either of the two spin-2 operators gives equivalent relations,
\begin{align}
\begin{aligned}
\text{conservation at $x_{1}$: }&\quad
a_{2}=\tfrac{2(2+3\Delta_{2}-3\Delta_{3})}{-6+\Delta_{2}-\Delta_{3}}a_{1}\,,\quad 
a_{3}=\tfrac{2(-8+3\Delta_{2}^{2}-6\Delta_{2}\Delta_{3}+3\Delta_{3}^{2})}{(-6+\Delta_{2}-\Delta_{3})(-4+\Delta_{2}-\Delta_{3})}a_{1}\,,\\
\text{conservation at $x_{2}$: }&\quad
a_{2}=\tfrac{2(2+3\Delta_{1}-3\Delta_{3})}{-6+\Delta_{1}-\Delta_{3}}a_{1}\,,\quad 
a_{3}=\tfrac{2(-8+3\Delta_{1}^{2}-6\Delta_{1}\Delta_{3}+3\Delta_{3}^{2})}{(-6+\Delta_{1}-\Delta_{3})(-4+\Delta_{1}-\Delta_{3})}a_{1}\,,
\end{aligned}
\end{align}
If both the spin-2 operators are conserved, the correlator is fixed upto an overall coefficient, and depends only on the conformal dimension $\Delta_{3}$ of the scalar,
\begin{align}
\langle J_{2}J_{2}O_{0}\rangle=
\frac{1}{|x_{12}|^{4-\Delta_{3}}|x_{23}|^{\Delta_{3}}|x_{31}|^{\Delta_{3}}}
\left(
Q_{1}^{2}Q_{2}^{2}+\tfrac{2(-14+3\Delta_{3})}{2+\Delta_{3}} Q_{1}Q_{2}\hat P_{12}+ 
\tfrac{2(40-24 \Delta_{3}+3\Delta_{3}^{2})}{\Delta_{3}(2+\Delta_{3})}\hat P_{12}^{2}
\right)\,.
\end{align}

\subsubsection*{$\langle O_{1}O_{1}O_{1}\rangle$}

This is the 3-point correlator of three identical spin-1 operators. This is also the first example where the parity-odd invariant $P_{123+}$ is allowed. The possible parity-even terms are all the ones listed in the relation Eq. (\ref{eq:GPY4d}). However, there is complete symmetry under exchange of any pair of operators $1\leftrightarrow2\leftrightarrow3$, which is not preserved by any of the parity-even structures. The only allowed invariant structure is parity-odd, and the non-conserved correlator is
\begin{align}
\langle O_{1}O_{1}O_{1}\rangle=
\frac{1}{|x_{12}|^{\tau_{12,3}}|x_{23}|
^{\tau_{23,1}}|x_{31}|^{\tau_{31,2}}}\,P_{123+}\,.
\end{align}

The above correlator is trivially conserved for each of the operators, i.e. imposing the conservation condition on any $x_{i}$ yields zero without any constraints, hence the conserved correlator is fixed upto a single parity-odd structure,
\begin{align}
\langle J_{1}J_{1}J_{1}\rangle=
\frac{1}{x_{12}^{2}x_{23}^{2}x_{31}^{2}}
P_{123+}\,.
\end{align}

This is an important result and was first computed in \cite{Schreier1971}.

\subsubsection*{$\langle O_{2}O_{2}O_{2} \rangle$}

This is the 3-point correlator of the spin-2 energy-momentum tensor. This correlator is also completely symmetric under exchange of $1\leftrightarrow2\leftrightarrow3$. There is no parity-odd structure that preserves this symmetry. The correlator with the allowed linearly independent parity-even terms is
\begin{align}
\begin{aligned}
\langle O_{2}O_{2}O_{2} \rangle=\frac{1}{|x_{12}|^{\tau_{12,3}}|x_{23}|
^{\tau_{23,1}}|x_{31}|^{\tau_{31,2}}}
\left[
a_{1}Q_{1}^{2}Q_{2}^{2}Q_{3}^{2}+
a_{2}\sum_{\rm cyc}Q_{1}^{2}Q_{2}Q_{3}\hat P_{23}+
a_{3}\sum_{\rm cyc}Q_{1}^{2}\hat P_{23}^{2}\right.+&\\
\left.
a_{4}\sum_{\rm cyc}Q_{1}Q_{2}\hat P_{23}\hat P_{31}+
a_{5}\hat P_{12}\hat P_{23}\hat P_{31}
\right]\,.&
\end{aligned}
\end{align}
When \textit{any one} of the operators are conserved, we get equivalent relations, fixing the correlator in terms of 3 undetermined coefficients, while the conformal dimension of the other two operators are constrained to be arbitrary but equal, i.e.
\begin{align}
\text{conservation at $x_{1}$:}\qquad \Delta_{3}=\Delta_{2}\,,\qquad
a_{4}=a_{1}-a_{2}+3a_{3}\,,\ a_{5}=\frac13(-3a_{1}+5a_{2}-a_{3})\,.
\end{align}
and similarly for $x_{2},x_{3}$.

Thus, the fully conserved correlator $\langle J_{2}J_{2}J_{2}\rangle$ is fixed in terms of 3 undetermined coefficients,
\begin{align}
\Delta_{1}=\Delta_{2}=\Delta_{3}=4\,,\qquad a_{4}=a_{1}-a_{2}+3a_{3}\,,\ a_{5}=\frac13(-3a_{1}+5a_{2}-a_{3})\,.
\end{align}

Thus we obtain the result that the 3-point correlator of the energy-momentum tensor in $4d$ CFT contains contributions from the free boson, free fermion and free vector theories. This was first shown in \cite{Stanev1988}.

For further results, the reader is referred to Appendix \ref{appsec:cftconservation}.

\section{Superspace building blocks}
\label{sec:scftbb}

The last section focussed on understanding the auxiliary spinor formalism for $4d$ CFT. We constructed conformally invariant structures, and computed the form of various 3-point correlators, and in doing so, verified some results in the existing literature and set the stage for the SCFT analysis. In this and the subsequent sections, we will study 4-dimensional ${\cal N}=1$ SCFTs using superspace methods and  polarization spinors. Like the CFT case, the analogous superinversion (the supersymmetric counterpart of inversion) operation will be important to build superspace invariants.

Superconformal symmetry is realized as the invariance under super-Poincar\'e (supertranslations$+$Lorentz) transformations, dilatations, $R$-symmetry and special superconformal transformations. For $4d$ ${\cal N}=1$ SCFT, the superconformal group is $SU(2,2|1)$, and there are 4 Poincar\'e (${\cal Q},\bar{\cal Q}$), and 4 conformal (${\cal S},\bar{\cal S}$) supercharges. The Lorentz group is $SO(3,1)$ and as done for the conformal theory, we use its $Spin$ group realization ($SO(3,1)\cong SL(2,\mathbb{C})$) and work throughout with bispinor matrices. The conventions for the Lorentz group are listed in Appendix \ref{appssec:4dconv-cft}.

The formalism that we adopt for the superconformal theory is relatively similar to the non-supersymmetric case. One obvious change is the addition of grassmanian coordinates $\theta,\bar\theta$ (see below), which will complicate the construction of superconformal invariants. Analogous to the conformal case, special superconformal transformations can be written as a composition ${\cal S}={\cal I}\cdot {\cal Q}\cdot {\cal I}$, where ${\cal I}$ denotes superinversion. In this section, making liberal use of the superspace formalism, we construct 2-point and 3-point superconformal building blocks that transform homogeneously under superconformal transformations and superinversion. 
All of these covariant structures were first constructed by Osborn \cite{Osborn4dN1SCFT1998} and Park \cite{Park4dSCFT1999} by considering a supermatrix realization of the superconformal group. 

The $4d$ Minkowski superspace is denoted as $\mathbb{R}^{4|4}$ with standard  superspace coordinates  $z^{A}=\{x^{\mu},\theta^{\alpha},\bar \theta_{\dot \alpha}\}$, where $\mu=0,1,2,3$, $\alpha=1,2$, $\dot\alpha=\bar 1,\bar 2$, and $\theta,\bar\theta$ are grassmanian left/right-handed Weyl spinors with $(\theta^{\alpha})^{*}=\bar \theta^{\dot\alpha}$. We decompose the superspace coordinates into chiral-antichiral coordinates, $z_{+}=(x^{\mu}_{+},\theta^{\alpha})$, $z_{-}=(x^{\mu}_{-},\bar\theta_{\dot\alpha})$, where $x_{i\pm}^{\mu}=x_{i}^{\mu}\pm i \theta_{i}\sigma^{\mu}\bar\theta_{i}$. In spinor notation,
\begin{align}
(X_{i\pm})_{\alpha\dot\alpha}=(X_{i})_{\alpha\dot\alpha}\mp 2i \theta_{i\alpha}\bar\theta_{i\dot\alpha}\,,\qquad 
(\tilde X_{i\pm})^{\dot\alpha\alpha}=(\tilde X_{i})^{\dot\alpha\alpha} \pm 2i \bar\theta_{i}^{\dot\alpha}\theta_{i}^{\alpha}\,,
\end{align}
such that $X_{i\pm}^{\dagger}=X_{i\mp}$. As we will see, this superspace decomposition ties in neatly with our formalism, since superconformal transformations leave the (anti-)chiral subspaces invariant, while superinversion transforms them into each other.

We also have the chiral-antichiral supercovariant derivatives
\begin{align}
D_{\alpha}=\frac{\partial}{\partial \theta^{\alpha}}+i\sigma^{\mu}_{\alpha \dot\alpha} \bar \theta^{\dot \alpha}\frac{\partial}{\partial x^{\mu}},\qquad 
\bar D_{\dot \alpha}=-\frac{\partial}{\partial \bar\theta^{\dot\alpha}} - i \theta^{ \alpha}\sigma^{\mu}_{\alpha \dot\alpha}\frac{\partial}{\partial x^{\mu}}\,,
\label{eq:supercovderiv}
\end{align}
such that they respectively annihilate the antichiral-chiral subspaces, $D_{\alpha}z_{-}=0=\bar D_{\dot\alpha}z_{+}$. These derivatives will be employed to implement the conservation constraint on superfield operators in Section \ref{sec:scftconscorr}.

Note that we can make Lorentz scalars $\tilde X_{i\pm}\cdot X_{\pm}=-x_{i\pm}^{2}\mathds{1}= X_{i\pm}\cdot \tilde X_{i\pm}$, where
\begin{align}
x_{i\pm}^{2}=x_{i}^{2}-2\theta_{i}^2\bar\theta_{i}^2\pm 2i\theta_{i}^{\alpha}(X_{i})_{\alpha\dot\alpha}\bar\theta_{i}^{\dot\alpha}\,.
\end{align}
Using the above, one defines the inverses
\begin{align}
(X_{i\pm}^{-1})^{\dot\alpha\alpha}=-\frac{1}{x_{i\pm}^{2}}(\tilde X_{i\pm})^{\dot\alpha\alpha},\qquad
(\tilde X_{i\pm}^{-1})_{\alpha\dot\alpha}=-\frac{1}{x_{i\pm}^{2}}(X_{i\pm})_{\alpha\dot\alpha}\,.
\label{eq:scft1ptinv}
\end{align}
For convenience, $X_{i\pm},X_{i\pm}^{-1},\tilde X_{i\pm},\tilde X_{i\pm}^{-1}$ are henceforth referred to as superconformal 1-point objects.

\subsection{2-point building blocks}
\label{ssec:scftbb-2pt}

We consider 2-point structures that are invariant under supertranslations \cite{Osborn4dN1SCFT1998, Park4dSCFT1999}
\begin{align}
x^{\mu}_{\bar \imath j}=x_{i-}^{\mu}-x_{j+}^{\mu}+2i\theta_{j}\sigma^{\mu}\bar \theta_{i}=-x^{\mu}_{j\bar \imath},\qquad \theta_{ij}^{\alpha}=\theta_{i}^{\alpha}-\theta_{j}^{\alpha},\quad \bar\theta_{ij}^{\dot\alpha}=\bar\theta_{i}^{\dot\alpha}-\bar\theta_{j}^{\dot\alpha}.
\end{align}
The notation means that $x_{\bar \imath j}^{\mu}$ is antichiral in  $z_{i}$ and chiral in $z_{j}$. 

The spinor counterparts of these structures will serve as the 2-point building blocks of our superconformal invariants,
\begin{gather}
(\tilde X_{\bar \imath j})^{\dot\alpha\alpha}=x^{\mu}_{\bar \imath j}(\tilde\sigma_{\mu})^{\dot\alpha\alpha}=(\tilde X_{i-})^{\dot\alpha\alpha}-(\tilde X_{j+})^{\dot\alpha\alpha}+4i\bar \theta_{i}^{\dot\alpha}\theta_{j}^{\alpha}\,,\\
(X_{j\bar \imath})_{\alpha\dot\alpha}=x^{\mu}_{ j\bar \imath}(\sigma_{\mu})_{\alpha\dot\alpha}= (X_{j+})_{\alpha\dot\alpha}-( X_{i-})_{\alpha\dot\alpha}+4i \theta_{j\alpha}\bar\theta_{i\dot\alpha}\,,
\end{gather}
where $X_{j\bar \imath}$ is defined such that $(X_{j\bar \imath})_{\alpha\dot\alpha}=-\epsilon_{\alpha\beta}\epsilon_{\dot\alpha\dot\beta}(\tilde X_{\bar \imath j})^{\dot\beta\beta}$, and $\tilde X_{\bar \imath j}^{\dagger}=-\tilde X_{\bar \jmath i}$. 

Also, one defines \cite{Osborn4dN1SCFT1998}
\begin{align}
x^{\mu}_{\bar \imath j}=y^{\mu}_{ij}-i\theta_{ij}\sigma^{\mu}\bar\theta_{ij}\,,\quad \text{where}\quad 
y^{\mu}_{ij}=x^{\mu}_{i}-x^{\mu}_{j}-i\theta_{i}\sigma^{\mu}\bar\theta_{j}+i\theta_{j}\sigma^{\mu}\bar\theta_{i}=-y_{ji}^{\mu}\,.
\label{eq:scfty12}
\end{align}
This helps us define the Poincar\'e scalars $\tilde X_{\bar \imath j}\cdot X_{j\bar \imath}= x_{\bar \imath j}^{2}\mathds{1}=x_{j \bar \imath}^{2}\mathds{1}$, which gives us the inverses
\begin{align}
(\tilde X_{\bar \imath j}^{-1})_{\alpha\dot\alpha}=\frac{1}{x_{\bar \imath j}^{2}}(X_{j\bar \imath})_{\alpha\dot\alpha}, \quad 
(X_{i\bar \jmath}^{-1})^{\dot\alpha\alpha}=\frac{1}{x_{i\bar \jmath}^{2}}(\tilde X_{\bar \jmath i})^{\dot\alpha\alpha}\,.
\label{eq:2ptinv}
\end{align}
As \cite{Osborn4dN1SCFT1998} and \cite{Park4dSCFT1999} show, these building blocks transform homogeneously under infinitesimal superconformal transformations. For finite transformations, we will study their transformations under superinversion (in Section \ref{ssec:scftbb-superinv}).

\subsection{3-point building blocks}
\label{ssec:scftbb-3pt}

Similarly, we consider structures made from three superspace points \cite{Osborn4dN1SCFT1998, Park4dSCFT1999},
\begin{align}
{\frak X}_{\bar 2 3}=\tilde X_{\bar 2 1}^{-1}\tilde X_{\bar 2 3}\tilde X_{\bar 13}^{-1}\,,\qquad 
{\frak X}_{\bar 3 2}=\tilde X_{\bar 3 1}^{-1}\tilde X_{\bar 3 2}\tilde X_{\bar 1 2}^{-1}=-{\frak X}_{\bar 23}^{\dagger}\,,
\label{eq:3ptblocks}
\end{align}
and their permutations. Recall, $\tilde X_{\bar \imath j}=-\tilde X_{\bar \jmath i}^{\dagger}$.\footnote{We are following the index notation $\tilde A \sim \tilde A^{\dot\alpha\alpha},\ A \sim A_{\alpha\dot\alpha}$.}

Equivalently, we have
\begin{align}
\tilde{\frak X}_{3\bar 2}=X_{3\bar 1}^{-1}X_{3\bar 2}X_{1\bar 2}^{-1},\qquad
\tilde {\frak X}_{2\bar 3}=X_{2\bar 1}^{-1}X_{2\bar 3}X_{1\bar 3}^{-1}=-\tilde{\frak X}_{3\bar 2}^{\dagger}\,.
\label{eq:3ptblockss}
\end{align}

We can also define grassmann-valued (anti-commuting) 3-point structures
\begin{align}
\Theta_{1\alpha}=i
\left(
	(\tilde X_{\bar21}^{-1}\bar\theta_{21})_{\alpha}-
(\tilde X_{\bar31}^{-1}\bar\theta_{31})_{\alpha}
\right)\,,&\qquad
\Theta_{1}^{\alpha}=\epsilon^{\alpha\beta}\Theta_{1\beta}=-i
\left(
	(\bar\theta_{12}X^{-1}_{1\bar2})^{\alpha}-(\bar\theta_{13}X^{-1}_{1\bar3})^{\alpha}
\right)\,,\\
\bar\Theta_{1}^{\dot\alpha}=i
\left(
	(X_{2\bar1}^{-1}\theta_{21})^{\dot\alpha}-(X_{3\bar1}^{-1}\theta_{31})^{\dot\alpha}
\right)\,,&\qquad
\bar\Theta_{1\dot\alpha}=\epsilon_{\dot\alpha\dot\beta}\bar\Theta_{1}^{\dot\beta}=-i
\left(
	(\theta_{12}\tilde X_{\bar12}^{-1})_{\dot\alpha}-(\theta_{13}\tilde X_{\bar13}^{-1})_{\dot\alpha}
\right)\,,
\end{align}
and their permutations. Just like the 2-point blocks, all of these 3-point structures transform homogeneously under infinitesimal superconformal transformations \cite{Osborn4dN1SCFT1998, Park4dSCFT1999}.

Note that the 3-point bosonic and grassmanian covariant structures are related,
\begin{align}
(\frak X_{\bar23}+\frak X_{\bar32})_{\alpha\dot\alpha}=-4i\Theta_{1\alpha} \bar\Theta_{1\dot\alpha}\,,\quad
(\tilde{\frak X}_{2\bar3}+\tilde{\frak X}_{3\bar2})^{\dot\alpha\alpha}=-4i\bar\Theta_{1}^{\dot\alpha}\Theta_{1}^{\alpha}\,,\quad \text{and perm.}
\end{align}

One can also define the scalar objects
\begin{align}
\frak X_{1}^{2}=\frac12{\rm tr}(\frak X_{\bar23}\tilde{\frak X}_{3\bar2})=\frac{x_{\bar23}^{2}}{x_{\bar21}^{2}x_{\bar13}^{2}}\,,\qquad
\frak X_{1}^{\dagger 2}=\frac12{\rm tr}(\frak X_{\bar32}\tilde{\frak X}_{2\bar3})=\frac{x_{\bar32}^{2}}{x_{\bar31}^{2}x_{\bar12}^{2}}\,,\quad \text{and perm.}
\label{eq:frakXsq}
\end{align}

\subsection{Superinversion}
\label{ssec:scftbb-superinv}

As mentioned before, special superconformal transformations require a composition of superinversion and supertranslations. Consequently, in this subsection, we analyze the transformations of 2-point and 3-point building blocks under superinversion. This will be subsequently used to construct superconformal invariants in the next section.

The superinversion transformation of various 1-point superspace structures is \cite{Osborn4dN1SCFT1998, Park4dSCFT1999, BuchKuzSUGRA1998}
\begin{align}
x^{\mu }_{\pm}\rightarrow
\frac{x^{\mu}_{\mp}}{x_{\mp}^{2}},
\qquad
\theta^{\alpha}\rightarrow 
-i\frac{x^{\mu}_{-}(\bar\theta\tilde\sigma_{\mu})^{\alpha}}{x_{-}^{2}}\,.
\end{align}
Since superinversion is idempotent, one can use the above to obtain the transformations of all superspace coordinates and their spinor counterparts,
\begin{gather}
\begin{gathered}
\tilde X_{\pm}\rightarrow - X_{\mp}^{-1}\,,\quad
X_{\pm}\rightarrow - \tilde X_{\mp}^{-1}\,,\quad
X_{\pm}^{-1}\rightarrow -\tilde X_{\mp}\,,\quad 
\tilde X_{\pm}^{-1}\rightarrow -X_{\mp}\,,\qquad 
x_{\pm}^{2}\rightarrow \frac{1}{x_{\mp}^{2}}\\
\theta^{\alpha}\rightarrow i(\bar\theta X_{-}^{-1})^{\alpha}\,,\quad
\theta_{\alpha}\rightarrow
-i(\tilde X_{-}^{-1}\bar\theta)_{\alpha},\quad
\bar\theta^{\dot\alpha}\rightarrow
-i (X_{+}^{-1}\theta)^{\dot\alpha}\,,\quad
\bar\theta_{\dot\alpha}\rightarrow
i(\theta \tilde X_{+}^{-1})_{\dot\alpha}\,.
\end{gathered}
\label{eq:1ptsuperinv}
\end{gather}

This leads to
\begin{align}
\begin{gathered}
\tilde X_{\bar \imath j}\rightarrow
X_{i+}^{-1}\,X_{i\bar \jmath}\,X_{j-}^{-1}\,,
\qquad
X_{i\bar \jmath}\rightarrow
\tilde X_{i-}^{-1}\,\tilde X_{\bar \imath j}\,\tilde X^{-1}_{j+}\,,
\\[.5em	]
\tilde X_{\bar \imath j}^{-1}\rightarrow X_{j-} X_{i\bar \jmath}^{-1}X_{i+}\,,
\qquad
X_{i\bar \jmath}^{-1}\rightarrow \tilde X_{j+} \tilde X_{\bar \imath j}^{-1}\tilde X_{i-}\,,
\end{gathered}\qquad\qquad
x^{2}_{\bar \imath j}\rightarrow \frac{x^{2}_{ i \bar \jmath}}{x_{i+}^{2}x_{j-}^{2}}\,,
\label{eq:2ptsuperinv}
\end{align}
which is the superinversion of the 2-point building blocks. 

Eq. (\ref{eq:2ptsuperinv}) suggests that the property of `homogeneity upto conjugation' (refer Section \ref{ssec:cftreview-bb}) is repeated for the supersymmetric case. Since the superspace is decomposed into chiral and anti-chiral subspaces, superinversion truly transforms each subspace into its conjugate. The 2-point blocks transform homogeneously upto conjugation under superinversion, but homogeneously under the full superconformal group (conjugation disappears since superinversion is applied twice). Given Eq. (\ref{eq:2ptsuperinv}), one easily determines the superinversion of the bosonic 3-point building blocks as well,
\begin{align}
&{\frak X}_{\bar 23}\rightarrow X_{1-}\,\tilde{\frak X}_{2\bar 3}\,X_{1+}\,,\qquad 
\frak X_{\bar 32}\rightarrow  X_{1-}\,\tilde{\frak X}_{3\bar 2}\,X_{1+}\,,\\[.5em]
&\tilde{\frak X}_{3\bar 2}\rightarrow \tilde X_{1+}\,\frak X_{\bar 32}\, \tilde X_{1-}\,,\qquad
\tilde{\frak X}_{2\bar 3}\rightarrow \tilde X_{1+}\,\frak X_{\bar 23}\, \tilde X_{1-}\,,
\label{eq:3ptsuperinv}
\end{align}
and permutations.

While the superinversion of the grassmanian 3-point blocks $\Theta,\bar \Theta$ is also straightforward since we know the transformation of all of its constituents, writing it in a covariant form is not so simple. This happens to be the case as $\Theta,\bar\Theta$ are charged under the $U(1)$ $R$-symmetry transformations. To this end, we define \cite{Park4dSCFT1999}
\begin{align}
V_{i}= 1- 4 i \theta_{i}^{\alpha} (\tilde X_{i+}^{-1})_{\alpha\dot\alpha}\bar\theta_{i}^{\dot\alpha}\,,
\label{eq:Vterm}
\end{align}	
and the inverse
\begin{align}
V_{i}^{-1}= 1+ 4 i \theta_{i}^{\alpha} (\tilde X_{i-}^{-1})_{\alpha\dot\alpha}\bar\theta_{i}^{\dot\alpha}\,,
\label{eq:Vterminv}
\end{align}
such that $V_{i}\cdot V_{i}^{-1}=1$. The $V$'s represent the $R$-symmetry part of the superconformal transformations, and are made from the 1-point objects.\footnote{For ${\cal N}\geq 2$ extended supersymmetry, $V_{i}$ is represented by an ${\cal N}\times{\cal N}$ matrix.}

The $V$'s facilitate the superinversion of $\Theta,\bar\Theta$ in a covariant form,
\begin{align}
\begin{aligned}
\Theta_{1\alpha}\rightarrow -i (X_{1-})_{\alpha\dot\alpha}\bar\Theta_{1}^{\dot\alpha}\,V_{1}\,,&\qquad
\bar\Theta_{1}^{\dot\alpha}\rightarrow -i(\tilde X_{1+})^{\dot\alpha\alpha}\Theta_{1\alpha}\,V_{1}^{-1}\,,\\
\Theta_{1}^{\alpha}\rightarrow i V_{1}\,\bar \Theta_{1\dot\alpha}(\tilde X_{1-})^{\dot\alpha\alpha}\,,&\qquad
\bar\Theta_{1\dot\alpha}\rightarrow iV_{1}^{-1}\,\Theta_{1}^{\alpha}(X_{1+})_{\alpha\dot\alpha}\,,
\end{aligned}
\label{eq:Thetasuperinv}
\end{align}
and permutations. It is also easy to check that $V,V^{-1}$ are invariant under superinversion. 

Finally, we also get the superinversion transformation of the scalars defined in Eq. (\ref{eq:frakXsq}),
\begin{align}
\frak X_{i}^{2}\rightarrow x_{i+}^{2}x_{i-}^{2} \frak X^{\dagger 2}_{i}
\,, \qquad
\frak X^{\dagger 2}_{i}\rightarrow  x_{i+}^{2}x_{i-}^{2} \frak X_{i}^{2}
\,.
\end{align}
Due to this behavior under superinversion, the scalars $\frak X^{2}_{i},\frak X^{\dagger2}_{i}$ will help us normalize the superconformal covariants to give us the desired invariant structures in Section \ref{sec:scftinv}.

\subsection{Polarization spinors}
\label{ssec:scftbb-polars}

Analogous to the conformal case, we will utilize the auxiliary commuting polarization spinors $\lambda,\bar\lambda$ to contract the Lorentz covariant 2-point and 3-point building blocks derived above to build superconformal invariants, and to write the spinning superfield operators in an index-free form.

We augment the superspace with commuting auxiliary polarization spinors $\lambda,\bar\lambda$,
\begin{align}
\check z_{i}=\{z_{i},\lambda^{\beta}_{i},\bar\lambda_{i}^{\dot\beta}\}=\{x_{i}^{\mu},\theta_{i}^{\alpha},\bar\theta_{i}^{\dot\alpha},\lambda_{i}^{\beta},\bar\lambda^{\dot\beta}_{i}\}\,,\qquad
\begin{tabular}{l}
\small$\mu=0,1,2,3$\\[.5em]
\small$\alpha,\dot\alpha,\beta,\dot\beta=1,2$
\end{tabular}
\end{align}
where $\check z_{i}$ denotes the $i^{\rm th}$ augmented superspace point.

The polarization spinors are commuting left/right-handed Weyl spinors, and thus have a similar behavior under superinversion as the $\theta,\bar\theta$,
\begin{align}
\lambda^{\alpha}\rightarrow i \bar\lambda_{\dot\alpha}(X_{-}^{-1})^{\dot\alpha\alpha}\,,\quad
\lambda_{\alpha}\rightarrow -i (\tilde X_{-}^{-1})_{\alpha\dot\alpha}\bar\lambda^{\dot\alpha}\,,\qquad 
\bar\lambda^{\dot\alpha}\rightarrow -i (X_{+}^{-1})^{\dot\alpha\alpha}\lambda_{\alpha}\,,\quad
\bar\lambda_{\dot\alpha}\rightarrow i \lambda^{\alpha}(\tilde X_{+}^{-1})_{\alpha\dot\alpha}\,.
\label{eq:lamsuperinv}
\end{align}
Note that the non-supersymmetric version of these transformations have the same form as conformal inversion of $\lambda,\bar\lambda$ in Eq. (\ref{eq:laminversion}).

To work with index-free superfield operators, we consider a spinning primary superfield ${\cal O}^{A}(z)$ at superspace point $z$, where $A$ denotes vector or spinor indices. Recall that a superfield operator is a superconformal multiplet labeled by $j,\bar\jmath$ and $q,\bar q$, such that the lowest superconformal primary in the multiplet transforms in the $(j/2,\bar\jmath/2)$ rep of the $4d$ Lorentz group $SL(2,\mathbb{C})$, and has a scaling dimension $\Delta=q+\bar q$ and $R$-charge proportional to $q-\bar q$. The spin-$s$ superconformal multiplet in $4d$ ${\cal N}=1$ SCFT contains conformal primaries of spins $\{s,s+1/2,s+1\}$. Since we will only consider bosonic superfield operators with integer spin $s$ and zero $R$-charge, we have
\begin{align}
q=\bar q\,,\quad j=\bar\jmath\,.
\end{align}
Thus, we can write the superfield operator in terms of spinor indices,
\begin{align}
{\cal O}^{A}(z)={\cal O}^{\alpha(s)\dot\alpha(s)}(z)={\cal O}^{(\alpha_{1}\hdots \alpha_{s})(\dot\alpha_{1}\hdots\dot\alpha_{s})}\,,
\end{align}
where $\alpha(s),\dot\alpha(s)$ denote symmetric sequences of $s$ spinor indices each.

Following Eq. (\ref{eq:indexfree}), we can write the spinning superfield operator in a index-free form,
\begin{align}
{\cal O}_{s}(\check z)=\lambda_{\alpha_{1}}\hdots\lambda_{\alpha_{s}}\ 
{\cal O}^{(\alpha_{1}\hdots \alpha_{s})(\dot\alpha_{1}\hdots\dot\alpha_{s})}
\bar\lambda_{\dot\alpha_{1}}\hdots\bar\lambda_{\dot\alpha_{s}}\,, 
\end{align}
where $\check z$ is the augmented superspace point. It is apparent that the value of spin $s$ of the operator is encoded in the number of $\lambda,\bar\lambda$ in ${\cal O}_{s}$.

\section{Superconformal invariants}
\label{sec:scftinv}

In this section, we construct independent superconformal and superinversion invariants out of three superspace points, that will suffice for all 3-point correlators (of integer-spin superfield operators) in $4d$ ${\cal N}=1$ SCFTs. It will become evident in this section that the construction of superconformal invariants in $4d$ SCFTs has some unique features when compared to $3d$ SCFTs \cite{Nizami2013, JainNizami2022}. While superconformally invariant objects can be constructed aplenty through the building blocks we obtained in the last section, the important task at hand is to determine a minimal set of independent invariants that would serve as a complete basis for expressing 3-point spinning correlators.

In Section \ref{ssec:scftinv-cons}, we use the same strategy as employed for the conformal case in Section \ref{ssec:cftreview-inv}, appropriately calibrated to include the grassmanian building blocks. The approach is as follows: the 2-point and 3-point building blocks are invariant under supertranslations, and can be made super-Poincar\'e invariant by dotting with the polarization spinors $\lambda,\bar\lambda$, and/or the grassmanian blocks $\Theta,\bar\Theta$. These objects have a definite scaling weight. For invariance under special superconformal and $R$-symmetry transformations, we can simply analyse their transformation under superinversion. Structures (made out of the building blocks) transforming invariantly under superinversion will automatically turn out to be superconformally invariant. Interestingly, using $\Theta,\bar\Theta$, not only can we construct purely grassmanian superconformal invariants that have no conformal counterparts, we can also obtain invariants which have no $\lambda,\bar\lambda$, and hence can be used to express correlators with scalar superfields.

As it stands, the construction strategy described above will turn out to be insufficient in building all the possible superconformally invariant structures. This will be manifest when we look at superconformal extensions to relations between conformal invariants in Section \ref{ssec:scftinv-grass}. The analysis there will lead us to novel superconformal invariants. We also exhaustively study the relations between all possible (products of) invariants, which is summarized in Section \ref{ssec:scftinv-relation}. Finally, we enumerate the minimal list of independent superconformal invariants, and any non-linear relations between them. The section is closed by considering the permutation symmetry of these invariants under exchange of augmented superspace points. This is of relevance when considering correlators containing two or more identical superfield operators.

It is also worth noting that the behavior of the superconformal invariants under a parity transformation is dictated by their behavior under superinversion.\footnote{A detailed discussion can be found for superinversion in \cite{Buch4dSCFTGen2024}, and for inversion in \cite{Buch4dCFTGen2023}.} Invariants transforming with a $+/-$ sign under superinversion also transform with a $+/-$ sign under parity, and are hence classified as parity-even and parity-odd, exactly like the CFT case (refer Section \ref{ssec:cftreview-inv}). This classification will help us distinguish parity-preserving (even) and parity-violating (odd) structures present in a general 3-point correlator in Section \ref{sec:scftcorrstr}. We will observe in Section \ref{sec:scftconscorr} that the parity-even (or parity-odd) contributions to a correlator can vanish based on the permutation symmetry of the correlator.

\subsection{Construction via building blocks}
\label{ssec:scftinv-cons}

Considering the superinversion of the building blocks in Eqs. (\ref{eq:2ptsuperinv}, \ref{eq:3ptsuperinv}, \ref{eq:Thetasuperinv}), and the polarization spinors in Eq. (\ref{eq:lamsuperinv}), it is straightforward to build structures that will transform covariantly under superinversion. The general strategy for the construction of these covariant structures is to take products (with indices contracted consistently) of the 2-point building blocks, polarization spinors and $\Theta$'s (as well as their conjugates).

The following covariant structures have been constructed systematically by considering products of increasing number of 2-point blocks.\footnote{Note that the 3-point blocks in Eq. (\ref{eq:3ptblocks}, \ref{eq:3ptblockss}) are constructed out of products of 2-point blocks as well.}  Their transformation under superinversion is also given. 
\begin{align}
\label{eq:superconf0}
\begin{tabular}{c}
zero\\
2-point blocks
\end{tabular}:\qquad&
\begin{tabular}{l}
$\psi_{1}=\lambda_{1}^{\alpha}\Theta_{1\alpha}\quad \text{and 2 perm,}$\\[.3em]
$\bar\psi_{1}=\bar\lambda_{1\dot\alpha}\bar\Theta_{1}^{\dot\alpha}\quad \text{and 2 perm,}$\\[0.3em]
$\Psi_{1}=\Theta_{1}^{\alpha}\Theta_{1\alpha}\quad \text{and 2 perm,}$\\[.3em]
$\bar\Psi_{1}=\bar\Theta_{1\dot\alpha}\bar\Theta_{1}^{\dot\alpha}\quad \text{and 2 perm,}$
\end{tabular}&&\quad
\begin{tabular}{l}
$\psi_{1}\rightarrow \bar\psi_{1} V_{1}\,,$\\[.3em]
$\bar\psi_{1}\rightarrow V_{1}^{-1}\psi_{1}\,,$\\[.3em]
$\Psi_{1}\rightarrow -x_{1-}^{2}V_{1}^{2}\bar\Psi_{1}\,,$\\[.3em]
$\bar\Psi_{1}\rightarrow -x_{1+}^{2} V_{1}^{-2}\Psi_{1}\,.$
\end{tabular}\qquad\\[.5em]
\label{eq:superconf1}
\begin{tabular}{c}
one\\
2-point block
\end{tabular}:\qquad&
\begin{tabular}{l}
$P_{12}=\lambda_{1}^{\alpha}\tilde X_{\bar 21\alpha\dot\alpha}^{-1}\bar\lambda_{2}^{\dot\alpha}\quad \text{and 5 perm,}$\\[.3em]
$\pi_{12}=\lambda_{1}^{\alpha}\tilde X_{\bar 21\alpha\dot\alpha}^{-1}\bar\Theta_{2}^{\dot\alpha}\quad \text{and 5 perm,}$\\[.3em]
$\bar \pi_{12}=\bar\lambda_{1\dot\alpha} X_{ j\bar i}^{-1\dot\alpha\alpha}\Theta_{2\alpha}\quad \text{and 5 perm,}$\\[.3em]
$\Pi_{12}=\Theta_{1}^{\alpha}\tilde X_{\bar 21\alpha\dot\alpha}^{-1}\bar\Theta_{2}^{\dot\alpha}\quad \text{and 5 perm,}$
\end{tabular}&&\quad
\begin{tabular}{l}
$P_{12}\rightarrow -P_{21}\,.$\\[.3em]
$\pi_{12}\rightarrow-x_{2+}^{2}\bar\pi_{12} V_{2}^{-1}\,,$\\[.3em]
$\bar\pi_{12}\rightarrow - x_{2-}^{2}\pi_{12}V_{2}\,,$\\[.3em]
$\Pi_{12}\rightarrow x_{1-}^{2}x_{2+}^{2}V_{1}\Pi_{21} V_{2}^{-1}\,.$
\end{tabular}\\[.5em]
\label{eq:superconf2}
\begin{tabular}{c}
two\\
2-point blocks
\end{tabular}:\qquad&
\begin{tabular}{l}
$S_{13}=\lambda_{1}^{\alpha} \tilde X_{\bar 21\alpha\dot\beta}^{-1}\tilde X_{\bar 23}^{\dot\beta\beta}\lambda_{3\beta}\quad \text{and 5 perm,}$\\[.3em]
$\bar S_{13}=\bar\lambda_{1\dot\alpha} X_{ 2\bar 1}^{-1\dot\alpha\beta}X_{2\bar 3\beta\dot\beta}\bar\lambda_{3}^{\dot\beta}\quad \text{and 5 perm,}$\\[.3em]
$\sigma_{13}=\lambda_{1}^{\alpha} \tilde X_{\bar 21\alpha\dot\beta}^{-1}\tilde X_{\bar 23}^{\dot\beta\beta}\Theta_{3\beta}\quad \text{and 5 perm,}$\\[.3em]
$\bar \sigma_{13}=\bar\lambda_{1\dot\alpha} X_{ 2\bar 1}^{-1\dot\alpha\beta}X_{2\bar 3\beta\dot\beta}\bar\Theta_{3}^{\dot\beta}\quad \text{and 5 perm,}$\\[.3em]
$\Sigma_{13}=\Theta_{1}^{\alpha} \tilde X_{\bar 21\alpha\dot\beta}^{-1}\tilde X_{\bar 23}^{\dot\beta\beta}\Theta_{3\beta}\quad \text{and 5 perm,}$\\[.3em]
$\bar \Sigma_{13}=\bar\Theta_{1\dot\alpha} X_{ 2\bar 1}^{-1\dot\alpha\beta}X_{2\bar 3\beta\dot\beta}\bar\Theta_{3}^{\dot\beta}\quad \text{and 5 perm.}$
\end{tabular}&&\quad
\begin{tabular}{l}
$S_{13}\rightarrow -\frac{1}{x_{3-}^{2}}\bar S_{13}\,,$\\[.5em]
$\bar S_{13}\rightarrow -\frac{1}{x_{3+}^{2}}S_{13}\,.$\\[.5em]	
$\sigma_{13}\rightarrow \bar\sigma_{13} V_{3}\,,$\\[.5em]
$\bar\sigma_{13}\rightarrow \sigma_{13}V_{3}^{-1}\,,$\\[.5em]
$\Sigma_{13}\rightarrow -x_{1-}^{2}V_{1} \bar\Sigma_{13} V_{3}\,,$\\[.5em]
$\Sigma_{13}\rightarrow -x_{1+}^{2}V_{1}^{-1} \Sigma_{13} V_{3}^{-1}\,.$
\end{tabular}\\[.5em]
\label{eq:superconf3}
\begin{tabular}{c}
three\\
2-point blocks
\end{tabular}:\qquad&
\begin{tabular}{l}
$Q_{1}=\lambda_{1}^{\alpha}\frak X_{\bar 23\alpha\dot\alpha}\bar\lambda_{1}^{\dot\alpha}\quad \text{and 2 perm,}$\\[.3em]
$\bar Q_{1}=\bar \lambda_{1\dot\alpha}\tilde{\frak X}_{2\bar 3}^{\dot\alpha\alpha}\lambda_{1\alpha}\quad \text{and 2 perm,}$\\[.3em]
$\Omega_{1}=\Theta_{1}^{\alpha}\frak X_{\bar 23\alpha\dot\alpha}\bar\Theta_{1}^{\dot\alpha}\quad \text{and 2 perm,}$\\[.3em]
$\bar \Omega_{1}=\bar \Theta_{1\dot\alpha}\tilde{\frak X}_{2\bar 3}^{\dot\alpha\alpha}\Theta_{1\alpha}\quad \text{and 2 perm,}$\\[.3em]
$\omega_{1}=\lambda_{1}^{\alpha}\frak X_{\bar 23\alpha\dot\alpha}\bar\Theta_{1}^{\dot\alpha}\quad \text{and 2 perm,}$\\[.3em]
$\bar \omega_{1}=\bar \lambda_{1\dot\alpha}\tilde{\frak X}_{2\bar 3}^{\dot\alpha\alpha}\Theta_{1\alpha}\quad \text{and 2 perm,}$
\end{tabular}&&\quad
\begin{tabular}{l}
$Q_{1}\rightarrow \bar Q_{1}\,,$\\[.3em]
$\bar Q_{1}\rightarrow Q_{1}\,,$\\[.3em]
$\Omega_{1}\rightarrow x_{1-}^{2}x_{1+}^{2}\bar \Omega_{1}\,,$\\[.3em]
$\bar\Omega_{1}\rightarrow x_{1-}^{2}x_{1+}^{2} \Omega_{1}\,,$\\[.3em]
$\omega_{1}\rightarrow -x_{1-}^{2}\bar \omega_{1}V_{1}^{-1}\,,$\\[.3em]
$\bar\omega_{1}\rightarrow -x_{1+}^{2} \omega_{1}V_{1}\,.$
\end{tabular}\\[.5em]
\label{eq:superconf5}
\begin{tabular}{c}
five\\
2-point blocks
\end{tabular}:\qquad&
\begin{tabular}{l}
$Y_{13}=\lambda_{1}^{\alpha}\frak X_{\bar 23\alpha\dot\beta}\tilde X_{\bar 12}^{\dot\beta\beta}\tilde X_{\bar 3 2\beta\dot\alpha}^{-1}\bar\lambda_{3}^{\dot\alpha}\quad \text{and 5 perm,}$
\end{tabular}&&\quad
\begin{tabular}{l}
$Y_{13}\rightarrow -Y_{31}\,.$
\end{tabular}
\end{align}
Here, if an object is not followed by its conjugate (barred version), then the conjugate is derivable by taking a permutation of the indices. For example, the conjugate of $P_{12}$ is $P_{21}$. Also, we have employed the notation wherein objects containing only $\lambda,\bar\lambda$ are denoted by uppercase Latin alphabets, objects with one $\Theta$ or $\bar\Theta$ by lowercase Greek, and objects with two $\Theta,\bar\Theta$ by uppercase Greek letters.

Note that the table presented above is not completely exhaustive. For instance, in (\ref{eq:superconf5}) one can build more structures using $\Theta,\bar\Theta$. Also, we have omitted the objects when there are four 2-point blocks involved as well as objects with six, seven,... 2-point blocks in it. The reason for not including them is that such structures are redundant. Below, we will see that only a handful of the objects defined above will suffice to construct all the required superconformal invariants.

We are now ready to construct structures which are \textit{invariant} under superconformal and superinversion transformations via combinations of the covariant objects in (\ref{eq:superconf0}$-$\ref{eq:superconf5}).\footnote{Henceforth, invariance under superinversion is upto a sign.} Superinversion  invariants have three-fold advantages. Firstly, since we are interested in correlators of bosonic integer-spin superfield operators, we require objects invariant under superconformal as well as superinversion transformations. Secondly, the building blocks are super-Poincar\'e invariant, and homogeneous under scaling, which means superinversion invariance would imply superconformal invariance. Lastly, the superinversion invariants have a definite behavior under parity, and can be thus classified as parity-even and parity-odd structures. For brevity, we present only the independent superinversion (and thus, superconformal) invariants: 
\begin{align}
\begin{aligned}
\text{from $\Psi_{i},\bar\Psi_{i}$}:&\qquad
\hat\Psi=\frac{\Psi_{1}\bar\Psi_{1}}{\frak X_{1}\frak X_{1}^{\dagger}}\quad \text{(parity-even),}
\\[.5em]
\text{from $P_{ij}$}:&\qquad\!\!\!
\begin{tabular}{l}
$\hat P_{12}=P_{12}P_{21}\quad \text{and 2 perm (parity-even),}$\\[.3em]
$P_{123-}=P_{12}P_{23}P_{31}-P_{13}P_{32}P_{21}\quad \text{(parity-even),}$\\[.3em]
$P_{123+}=P_{12}P_{23}P_{31}+P_{13}P_{32}P_{21}\quad \text{(parity-odd),}$
\end{tabular}
\\[.5em]
\text{from $S_{ij},\bar S_{ij}$}:&\qquad
\hat S_{12}=S_{12}\bar S_{12}\,\frak X_{2}\frak X_{2}^{\dagger} =S_{21}\bar S_{21}\,\frak X_{1}\frak X_{1}^{\dagger}\quad \text{and 2 perm (parity-even),}
\\[.5em]
\text{from $Q_{i},\bar Q_{i}$}:&\qquad\!\!\!
\begin{tabular}{l}
$Q_{1+}=\frac12(Q_{1}+\bar Q_{1})\quad \text{and 2 perm (parity-even),}$\\[.3em]
$Q_{1-}=\frac12(Q_{1}-\bar Q_{1})\quad \text{and 2 perm (parity-odd),}$
\end{tabular}
\\[.5em]
\text{from $\Omega_{i},\bar \Omega_{i}$}:&\qquad
\Omega_{-}=\frac12\frac{\Omega_{1}-\bar\Omega_{1}}{\frak X_{1}\frak X_{1}^{\dagger}}\quad \text{(parity-odd),}
\\[.5em]
\text{from $Y_{ij}$}:&\qquad
Y_{123-}=Y_{12}Y_{23}Y_{31}-Y_{13}Y_{32}Y_{21}\quad \text{(parity-even).}
\end{aligned}
\label{eq:invariantsconst}
\end{align}
We notice that some of these have the same form for different permutations of points, e.g. $\hat \Psi$. 

Clearly, there are more invariants that one can build, but they are all related to the independent list we've constructed above. For instance, consider the objects
\begin{align}
&\hat \psi_{1}=\psi_{1}\bar\psi_{1}\quad
\text{and 2 perm (parity-odd),}\\
&\hat \Sigma=\frac{\Sigma_{13}\bar\Sigma_{13}}{\frak X_{1}\frak X_{1}^{\dagger}}\quad
\text{(parity-even).}
\end{align}
Using (\ref{eq:superconf0}, \ref{eq:superconf2}) it is easy to check that $\hat\psi_{i},\hat\Sigma$ transform invariantly under superinversion. But one finds the relations
\begin{gather}
\hat\psi_{1}=\frac i4 Q_{1-}\,,\qquad
\hat \Sigma=\hat\Psi\,.
\end{gather}
If done systematically, there are numerous other superconformal invariants that one could build, but all of them are related to the ones listed in (\ref{eq:invariantsconst}). For convenience, we will use the notation $R'=\hat \Psi$, and $T'=\Omega_{-}$, henceforth.

We further classify the independent invariants in (\ref{eq:invariantsconst}) as bosonic or grassmanian, based on their non-supersymmetric ($\theta,\bar\theta=0$) form.\footnote{As noted previously, grassmanian invariants vanish when $\theta,\bar\theta=0$, while bosonic invariants do not.}
\begin{align}
\begin{aligned}
\text{bosonic:}\quad&
Q_{i+}\,,\ \hat P_{ij}\,,\ \hat S_{ij}\,,\ 
P_{123-}\,,\ P_{123+}\,,\ Y_{123-}\,,\\
\text{grassmanian:}\quad&
Q_{i-}\,,\ R'\,,\ T'\,.
\label{eq:scftinvdep}
\end{aligned}
\end{align}
Next, to know if this list of invariants is truly exhaustive or if new invariants are required, we review and extend some of the analysis done in Section \ref{sec:cftreview}.

\subsection{Using CFT relations for grassmanian invariants}
\label{ssec:scftinv-grass}

One of the reasons for doing a comprehensive analysis for $4d$ CFT in Section \ref{sec:cftreview} is that it aids in building new grassmanian superconformal invariants that our construction strategy in Section \ref{ssec:scftinv-cons} fails to build. Let us understand why we need more invariants and how the CFT analysis is utilized to obtain them.

Bosonic superconformally invariant objects made out of superspace variables must reduce to standard position space conformally invariant objects when the grassmanian coordinates $\theta,\bar\theta$ are put to zero. This in turn implies that the relations between conformal invariants must have superconformal extensions. Simply put, if there is a combination $\{A\}$ of bosonic superconformal invariants which vanishes for the non-supersymmetric case, i.e.
\begin{align}
\sum_{\{A\}} \!\!\!\!\text{\footnotesize
\begin{tabular}{c}
bosonic\\
superconformal\\
invariants
\end{tabular}
}\!\!\!\!\!\Bigg|_{\theta,\bar\theta=0}=\sum_{\{A\}} \!\!\!\text{\footnotesize
\begin{tabular}{c}
conformal\\
invariants
\end{tabular}
}\!\!\!\!=0\,,
\end{align}
then there must be a combination $\{B\}$ of grassmanian superconformal invariants such that we get the superconformal relation 
\begin{align}
\sum_{\{A\}} \!\!\!\!\text{\footnotesize
\begin{tabular}{c}
bosonic\\
superconformal\\
invariants
\end{tabular}
}\!\!\!\!\!+
\sum_{\{B\}} \!\!\!\!\text{\footnotesize
\begin{tabular}{c}
grassmanian\\
superconformal\\
invariants
\end{tabular}
}\!\!\!\!=0\,.
\end{align}
Each term in $\{B\}$ is purely grassmanian, and thus individually vanishes for $\theta,\bar\theta=0$.

This will become clearer by looking at specific examples.

\begin{enumerate}[label=$\blacktriangleright$]

	\item At ${\cal O}(\lambda_{1}\lambda_{2}\bar\lambda_{1}\bar\lambda_{2})$
	\begin{align}
	\hat S_{ij}	-\hat P_{ij}-  Q_{i+} Q_{j+}\ \bigg|_{\theta,\bar\theta=0}=0
	\end{align}
	This is Eq. (\ref{eq:s12cft}), written in terms of bosonic superconformal invariants listed in (\ref{eq:scftinvdep}). This relation effectively removed $\hat S_{ij}$ as an independent invariant for the conformal case. The superconformal version ($\theta,\bar\theta\neq0$) of this relation does not hold true, and needs corrections through purely grassmanian invariants. As such, no invariants that we constructed in the last subsection can be used to correct the superconformal version. Hence, we use this expression as the definition of a new grassmanian invariant $\hat R_{ij}$,
	\begin{align}
	\hat R_{12}= \hat S_{12}	-\hat P_{12}-  Q_{1+} Q_{2+} \quad \text{and 2 perm (parity even)}
	\end{align}
	Thus, the definition of $\hat R_{ij}$'s
	\begin{align}
	\hat S_{ij}	-\hat P_{ij}-  Q_{i+} Q_{j+}-\hat R_{ij}=0\,,
	\end{align}
	can be interpreted as the ``corrected" superconformal relation, and can be used instead of $\hat S_{ij}$ as a new invariant. Note that unlike $\hat S_{ij}$'s, $\hat R_{ij}$'s are grassmanian invariants.\\
	
	\item  At ${\cal O}(\lambda_{1}\lambda_{2}\lambda_{3}\bar\lambda_{1}\bar\lambda_{2}\bar\lambda_{3})$
	\begin{align}
	P_{123-}+ Y_{123-}\ \bigg|_{\theta,\bar\theta=0}=0
	\end{align}
	The invariant $ Y_{123-}$ was not constructed for the CFT case as it has the same structure as $ P_{123-}$ (modulo a minus sign), which is precisely why we get this relation. In the superconformal theory, the relation does not hold, and $ P_{123-}$ and $Y_{123-}$ , both of which have a non-zero bosonic part, differ by a grassmanian structure. Again, the invariants constructed in the last subsection, along with $\hat R_{ij}$'s are not enough to fix this relation. Therefore, we define another grassmanian superconformal invariant using this equation,
\begin{align}
R_{123}= P_{123-}+ Y_{123-}
\end{align}
This `fixes' the superconformal relation
\begin{align}
P_{123-}+ Y_{123-}- R_{123}=0.
\end{align}
Note that $R_{123}$ is again a purely grassmanian invariant.\\
	
	\item  We have another relation at ${\cal O}(\lambda_{1}\lambda_{2}\lambda_{3}\bar\lambda_{1}\bar\lambda_{2}\bar\lambda_{3})$
	\begin{align}
    	P_{123-}-Q_{1+} Q_{2+} Q_{3+}-\sum_{\rm cyc} Q_{1+}\hat P_{23}\ \bigg|_{\theta,\bar\theta=0}=0
	\end{align}
	This is Eq. (\ref{eq:GPY4d}) written in terms of superconformal invariants. It turns out that the superconformal extension of this relation is completely fixed by the new grassmanian invariants constructed above, viz. $\hat R_{ij}$'s and $R_{123}$. The corrected superconformal relation is\footnote{We have defined $Q_{ij+}\equiv Q_{i+}Q_{j+}$, \,\,$Q_{123+}\equiv Q_{1+}Q_{2+}Q_{3+}$ for brevity.}
	\begin{align}
	P_{123-}-Q_{123+}-\sum_{\rm cyc} Q_{1+}\hat P_{23}-	\frac13\sum_{\rm cyc}\hat Q_{1+}\hat R_{23}	-\frac16 {R_{123}}-\left(2Q_{123+}+ \frac23\sum_{\rm cyc}Q_{1+}\hat P_{23}\right)R'=0
	\label{eq:GPYscft}
	\end{align}
	
 Just like the CFT case, this equation removes $P_{123-}$ as an independent superconformal invariant. While Eq. (\ref{eq:GPYscft}) does not define any new grassmanian invariant, it does give us a robust consistency check of our list of independent parity-even superconformal invariants.

\end{enumerate}

\subsection{Relations between superconformal invariants}
\label{ssec:scftinv-relation}

The bosonic invariants (which have non-supersymmetric CFT counterparts) are
\begin{align}
\begin{aligned}
\text{parity-even}:\quad&
Q_{i+}\,,\ 
\hat P_{ij}\,,\\
\text{parity-odd}:\quad&
P_{123+}\,.
\end{aligned}
\label{eq:scftinvboson}
\end{align}
Note that $Q_{i+}$ reduces to $Q_{i}$ for the CFT case, while $\hat P_{ij},P_{123+}$ reduce to their namesakes. 

The grassmanian invariants (which vanish for the non-supersymmetric case) are
\begin{align}
\begin{aligned}
\text{parity-even}:\quad&
\hat R_{ij}\,,\ 
R_{123}\,,\ 
R'\,,\\
\text{parity-odd}:\quad&
Q_{i-}\,,\ 
T'\,.
\end{aligned}
\label{eq:scftinvgrass}
\end{align}

While the above listed invariants are linearly independent, one can still find non-linear relations between them due to their $\theta,\bar\theta$ content and/or their parity. These relations will ultimately pose restrictions for the allowed independent structures of SCFT correlators.

\begin{enumerate}[label=(R\arabic*)]

	\item \label{r1} Since $R'$ has the highest allowed degree of $\theta,\bar\theta$, the product of any grassmanian invariant (parity-even or odd) with $R'$ vanishes,
	\begin{align}
	\text{(grassmann)}\cdot R'=0\,,
	\label{eq:r1}
	\end{align}
	where (grassmann) could be any of the invariants listed in (\ref{eq:scftinvgrass}).\\
	
	\item \label{r2} The product of any two grassmanian invariants (parity-even or odd) either vanishes, or can be written as a combination of bosonic invariants multiplied with $R'$,
	\begin{align}
	\text{(grassmann)}\cdot\text{(grassmann)}=\sum 
	\left(\!\!\!
		\begin{tabular}{c}
		products of\\
		bosonic inv
		\end{tabular}\!\!\!
	\right)\cdot R'\,.
	\label{eq:r2}
	\end{align}
	An exhaustive list of these relations is provided in Appendix \ref{appsec:relations}. This relation in tandem with \ref{r1} implies that a grassmanian invariant can only occur linearly in a superconformal correlator.\footnote{\ref{r1} can actually be considered a special case of \ref{r2}.}\\
	
	\item \label{r3} The product of any two parity-odd invariants either vanishes, or can be written as a combination of parity-even invariants,
	\begin{align}
	\text{(odd)}\cdot \text{(odd)}=\sum \text{(even)}\,.
	\label{eq:r3}
	\end{align}
	Again, the full list can be found in Appendix \ref{appsec:relations}. This relation implies that a parity-odd invariant can only occur linearly in a correlator. Note that this was the case for the conformal theory as well since we had Eq. (\ref{eq:p123square}).
\end{enumerate}

We present below the bosonic and grassmanian superconformal invariants constructed in this section. They are classified based on their behavior under parity. The $\lambda,\bar\lambda$ content of each invariant is mentioned as well.
\begin{align}
\renewcommand{\arraystretch}{2}
\begin{tabular}{|c|c|c|}
\hline
& parity-even & parity-odd\\\hline
bosonic & 
$\begin{aligned}
Q_{i+}:&\ \lambda_{i}\bar\lambda_{i}\\
\hat P_{ij} :&\ \lambda_{i}\lambda_{j}\bar \lambda_{i}\bar\lambda_{j}
\end{aligned}$
& 
$P_{123+}:\ \lambda_{1}\lambda_{2}\lambda_{3}\bar \lambda_{1}\bar \lambda_{2}\bar\lambda_{3}$
\\\hline
grassmanian &
$\begin{aligned}
\hat R_{ij} :&\ \lambda_{i}\lambda_{j}\bar \lambda_{i}\bar\lambda_{j}\\
R_{123}:&\ \lambda_{1}\lambda_{2}\lambda_{3}\bar \lambda_{1}\bar \lambda_{2}\bar\lambda_{3}\\
R':&\  \text{no $\lambda,\bar\lambda$}
\end{aligned}$
& 
$\begin{aligned}
Q_{i-}:&\ \lambda_{i}\bar\lambda_{i}\\
T':&\  \text{no $\lambda,\bar\lambda$}
\end{aligned}$
\\\hline
\end{tabular}
\label{eq:scftinvariantsfinal}
\end{align}
This table provides the final list of  superconformal and superinversion invariants built out of 3 superspace points in $4d$ ${\cal N}=1$ SCFT, and after imposing the constraints in \ref{r1}, \ref{r2}, \ref{r3}, one can express every possible spinning 3-pt correlator (containing symmetric traceless superfield operators) in terms of multinomials of these invariants.

\subsection{Permutation symmetry}
\label{ssec:scftinv-pointswitch}

In the next section, we will exploit the point-switch symmetry (if any) of the correlator (as for the CFT case) to determine the invariant structures allowed for that correlator. To this end, we present below the transformation of each of the invariants listed above (parity-even and parity-odd) under a $\check z_{i}\leftrightarrow \check z_{j}$ swap, also denoted as $i\leftrightarrow j$.
\begin{center}
\begin{tikzpicture}
	\node (p12) {$\hat P_{12}$};
	\node (p1231)[below right = 1em and 2cm of p12] {$\hat P_{23}$};
	\node (p1212)[above right = 1em and 2cm of p12] {$\hat P_{12}$};
	\node (p1223)[right = 2cm of p12] {$\hat P_{31}$};
	\draw [->]  (p12.north east)|- node[midway,above,pos=0.75]{$1\leftrightarrow 2$}(p1212.west);
	\draw [->]  (p12.south east)|-node[midway,above,pos=0.75]{$3\leftrightarrow 1$}(p1231.west);
	\draw [->]  (p12.east)--node[midway,above]{$2\leftrightarrow 3$}(p1223.west);
\end{tikzpicture}$\quad$
\begin{tikzpicture}
	\node (q1) {$Q_{1+}$};
	\node (q131)[below right = 1em and 2cm of q1] {$-Q_{3+}$};
	\node (q112)[above right = 1em and 2cm of q1] {$- Q_{2+}$};
	\node (q123)[right = 2cm of q1] {$- Q_{1+}$};
	\draw [->]  (q1.north east)|- node[midway,above,pos=0.75]{$1\leftrightarrow 2$}(q112.west);
	\draw [->]  (q1.south east)|-node[midway,above,pos=0.75]{$3\leftrightarrow 1$}(q131.west);
	\draw [->]  (q1.east)--node[midway,above]{$2\leftrightarrow 3$}(q123.west);
\end{tikzpicture}$\quad$
\begin{tikzpicture}
	\node (p12) {$\hat R_{12}$};
	\node (p1231)[below right = 1em and 2cm of p12] {$\hat R_{23}$};
	\node (p1212)[above right = 1em and 2cm of p12] {$\hat R_{12}$};
	\node (p1223)[right = 2cm of p12] {$\hat R_{31}$};
	\draw [->]  (p12.north east)|- node[midway,above,pos=0.75]{$1\leftrightarrow 2$}(p1212.west);
	\draw [->]  (p12.south east)|-node[midway,above,pos=0.75]{$3\leftrightarrow 1$}(p1231.west);
	\draw [->]  (p12.east)--node[midway,above]{$2\leftrightarrow 3$}(p1223.west);
\end{tikzpicture}$\quad$
\begin{tikzpicture}
	\node (q1) {$Q_{1-}$};
	\node (q131)[below right = 1em and 2cm of q1] {$Q_{3-}$};
	\node (q112)[above right = 1em and 2cm of q1] {$Q_{2-}$};
	\node (q123)[right = 2cm of q1] {$Q_{1-}$};
	\draw [->]  (q1.north east)|- node[midway,above,pos=0.75]{$1\leftrightarrow 2$}(q112.west);
	\draw [->]  (q1.south east)|-node[midway,above,pos=0.75]{$3\leftrightarrow 1$}(q131.west);
	\draw [->]  (q1.east)--node[midway,above]{$2\leftrightarrow 3$}(q123.west);
\end{tikzpicture}\\[1em]
\begin{tikzpicture}
	\node (p12) {$P_{123+}$};
	\node (p1223)[right = 2cm of p12] {$P_{123+}$};
	\draw [->]  (p12.east)--node[midway,above]{any swap}(p1223.west);
\end{tikzpicture}\qquad
\begin{tikzpicture}
	\node (p12) {$R_{123}$};
	\node (p1223)[right = 2cm of p12] {$-R_{123}$};
	\draw [->]  (p12.east)--node[midway,above]{any swap}(p1223.west);
\end{tikzpicture}\\[.5em]
\begin{tikzpicture}
	\node (p12) {$R'$};
	\node (p1223)[right = 2cm of p12] {$R'$};
	\draw [->]  (p12.east)--node[midway,above]{any swap}(p1223.west);
\end{tikzpicture}\qquad
\begin{tikzpicture}
	\node (p12) {$T'$};
	\node (p1223)[right = 2cm of p12] {$-T'$};
	\draw [->]  (p12.east)--node[midway,above]{any swap}(p1223.west);
\end{tikzpicture}
\end{center}
The transformation for the remaining invariants can be obtained straightforwardly by permuting the indices.

\section{Structure of 3-point SCFT correlators}
\label{sec:scftcorrstr}

Correlators in an SCFT are superconformally invariant. In Section \ref{sec:scftinv} we constructed an exhaustive and independent list of invariant structures built out of 3 superspace points in $4d$ ${\cal N} =1$ SCFT. In this section, we express 3-point correlators of general traceless symmetric spinning superfield operators as multinomials of these invariants. The superfield operators ${\cal O}_{s_{i}}$ (written in their index-free form, refer Section \ref{ssec:scftbb-polars}) are superconformal multiplets with spin $s_{i}$, where $s_{i}$ is the spin of the lowest superconformal primary in the multiplet. The scaling dimension $\Delta_{i}$ of the superfield operator is simply the conformal dimension of the superconformal primary. Also, the superfield operator ${\cal O}_{s_{i}}$ is neutral under $R$-symmetry. Note that (anti-)chiral superfields do not transform as bosonic traceless symmetric representations of $SL(2,\mathbb{C})$, and are thus not considered in our work.

For a general 3-point correlator containing superfield operators ${\cal O}_{s_{i}}(\check z_{i})$ with arbitrary spin $s_{i}$ and conformal dimension $\Delta_{i}$, the allowed invariant structures are determined by the spin $s_{i}$ (which fixes the homogeneity in $\lambda,\bar\lambda$), and permutation symmetry if any, just as they did for the conformal case in Section \ref{ssec:cftreview-confcorr}. We can write the 3-point correlator\footnote{The expression looks very similar to Eq. (\ref{eq:cftcorrelator}), and reduces to it for the non-supersymmetric case.}
\begin{align}
\langle
{\cal O}_{s_{1}}(\check z_{1}){\cal O}_{s_{2}}(\check z_{2}){\cal O}_{s_{3}}(\check z_{3})
\rangle=
\frac{1}{x_{123}(\tau_{1},\tau_{2},\tau_{3})}
\left(
\sum_{m}a_{m}{\cal W}^{\rm even}_{m}+\sum_{n}b_{n}{\cal W}^{\rm odd}_{n}
\right)\,,
\label{eq:scftcorrelator}
\end{align}
where we have introduced
\begin{align}
\frac{1}{x_{123}(\tau_{1},\tau_{2},\tau_{3})}=
\frac{1}{(
|x_{1\bar 2}x_{\bar 12}|^{\tau_{12,3}}
|x_{2\bar 3}x_{\bar 23}|^{\tau_{23,1}}
|x_{3\bar 1}x_{\bar 31}|^{\tau_{31,2}})^{\frac12}}\,,\  \text{and}\quad
\begin{gathered}
\tau_{ij,k}=\tau_{i}+\tau_{j}-\tau_{k}\,,\\
\tau_{i}=\Delta_{i}-s_{i}\,.
\end{gathered}
\label{eq:scftbehavscaling}
\end{align}
Of course, each ${\cal W}^{\rm even}_{m}$  (or ${\cal W}^{\rm odd}_{n}$) is made out of parity-even (odd) superconformal invariants listed in Table \ref{eq:scftinvariantsfinal}, and must have the appropriate homogeneity in $\lambda,\bar\lambda$ based on the $s_{i}$'s. 
Each of these terms is also subjected to the restrictions posed by Eqs. (\ref{eq:r1}$-$\ref{eq:r3}). Note that since the invariant structures have a scaling weight of zero, the overall factor $1/x_{123}(\tau_{1},\tau_{2},\tau_{3})$ takes care of the transformation of the full correlator under superinversion.\footnote{Here we have used the superconformally covariant symmetric scalar $x_{i\bar \jmath}x_{\bar \imath j}=(y_{ij}^2+\theta_{ij}^2\bar\theta_{ij}^2)$ defined using Eq. (\ref{eq:scfty12}).}

We next present some examples below. For correlators containing a scalar superfield operator (at least one $s_{i}=0$), the invariants that involve $\lambda,\bar{\lambda}$ at all three points, namely $R_{123}$ (even) and $P_{123+}$ (odd) can not be used.  Note that we'll omit the obvious superspace label $\check z_{i}$ and just write the operator with its spin. The spins will obey $s_{1}\geq s_{2}\geq s_{3}$.

\subsubsection*{$\langle {\cal O}_{0}{\cal O}_{0}{\cal O}_{0} \rangle$}

The simplest 3-point correlator one can construct in a superconformal theory contains three scalar superfield operators. For the (non-supersymmetric) conformal theory, the form of this correlator is trivial, since it does not carry any tensor structure. In the superconformal theory, since we have invariants which do not contain any $\lambda,\bar\lambda$, viz. $R',T'$, we can have the following possible contributions to this correlator,
\begin{align}
\begin{aligned}
{\rm even}:\quad & 
1\,,\ R'\,,\\
{\rm odd}:\quad & 
T'\,,
\end{aligned}
\end{align}
where we have included the trivial structure `1'. Moreover, this correlator is completely symmetric under any of the $1\leftrightarrow2\leftrightarrow3$ swaps. Owing to point-switch transformation of the invariants in Section \ref{ssec:scftinv-pointswitch}, only the parity-even structure $R'$ preserves this symmetry,
\begin{align}
R'\xrightarrow{\text{any } 1\leftrightarrow2\leftrightarrow3}R'\,,\qquad 
T'\xrightarrow{\text{any } 1\leftrightarrow2\leftrightarrow3}-T'\,.
\end{align}
Hence the correlator has the form
\begin{align}
\langle
{\cal O}_{0}{\cal O}_{0}{\cal O}_{0}
\rangle=
\frac{1}{x_{123}(\tau_{1},\tau_{2},\tau_{3})}\,
\left(a_{1}+a_{2}R'\right)\,.
\label{eq:scfto0o0o0}
\end{align}

\subsubsection*{$\langle {\cal O}_{1}{\cal O}_{0}{\cal O}_{0} \rangle$}

Since scalar superfield operators have no $\lambda,\bar\lambda$ contractions, this correlator containing one spin-$1$ and two scalar operators has homogeneity $\lambda_{1}\bar\lambda_{1}$. The possible structures are
\begin{align}
\begin{aligned}
{\rm even}:\quad & 
Q_{1+}\,,\ Q_{1+}R'\,,\\
{\rm odd}:\quad & 
{Q_{1-}}\,,\  Q_{1+}{T'}\,.
\end{aligned}
\end{align}
Notice that we have not included the term $Q_{1-}R'$, since $Q_{1-}$ is grassmanian, and thus this structure vanishes due to Eq. (\ref{eq:r1}).

This correlator has a point-switch symmetry under a $2\leftrightarrow 3$ swap, but we note that all the parity-even structures are anti-symmetric under the swap, i.e. the even part vanishes for this correlator. 
\begin{align}
Q_{1+}\xrightarrow{2\leftrightarrow3}-Q_{1+}\,,\qquad 
Q_{1+}R'\xrightarrow{2\leftrightarrow3}(-Q_{1+})R'\,.
\end{align}
The parity-odd terms preserve the symmetry. 
\begin{align}
Q_{1-}\xrightarrow{2\leftrightarrow3}Q_{1-}\,,\qquad 
Q_{1+}T'\xrightarrow{2\leftrightarrow3}(-Q_{1+})(-T')=Q_{1+}T'\,.
\end{align}

Hence, the correlator containing non-conserved superfield operators takes the form
\begin{align}
\langle
{\cal O}_{1}{\cal O}_{0}{\cal O}_{0}
\rangle=
\frac{1}{x_{123}(\tau_{1},\tau_{2},\tau_{3})}
\left(
b_{1}Q_{1-}+b_{2}Q_{1+}T'
\right)\,.
\label{eq:scfto1o0o0}
\end{align}
\subsubsection*{$\langle {\cal O}_{s}{\cal O}_{0}{\cal O}_{0} \rangle$}

This correlator has homogeneity $\lambda^{s}_{1}\bar\lambda^{s}_{1}$. The possible structures are
\begin{align}
\begin{aligned}
{\rm even}:\quad & 
Q_{1+}^{s}\,,\ Q_{1+}^{s}R'\,,\\
{\rm odd}:\quad & 
Q_{1}^{s-1}{Q_{1-}}\,,\  Q_{1+}^{s}{T'}\,.
\end{aligned}
\end{align}
The point-switch symmetry under $2\leftrightarrow 3$ continues to hold, but it is easy to check that for even/odd values of $s$, only the parity-even/odd structures satisfy that symmetry.

Thus, for $s=\text{even}$, the allowed structures are all parity-even, and the correlator looks like
\begin{align}
\langle
{\cal O}_{s={\rm even}}{\cal O}_{0}{\cal O}_{0}
\rangle=
\frac{1}{x_{123}(\tau_{1},\tau_{2},\tau_{3})}
Q_{1+}^{s}\left(a_{1}+a_{2}R'
\right)\,.
\label{eq:scftoso0o0even}
\end{align}
For $s=\text{odd}$, the allowed structures are all parity-odd, and the correlator is
\begin{align}
\langle
{\cal O}_{s={\rm odd}}{\cal O}_{0}{\cal O}_{0}
\rangle=
\frac{1}{x_{123}(\tau_{1},\tau_{2},\tau_{3})}
Q_{1+}^{s-1}\left(
b_{1}Q_{1-}+b_{2}Q_{1+}T'
\right)\,.
\label{eq:scftoso0o0odd}
\end{align}

\subsubsection*{$\langle {\cal O}_{1}{\cal O}_{1}{\cal O}_{0} \rangle$}

The correlator has homogeneity $\lambda_{1}\lambda_{2}\bar \lambda_{1}\bar\lambda_{2}$. The possible structures are
\begin{align}
\begin{aligned}
{\rm even:\quad }& 
Q_{12+}\,,\ 
\hat P_{12}\,,\ 
Q_{12+}R'\,,\ 
\hat P_{12} R'\,,\ 
\hat R_{12}\,,
\\{\rm odd:\quad }& 
Q_{12+}T'\,,\ 
\hat P_{12} T'\,,\ 
Q_{1+}Q_{2-}\,,\ 
Q_{2+}Q_{1-}\,.
\end{aligned}
\end{align}
This correlator is symmetric under a $1\leftrightarrow 2$ swap, and all the parity-even structures follow that symmetry. The only parity-odd term that survives the point-switch symmetry is
\begin{align*}
Q_{1+}Q_{2-}-Q_{2+}Q_{1-}\,.
\end{align*}
All the other parity-odd terms are anti-symmetric under the swap, and hence do not contribute to the correlator. The non-conserved correlator allows 5 parity-even and 1 parity-odd structures,
\begin{align}
\langle
{\cal O}_{1}{\cal O}_{1}{\cal O}_{0}
\rangle=
\frac{1}{x_{123}(\tau_{1},\tau_{2},\tau_{3})}
\left[
Q_{12+}\left(
	a_{1}+a_{2}R'
\right)+
\hat P_{12}\left(
	a_{3}+a_{4}R'
\right)+
a_{5}\hat R_{12}+
b_{1}
\left(
	Q_{1+}Q_{2-}-
Q_{2+}Q_{1-}
\right)
\right]\,.
\label{eq:scfto1o1o0}
\end{align}

\subsubsection*{$\langle {\cal O}_{s}{\cal O}_{1}{\cal O}_{0} \rangle$}

The homogeneity of this correlator is $\lambda_{1}^{s}\lambda_{2}\bar \lambda_{1}^{s}\bar\lambda_{2}$. For $s>1$, there is no point-switch symmetry, and the correlator contains 5 parity-even and 5 parity-odd terms,
\begin{align}
\begin{aligned}
\langle
{\cal O}_{s}{\cal O}_{1}{\cal O}_{0}
\rangle=
\frac{1}{x_{123}(\tau_{1},\tau_{2},\tau_{3})}
\left[
Q_{1+}^{s}Q_{2+}\left(
	a_{1}+a_{2}R'
\right)+
Q_{1+}^{s-1}\hat P_{12}\left(
	a_{3}+a_{4}R'
\right)+
a_{5}Q_{1+}^{s-1}\hat R_{12}\right.+&\\
\left.
b_{1}Q_{1+}^{s}Q_{2+}T'+
b_{2}Q_{1+}^{s-1} \hat P_{12} T'+
b_{3}Q_{1+}^{s-2}\hat P_{12}Q_{1-}+
b_{4}Q_{1+}^{s-1}Q_{2+}Q_{1-}+
b_{5}Q_{1+}^{s}Q_{2-}
\right]&\,.
\end{aligned}
\label{eq:scftoso1o0}
\end{align}

\subsubsection*{$\langle {\cal O}_{2}{\cal O}_{2}{\cal O}_{0} \rangle$}

The possible structures are
\begin{align}
\begin{aligned}
{\rm even:\quad }& 
Q_{12+}^{2}\,,\ 
\hat P_{12}^{2}\,,\ 
Q_{12+}\hat P_{12}\,,\ 
Q_{12+}^{2}R'\,,\ 
\hat P_{12}^{2}R'\,,\ 
Q_{12+}\hat P_{12}R'\,,\ 
Q_{12+}\hat R_{12}\,,\ 
\hat P_{12}\hat R_{12}\,,
\\{\rm odd:\quad }& 
Q_{12+}^{2}T'\,,\ 
\hat P_{12}^{2}T'\,,\ 
Q_{12+}\hat P_{12}T'\,,\ 
Q_{1+}Q_{2+}^{2}Q_{1-}\,,\ 
Q_{1+}^{2}Q_{2+}Q_{2-}\,,\ 
Q_{1+}\hat P_{12}Q_{2-}\,,\ 
Q_{2+}\hat P_{12}Q_{1-}\,.
\end{aligned}
\end{align}
There is point-switch symmetry under a $1\leftrightarrow2$ swap, and all the 8 parity-even terms preserve it. However, only the following 2 parity-odd structures are symmetric under the swap,
\begin{align}
Q_{12+}\left(Q_{1+}Q_{2-}-
Q_{2+}Q_{1-}\right)\,,\ \hat P_{12}\left(Q_{1+}Q_{2-}-
Q_{2+}Q_{1-}\right) \,.
\end{align}
Hence, the correlator is
\begin{align}
\begin{aligned}
\langle
{\cal O}_{2}{\cal O}_{2}{\cal O}_{0}
\rangle=
\frac{1}{x_{123}(\tau_{1},\tau_{2},\tau_{3})}
\left[
Q_{12+}^{2}\left(
	a_{1}+a_{2}R'
\right)+
\hat P_{12}^{2}\left(
	a_{3}+a_{4}R'
\right)+
Q_{12+}\hat P_{12}\left(
	a_{5}+a_{6}R'
\right)+
a_{7}\hat P_{12}\hat R_{12}\right.+&\\
\left.
a_{8}Q_{12+}\hat R_{12}+
\left(b_{1}Q_{12+}+b_{2}\hat P_{12}\right)\left(
Q_{1+}Q_{2-}-Q_{2+}Q_{1-}
\right)
\right]&\,.
\end{aligned}
\label{eq:scfto2o2o0}
\end{align}

\subsubsection*{$\langle {\cal O}_{s}{\cal O}_{2}{\cal O}_{0} \rangle$}

For $s>2$, there is no point-switch symmetry, and there are 8 parity-even and 7 parity-odd structures
\begin{align}
\begin{aligned}
{\rm even:\quad }& 
Q_{1+}^{s}Q_{2+}^{2}\,,\ 
Q_{1+}^{s-2}\hat P_{12}^{2}\,,\ 
Q_{1+}^{s-1}Q_{2+}\hat P_{12}\,,\ 
Q_{1+}^{s}Q_{2+}^{2}R'\,,\ 
Q_{1+}^{s-2}\hat P_{12}^{2}R'\,,\ 
Q_{1+}^{s-1}Q_{2+}\hat P_{12}R'\,,\ \\&
Q_{1+}^{s-2}\hat P_{12}\hat R_{12}\,,\ 
Q_{1+}^{s-1}Q_{2+}\hat R_{12}\,.
\\[.5em]{\rm odd:\quad }& 
Q_{1+}^{s}Q_{2+}^{2}T'\,,\ 
Q_{1+}^{s-2}\hat P_{12}^{2}T'\,,\ 
Q_{1+}^{s-1}Q_{2+}\hat P_{12}T'\,,\ 
Q_{1+}^{s-1}Q_{2+}^{2}Q_{1-}\,,\ 
Q_{1+}^{s-3}\hat P_{12}^{2}Q_{1-}\,,\ \\&
Q_{1+}^{s-2}Q_{2+}\hat P_{12}Q_{1-}\,,\ 
Q_{1+}^{s}Q_{2+}Q_{2-}\,. 
\end{aligned}
\end{align}

The correlator is
\begin{align}
\begin{aligned}
\langle
{\cal O}_{s}{\cal O}_{2}{\cal O}_{0}
\rangle=&
\frac{1}{x_{123}(\tau_{1},\tau_{2},\tau_{3})}
\left[
Q_{1+}^{s}Q_{2+}^{2}(a_{1}+a_{2}R')+
Q_{1+}^{s-2}\hat P_{12}^{2}(a_{3}+a_{4}R')+
Q_{1+}^{s-1}Q_{2+}\hat P_{12}(a_{5}+a_{6}R')+
\right.
\\&\left.
a_{7}Q_{1+}^{s-2}\hat P_{12}\hat R_{12}+
a_{8}Q_{1+}^{s-1}Q_{2+}\hat R_{12}+
b_{1}Q_{1+}^{s}Q_{2+}^{2}T'+
b_{2}Q_{1+}^{s-2}\hat P_{12}^{2}T'+
b_{3}Q_{1+}^{s-1}Q_{2+}\hat P_{12}T'+
\right.
\\&\left.
b_{4}Q_{1+}^{s-1}Q_{2+}^{2}Q_{1-}+
b_{5}Q_{1+}^{s-3}\hat P_{12}^{2}Q_{1-}+
b_{6}Q_{1+}^{s-2}Q_{2+}\hat P_{12}Q_{1-}+
b_{7}Q_{1+}^{s}Q_{2+}Q_{2-} \right]\,.
\end{aligned}
\label{eq:scftoso2o0}
\end{align}

~

Next, we explore the more interesting case where all the operators in the 3-point function have non-zero spin $s_{i}$. Consequently, all the invariants that were constructed in Section \ref{sec:scftinv} are accessible. Some examples are listed below.

\subsubsection*{$\langle {\cal O}_{1}{\cal O}_{1}{\cal O}_{1} \rangle$}

The homogeneity of the correlator is $\lambda_{1}\lambda_{2}\lambda_{3}\bar\lambda_{1}\bar\lambda_{2}\bar\lambda_{3}$. The possible structures are
\begin{align}
\begin{aligned}
{\rm even:\quad }& 
Q_{123+}\,,\ 
Q_{1+}\hat P_{23}\,,\ 
Q_{2+}\hat P_{31}\,,\ 
Q_{3+}\hat P_{12}\,,\ 
Q_{123+}R'\,,\ 
Q_{1+}\hat P_{23}R'\,,\ \\&
Q_{2+}\hat P_{31}R'\,,\ 
Q_{3+}\hat P_{12}R'\,,\ 
Q_{1+}\hat R_{23}\,,\ 
Q_{2+}\hat R_{31}\,,\ 
Q_{3+}\hat R_{12}\,,\  
R_{123}\,,\ 
\\[.5em]{\rm odd:\quad }& 
P_{123+}\,,\ 
P_{123+}R'\,,\ 
Q_{12+}T'\,,\ 
Q_{1+}\hat P_{23}T'\,,\ 
Q_{2+}\hat P_{31}T'\,,\ 
Q_{3+}\hat P_{12}T'\,,\ \\&
Q_{12+}Q_{3-}\,,\ 
Q_{23+}Q_{1-}\,,\ 
Q_{31+}Q_{2-}\,,\ 
\hat P_{23}Q_{1-}\,,\ 
\hat P_{31}Q_{2-}\,,\ 
\hat P_{12}Q_{3-}\,,\ 
\end{aligned}
\end{align}
Note that all the allowed parity-even structures are present in Eq. (\ref{eq:GPYscft}) (except $P_{123-}$ of course, since it is removed by that equation). However, the correlator needs to be completely symmetric under any of the $1\leftrightarrow 2\leftrightarrow 3$ swap. It is easy to check that every one of the allowed parity-even structure is \textit{antisymmetric} under any of the $1\leftrightarrow 2\leftrightarrow 3$ swap, hence the even part vanishes. 
There is a significant reduction in the allowed structures for this correlator, and we are left with 6 parity-odd structures,
\begin{align}
\begin{aligned}
{\rm odd:\quad }& 
P_{123+}\,,\ 
P_{123+}R'\,,\ 
Q_{123+}T'\,,\ 
\left(\sum_{\rm cyc}\hat P_{12}Q_{3+}\right)T'\,,\ \sum_{\rm cyc}\hat P_{12}Q_{3-}\,,\ 
\sum_{\rm cyc}Q_{12+}Q_{3-}\,.
\end{aligned}
\end{align}
The non-conserved correlator has the form
\begin{align}
\begin{aligned}
\langle
{\cal O}_{1}{\cal O}_{1}{\cal O}_{1}
\rangle=
\frac{1}{x_{123}(\tau_{1},\tau_{2},\tau_{3})}
\left[
P_{123+}(b_{1}+b_{2}R')+
b_{3}Q_{123+}T'+
b_{4}\left(\sum_{\rm cyc}\hat P_{12}Q_{3+}\right)T'+\right.&\\
\left.
b_{5}\sum_{\rm cyc}\hat P_{12}Q_{3-}+
b_{6}\sum_{\rm cyc}Q_{12+}Q_{3-}
\right]&\,.
\end{aligned}
\label{eq:scfto1o1o1}
\end{align}

\subsubsection*{$\langle {\cal O}_{2}{\cal O}_{1}{\cal O}_{1} \rangle$}

For this correlator, after imposing symmetry under $2\leftrightarrow 3$ swap, the number of allowed terms are significantly restricted. For instance, there are 12 parity-even and 4 parity-odd terms that preserve the point-switch symmetry,
\begin{align}
\begin{aligned}
\langle
{\cal O}_{2}{\cal O}_{1}{\cal O}_{1}
\rangle\sim&\ 
Q_{1+}^{2}Q_{23+}\left(a_{1}+a_{2}R'\right)+
Q_{1+}^{2}\hat P_{23}\left(a_{3}+a_{4}R'\right)+
\hat P_{12}\hat P_{31}\left(a_{5}+a_{6}R'\right)+
\\&
\left(Q_{12+}\hat P_{31}+Q_{13+}\hat P_{12}\right)\left(a_{7}+a_{8}R'\right)+
a_{9}Q_{1+}^{2}\hat R_{23}+
a_{10}\left(Q_{12+}\hat R_{31}+Q_{31+}\hat R_{12}\right)+
\\&
a_{11}\left(\hat P_{12}\hat R_{31}+\hat P_{31}\hat R_{12}\right)+
a_{12}Q_{1+}R_{123}+
b_{1}\left(Q_{12+}\hat P_{31}-Q_{31+}\hat P_{12}\right)T'+
\\&
b_{2}Q_{1+}^{2}\left(Q_{2+}Q_{3-}-Q_{3+}Q_{2-}\right)+
b_{3}Q_{1+}\left(\hat P_{31}Q_{2-}-\hat P_{12}Q_{3-}\right)+
b_{4}\left(Q_{2+}\hat P_{31}-Q_{3+}\hat P_{12}\right)Q_{1-}\,.
\end{aligned}
\label{eq:scfto2o1o1}
\end{align}
Here and henceforth, the overall $x_{123}$ factor has been omitted for convenience.

\subsubsection*{$\langle {\cal O}_{3}{\cal O}_{1}{\cal O}_{1} \rangle$}

The correlator contains 4 parity-even and 12 parity-odd structures after implementing restrictions due to point-switch symmetry,
\begin{align}
\begin{aligned}
\langle
{\cal O}_{3}{\cal O}_{1}{\cal O}_{1}
\rangle\sim&\ 
Q_{1+}^{2}\left(Q_{2+}\hat P_{31}-Q_{3+}\hat P_{12}\right)(a_{1}+a_{2}R')+
a_{3}Q_{1+}^{2}\left(Q_{2+}\hat R_{31}-Q_{3+}\hat R_{12}\right)+\\
&
a_{4}Q_{1+}\left(\hat P_{12}\hat R_{31}-\hat P_{31}\hat R_{12}\right)+
Q_{1+}^{2}P_{123+}(b_{1}+b_{2}R')+
b_{2}Q_{1+}^{3}Q_{23+}T'+
b_{3}Q_{1+}^{3}\hat P_{23}T'+\\
&
b_{4}Q_{1+}^{2}\left(Q_{2+}\hat P_{31}+
b_{5}Q_{3+}\hat P_{12}\right)T'+
b_{6}Q_{1+}\hat P_{12}\hat P_{31}T'+
b_{7}\hat P_{12}\hat P_{31}Q_{1-}+
b_{8}Q_{1+}^{2}\hat P_{23}Q_{1-}+\\
&
b_{9}\left(Q_{12+}\hat P_{31}+Q_{31+}\hat P_{12}\right)Q_{1-}+
b_{10}Q_{1+}^{3}\left(Q_{2+}Q_{3-}+Q_{3+}Q_{2-}\right)+
\\&
b_{11}Q_{1+}^{2}\left(\hat P_{31}Q_{2-}+
\hat P_{12}Q_{3-}\right)+
b_{12}Q_{1+}^{2}Q_{23+}Q_{1-}\,.
\end{aligned}
\label{eq:scfto3o1o1}
\end{align}

\subsubsection*{$\langle {\cal O}_{s}{\cal O}_{1}{\cal O}_{1} \rangle$}

For even values of $s$, the allowed structures are easily derivable from $\langle O_{2}O_{1}O_{1} \rangle$ by multiplying with $Q_{1+}^{s-2}$. Thus, the non-conserved correlator is simply $Q_{1+}^{s-2}$ times the expression in Eq. (\ref{eq:scfto2o1o1}).

Similarly, when $s$ is odd, the correlator is just $Q_{1+}^{s-3}$ times the expression for $\langle O_{3}O_{1}O_{1} \rangle$ in Eq. (\ref{eq:scfto3o1o1}).

\subsubsection*{$\langle {\cal O}_{2}{\cal O}_{2}{\cal O}_{1} \rangle$}

After considering the $1\leftrightarrow2$ point-switch symmetry, there are 7 parity-even and 17 parity-odd allowed structures. The correlator is
\begin{align}
\begin{aligned}
\langle
{\cal O}_{2}{\cal O}_{2}{\cal O}_{1}
\rangle\sim&\ 
Q_{12+}\left(Q_{1+}\hat P_{23}-Q_{2+}\hat P_{31}\right)(a_{1}+a_{2}R')+
\hat P_{12}\left(Q_{1+}\hat P_{23}-Q_{2+}\hat P_{31}\right)(a_{3}+a_{4}R')+
\\&
a_{5}Q_{12+}\left(Q_{1+}\hat R_{23}-Q_{2+}\hat R_{31}\right)+
a_{6}\hat P_{12}\left(Q_{1+}\hat R_{23}-Q_{2+}\hat R_{31}\right)+
a_{7}\left(Q_{1+}\hat P_{23}-Q_{2+}\hat P_{31}\right)\hat R_{12}+
\\&
Q_{12+}P_{123+}(b_{1}+b_{2}R') +
\hat P_{12}P_{123+}(b_{3}+b_{4}R')+ 
b_{5}\hat R_{12}P_{123+}+
b_{6}Q_{123+}\hat P_{12}T'+
b_{7}Q_{3+}\hat P_{12}^{2}T'+
\\&
b_{8}Q_{12+}\left(Q_{1+}\hat P_{23}+Q_{2+}\hat P_{31}\right)T'+ 
b_{9}\hat P_{12}\left(Q_{1+}\hat P_{23}+Q_{2+}\hat P_{31}\right)T'+ 
b_{10}Q_{12+}\left(\hat P_{23}Q_{1-}+\hat P_{31}Q_{2-}\right)+\\
&
b_{11}\hat P_{12}\left(\hat P_{23}Q_{1-}+\hat P_{31}Q_{2-}\right)+ 
b_{12}Q_{12+}^{2}Q_{3-}+
b_{13}Q_{12+}\hat P_{12}Q_{3-}+
b_{14}\hat P_{12}^{2}Q_{3-}+
\\&
b_{15}Q_{123+}\left(Q_{1+}Q_{2-}+Q_{2+}Q_{1-}\right)+
b_{16}\hat P_{12}Q_{3+}\left(Q_{1+}Q_{2-}+Q_{2+}Q_{1-}\right)+
\\&
b_{17}\left(Q_{1+}^{2}\hat P_{23}Q_{2-}+
Q_{2+}^{2}\hat P_{31}Q_{1-}\right)\,.
\end{aligned}
\label{eq:scfto2o2o1}
\end{align}

\subsubsection*{$\langle {\cal O}_{2}{\cal O}_{2}{\cal O}_{2} \rangle$}

This correlator has homogeneity $\lambda_{1}^{2}\lambda_{2}^{2}\lambda_{3}^{2}\bar\lambda_{1}^{2}\bar\lambda_{2}^{2}\bar\lambda_{3}^{2}$. It also exhibits a point-switch symmetry under any of the $1\leftrightarrow 2\leftrightarrow3$ swaps, and there are no parity-odd structures that preserve that symmetry. The correlator contains 17 parity-even terms,
\begin{align}
\begin{aligned}
\langle
{\cal O}_{2}{\cal O}_{2}{\cal O}_{2}
\rangle\sim&\ 
Q_{123+}^{2}(a_{1}+a_{2}R')+ 
Q_{123+}\sum_{\rm cyc}Q_{1+}\hat P_{23}(a_{3}+a_{4}R')+
\sum_{\rm cyc}Q_{1+}^{2}\hat P_{23}^{2}(a_{5}+a_{6}R')+
\\&
\sum_{\rm cyc}Q_{12+}\hat P_{23}\hat P_{31}(a_{7}+a_{8}R')+ 
\hat P_{12}\hat P_{23}\hat P_{31}(a_{9}+a_{10}R')+
a_{11}\sum_{\rm cyc}\hat P_{12}\hat P_{23}\hat R_{31}+
\\&
a_{12}Q_{123+}R_{123}+
a_{13}Q_{123+}\sum_{\rm cyc}Q_{1+}\hat R_{23}+ 
a_{14}\sum_{\rm cyc}Q_{1+}\hat P_{23}R_{123}+ 
a_{15}\sum_{\rm cyc}Q_{1+}^{2}\hat P_{23}\hat R_{23}+
\\&
a_{16}\sum_{\rm cyc}Q_{12+}\hat P_{23}\hat R_{31}+
a_{17}\sum_{\rm cyc}\hat P_{12}\hat P_{31}\hat R_{23}\,.
\end{aligned}
\label{eq:scfto2o2o2}
\end{align}

Note that in all the above examples, the only constraints are of $\mathcal{N}=1$ superconformal invariance and permutation invariance (if any). The next section discusses further constraints arising due to conservation conditions on spinning operators.

\section{Conserved 3-point SCFT correlators}
\label{sec:scftconscorr}

In this section we will build on the results of Section \ref{sec:scftcorrstr}, and consider further constraints on the structure of 3-point correlators arising from one or more operator in the correlator saturating the unitarity bound, that is, satisfying a current conservation equation.

In the superconformal theory, an integer spin-$s$ conserved supercurrent is a symmetric traceless tensor ${\cal J}^{\alpha(s)\dot\alpha(s)}(z)$.\footnote{The non-conserved superfield operators are denoted as ${\cal O}$, while conserved supercurrents are denoted as ${\cal J}$.} Just like an ordinary superfield operator, the supercurrent is also a superconformal multiplet, but it also satisfies a shortening condition,
\begin{align}
D_{\alpha_{1}}\,{\cal J}^{\alpha(s)\dot\alpha(s)}(z)=\bar D_{\dot\alpha_{1}}\, {\cal J}^{\alpha(s)\dot\alpha(s)}(z)=0\,,
\end{align}
where $D,\bar D$ are the supercovariant derivatives defined in Eq. (\ref{eq:supercovderiv}). Consequently, this supermultiplet is referred to as a `short' multiplet, and contains conserved conformal currents.\footnote{It is straightforward to expand the superconformal multiplet into its $\theta,\bar\theta$ components form. For instance, the `short' superconformal multiplet ${\cal J}^{\alpha\dot\alpha}$, also called the Ferrara-Zumino multiplet \cite{FZmultiplet1975} contains spin-1, spin-$\frac32$ and spin-2 conserved conformal primaries.}

Since we have augmented superspace with polarization spinors $\lambda,\bar\lambda$, the shortening condition on a spin-$s$ index-free operator ${\cal J}_{s}$ takes the form
\begin{align}
\frac{\partial}{\partial \lambda_{\alpha}}D_{\alpha}\,{\cal J}_{s}(\check z)=0\,,\qquad
\frac{\partial}{\partial \bar\lambda_{\dot \alpha}}\bar D_{\dot\alpha}\, {\cal J}_{s}(\check z)=0\,.
\label{eq:shortening}
\end{align}
Note that when we consider the shortening of a scalar superfield operator, Eq. (\ref{eq:shortening}) is unusable since ${\cal J}_{0}$ does not contain $\lambda,\bar\lambda$. Hence, we resort to the free superfield conservation condition
\begin{align}
D_{\alpha}D^{\alpha}{\cal J}_{0}(\check z)=0\,,\qquad \bar D_{\dot\alpha}\bar D^{\dot\alpha}{\cal J}_{0}(\check z)=0\,.
\label{eq:shorteningscalar}
\end{align}

We have already enumerated the 3-point correlators as constrained by ${\cal N}=1$ superconformal invariance and permutation invariance (if any) in Section \ref{sec:scftcorrstr}. We will now see how the structure is further constrained when one or more spinning operators in the correlator is a conserved supercurrent. We also note that on the application of the shortening conditions Eq. (\ref{eq:shortening}) on an operator, the scaling dimension is fixed to be the canonical dimension of the superconformal primary. The first few examples provide the details of how the constant coefficients $a_m,\, b_n$ in the correlator in Eq. (\ref{eq:scftcorrelator}) get related. In the rest of the cases, the final answers are given. Appendix \ref{appsec:morescft3ptcorr} contains more examples of 3-point correlators where the spinning operator is not conserved.

\subsubsection*{$\langle{\cal J}_{0}{\cal J}_{0}{\cal J}_{0} \rangle$}

The simplest 3-point correlator containing three scalar superfield operators has the form given in Eq. (\ref{eq:scfto0o0o0}). Since there are no $\lambda,\bar\lambda$ in the expression, the shortening condition is not applicable, and we use superfield conservation given in Eq. (\ref{eq:shorteningscalar}). The conservation condition on each of the $\check z_{i}$ gives identical constraints:
\begin{align}
\text{conservation at $\check z_{1}$ }:\quad
\Delta_{1}=2,\qquad	
a_{2}=\left(1-\frac14(\Delta_{2}-\Delta_{3})^{2}\right)a_{1}\,,
\end{align}
and similarly for $\check z_{2}, \check z_{3}$.

Hence, the correlator with all three scalar superfields conserved is fixed upto an overall constant,
\begin{align}
\langle{\cal J}_{0}{\cal J}_{0}{\cal J}_{0} \rangle=
\frac{1}{x_{123}(2,2,2)}\left(1+R'\right)\,.
\end{align}

\subsubsection*{$\langle {\cal J}_{1}{\cal O}_{0}{\cal O}_{0} \rangle$}

The non-conserved correlator is given in Eq. (\ref{eq:scfto1o0o0}). Based on the point-switch symmetry, there are 2 parity-odd structures, and no surviving parity-even structures. After imposing the shortening condition Eq. (\ref{eq:shortening}) on ${\cal O}_1(\check z_1)$, we get the following constraints,
\begin{align}
\Delta_{3}=\Delta_{2}\,,\quad \Delta_{1}=3\,, \quad b_{2}=\frac i2 b_{1}\,.
\end{align}
The form of the correlator $\langle {\cal J}_{1}{\cal O}_{0}{\cal O}_{0} \rangle$ is obtainable by applying the above relations to Eq. (\ref{eq:scfto1o0o0}).

We observe that in all the examples, imposing the shortening condition on a spin-$s$ superfield operator always fixes its conformal dimension to its canonical value, $\Delta=2+s$.

\subsubsection*{$\langle {\cal O}_{1}{\cal J}_{0}{\cal O}_{0} \rangle$}

If in Eq. (\ref{eq:scfto1o0o0}), we instead impose the free superfield conservation condition on ${\cal O}_{0}(\check z_2)$  (with ${\cal O}_{1}$ non-conserved) we get the constraint,
\begin{align}
\Delta_{2}=2\,,\quad b_{2}=\frac{i(\Delta_{1}-\Delta_{3})}{-3+\Delta_{1}-\Delta_{3}}b_{1}\,,
\end{align}
and the correlator is again fixed upto a single structure. Since the two scalar operators are identical, a similar result follows for $\langle {\cal O}_{1}{\cal O}_{0}{\cal J}_{0} \rangle$, with $\Delta_2,\Delta_3$ interchanged.

In all the subsequent examples, we give the form of the correlator without imposing conservation on the scalar superfield operator.

\subsubsection*{$\langle {\cal J}_{2}{\cal O}_{0}{\cal O}_{0} \rangle$}

For $\langle {\cal O}_{2}{\cal O}_{0}{\cal O}_{0} \rangle$, the only structures that preserve the $2\leftrightarrow3$ point-switch symmetry are all parity-even, as shown in Eq. (\ref{eq:scftoso0o0even}). After imposing conservation on ${\cal O}_{2}$, we get the constraints
\begin{align}
a_{2}=3a_{1}\,,\quad \Delta_{3}=\Delta_{2}\,.
\end{align}
Hence, the correlator $\langle {\cal J}_{2}{\cal O}_{0}{\cal O}_{0} \rangle$ is fixed upto a single parity-even structure.

\subsubsection*{$\langle {\cal J}_{s}{\cal O}_{0}{\cal O}_{0} \rangle$}

For $s=$ even, the correlator only allows for 2 parity-even structures as given in Eq. (\ref{eq:scftoso0o0even}). The form of the correlator with conserved supercurrent ${\cal J}_{s}$ is fixed upto an overall unknown constant,
\begin{align}
\langle {\cal J}_{s={\rm even}}{\cal O}_{0}{\cal O}_{0} \rangle=\frac{1}{x_{123}(2,\Delta,\Delta)}Q_{1+}^{s}\left(1+(2s-1)R'
\right)\,.
\end{align}

For odd values of $s$, the correlator only has 2 parity-odd contributions as given in Eq. (\ref{eq:scftoso0o0odd}), and the form with conserved ${\cal J}_{s}$ is fixed upto an overall constant as well,
\begin{align}
\langle {\cal J}_{s={\rm odd}}{\cal O}_{0}{\cal O}_{0} \rangle=\frac{1}{x_{123}(2,\Delta,\Delta)}Q_{1+}^{s-1}\left(
Q_{1-}+\tfrac i2 s\,Q_{1+}T'
\right)\,.
\end{align}
In both the results above, $\Delta$ is the conformal dimension of the identical scalar operators on $\check z_2,\check z_3$.

\subsubsection*{$\langle {\cal J}_{1}{\cal O}_{1}{\cal O}_{0} \rangle$}

The non-conserved correlator containing two spin-1 and a scalar superfield operators has the form given in Eq. (\ref{eq:scfto1o1o0}).

Just like in the non-supersymmetric case, conservation on \textit{either} of the spin-1 supercurrents on $\check z_{1}, \check z_{2}$ gives identical constraints. Implementing the shortening condition at $\check z_{1}$ gives the constraints
\begin{align}
\begin{gathered}
b_{1}=0\,,\quad a_{2}= -\tfrac{(-5+\Delta_{2}-\Delta_{3})(3+\Delta_{2}-\Delta_{3})}{4}a_{1}\,,\quad
a_{3} = \tfrac{2 (\Delta_{2} - \Delta_{3})}{-3 + \Delta_{2} - \Delta_{3}}a_{1}\,,\\[.5em]
a_{4}= -\tfrac{(\Delta_{2}-\Delta_{3})(3+\Delta_{2}-\Delta_{3})}{2}a_{1}\,,\quad
a_{5}=-\tfrac{1+\Delta_{2}-\Delta_{3}}{2}a_{1}\,,
\end{gathered}
\end{align}
and the correlator $\langle {\cal J}_{1}{\cal O}_{1}{\cal O}_{0} \rangle$ is fixed upto an overall parity-even coefficient, while the parity-odd contribution drops out. An identical result follows if we instead impose conservation on $\check z_{2}$, giving the form of $\langle {\cal O}_{1}{\cal J}_{1}{\cal O}_{0} \rangle$.

\subsubsection*{$\langle {\cal J}_{1}{\cal J}_{1}{\cal O}_{0} \rangle$}

When both the spin-1 operators in Eq. (\ref{eq:scfto1o1o0}) are conserved, the correlator is fixed in terms of a single unknown coefficient $a_{1}$, and the conformal dimension of the scalar operator. The constraints for $\langle {\cal J}_{1}{\cal J}_{1}{\cal O}_{0} \rangle$ are
\begin{align}
\Delta_{3}=\Delta\,,\quad
b_{1}=0\,,\quad a_{2}=\tfrac{12+4\Delta-\Delta^{2}}{4}a_{1}\,,\ 
a_{3}=\tfrac{2(-3+\Delta)}{\Delta}a_{1}\,,\ 
a_{4}=-\tfrac{18-9\Delta+\Delta^{2}}{2}a_{1}\,,\ 
a_{5}=\tfrac{-4+\Delta}{2}a_{1}\,.
\end{align}
We note that this result matches that of \cite{Osborn4dN1SCFT1998}.

\subsubsection*{$\langle {\cal J}_{2}{\cal O}_{1}{\cal O}_{0} \rangle$}

The form of the non-conserved correlator $\langle {\cal O}_{2}{\cal O}_{1}{\cal O}_{0} \rangle$ is given in Eq. (\ref{eq:scftoso1o0}) with $s=2$. As expected, employing conservation individually on $\check z_{1}, \check z_{2}$ yields different constraints. In both the cases though, the correlator is fixed upto 1 parity-even and 1 parity-odd coefficient.

Imposing shortening at $\check z_1$, we get the constraints for $\langle {\cal J}_{2}{\cal O}_{1}{\cal O}_{0} \rangle$,
\begin{align}
\Delta_{3}=\Delta_{2}-1\,,\quad
\begin{gathered}
a_{2}=6 a_{1}\,,\ 
a_{3}=-3 a_{1}\,,\ 
a_{4}=-6 a_{1}\,,\ 
a_{5}=-2 a_{1}\,,\\[.5em]
b_{2}=4 b_{1}\,,\ 
b_{3}=i b_{1}\,,\ 
b_{4}=-2i b_{1}\,,\ 
b_{5}=-\tfrac {3i}2 b_{1}\,.
\end{gathered}
\end{align}
Interestingly, the dimension of the scalar operator has been fixed.

\subsubsection*{$\langle {\cal O}_{2}{\cal J}_{1}{\cal O}_{0} \rangle$}

When the conservation condition is implemented on $\check z_2$ in Eq. (\ref{eq:scftoso1o0}) with $s=2$, we get the constraints
\begin{align}
\begin{gathered}
a_{2}=\tfrac{(-6+\Delta_{1}-\Delta_{3})(4+\Delta_{1}-\Delta_{3})}{4}a_{1}\,,\ 
a_{3}=\tfrac{2(\Delta_{1}-\Delta_{3})}{-4+\Delta_{1}+\Delta_{3}}a_{1}\,,\ 
a_{4}=\tfrac{(\Delta_{1}-\Delta_{3})(4+\Delta_{1}-\Delta_{3})}{2}a_{1}\,,\ 
a_{5}=-\tfrac{2+\Delta_{1}-\Delta_{3}}{2}a_{1}\,,\\[.5em]
b_{2}=\tfrac{2-3\Delta_{1}-3\Delta_{3}}{2(-1+\Delta_{1}-\Delta_{3})}b_{1}\,,\  
b_{3}=-\tfrac{i(-8+3(\Delta_{1}-\Delta_{3})^{2})}{2(-4+\Delta_{1}-\Delta_{3})(-1+\Delta_{1}-\Delta_{3})}b_{1}\,,\ 
b_{4}=\tfrac{i(2+3\Delta_{1}-3\Delta_{3})}{4(-1+\Delta_{1}-\Delta_{3})}b_{1}\,,\ 
b_{5}=-\tfrac{i(4+\Delta_{1}-\Delta_{3})}{4(-1+\Delta_{1}-\Delta_{3})}b_{1}\,.
\end{gathered} 
\end{align}
We find that $\langle {\cal O}_{2}{\cal J}_{1}{\cal O}_{0} \rangle$ is fixed upto 1 parity-even and 1 parity-odd coefficient in terms of the dimensions of spin-2 and scalar operators.

\subsubsection*{$\langle {\cal J}_{2}{\cal J}_{1}{\cal O}_{0} \rangle$}

Finally, with both the supercurrents at $\check z_{1}, \check z_{2}$ in $\langle {\cal O}_{2}{\cal O}_{1}{\cal O}_{0} \rangle$ conserved, we get the following relations,
\begin{align}
\Delta_{1}=4\,,\quad \Delta_{2}= 3\,,\quad \Delta_{3}=2\,,\quad
\begin{gathered}
a_{2}=6 a_{1}\,,\ 
a_{3}=-3 a_{1}\,,\ 
a_{4}=-6 a_{1}\,,\ 
a_{5}=-2 a_{1}\,,\\[.5em]
b_{2}=4 b_{1}\,,\ 
b_{3}=i b_{1}\,,\ 
b_{4}=-2i b_{1}\,,\ 
b_{5}=-\tfrac {3i}2 b_{1}\,.
\end{gathered}
\end{align}
On applying these relations to Eq. (\ref{eq:scftoso1o0}) with $s=2$, we get the fully fixed correlator $\langle {\cal J}_{2}{\cal J}_{1}{\cal O}_{0} \rangle$ in terms of 1 parity-even and 1 parity-odd coefficients.

\subsubsection*{$\langle {\cal J}_{3}{\cal J}_{1}{\cal O}_{0} \rangle$}

The non-conserved correlator $\langle {\cal O}_{3}{\cal O}_{1}{\cal O}_{0} \rangle$ has the form given in Eq. (\ref{eq:scftoso1o0}) with $s=3$. The shortening condition gives results similar to the $s=2$ case. The intermediary results are relegated to Appendix \ref{appsec:morescft3ptcorr}, and we present the final set of relations when both the spinning operators are conserved,
\begin{align}
\Delta_{1}=5\,,\quad \Delta_{2}= 3\,,\quad \Delta_{3}=2\,,\quad
\begin{gathered}
a_{2}=8 a_{1}\,,\ 
a_{3}=-3 a_{1}\,,\ 
a_{4}=-12 a_{1}\,,\ 
a_{5}=-3 a_{1}\,,\\[.5em]
b_{2}=3 b_{1}\,,\ 
b_{3}=\tfrac{3i}2 b_{1}\,,\ 
b_{4}=-\tfrac{3i}2  b_{1}\,,\ 
b_{5}=-i b_{1}\,,
\end{gathered}
\end{align}
i.e. the correlator $\langle {\cal J}_{3}{\cal J}_{1}{\cal O}_{0} \rangle$ is fixed upto 1 parity-even and 1 parity-odd term. Also note the similarity with the results of $\langle {\cal J}_{2}{\cal J}_{1}{\cal O}_{0} \rangle$.

\subsubsection*{$\langle {\cal J}_{2}{\cal J}_{2}{\cal O}_{0} \rangle$}

The non-conserved correlator is given in Eq. (\ref{eq:scfto2o2o0}). Analogous to the case $\langle {\cal J}_{1}{\cal J}_{1}{\cal O}_{0} \rangle$, imposing conservation on any one of the two spin-2 operators gives equivalent constraints and fixes the correlator in terms of 1 parity-even structure and the scaling dimensions of the other two operators. Again, the intermediary result $\langle {\cal J}_{2}{\cal O}_{2}{\cal O}_{0} \rangle$ is discussed in Appendix \ref{appsec:morescft3ptcorr}.

Imposing conservation on both the spinning operators gives the constraints,
\begin{align}
\begin{gathered}
b_{1}=b_{2}=0\,,\quad
a_{2}=\tfrac{32+4\Delta_{3}-\Delta_{3}^{2}}{4}a_{1}\,,\quad
a_{3}=\tfrac{2(40-24\Delta_{3}+3\Delta_{3}^{2})}{\Delta_{3}(2+\Delta_{3})}a_{1}\,,\quad
a_{4}=\tfrac{(248-196 \Delta_{3}+44\Delta_{3}^{2}+3\Delta_{3}^{3})}{2(2+\Delta_{3})}a_{1}\,,\\
a_{5}=\tfrac{2(-14+3\Delta_{3})}{2+\Delta_{3}}a_{1}\,,\quad
a_{6}=-\tfrac{(100-36\Delta_{3}+3\Delta_{3}^{2})}{2}a_{1}\,,\quad
a_{7}=\tfrac{2(18-9\Delta_{3}+\Delta_{3}^{2})}{2+\Delta_{3}}a_{1}\,,\quad
a_{8}={\scriptstyle (-6+\Delta_{3})}a_{1}\,.
\end{gathered}
\end{align}
It is worth noting that the value of $\Delta_{3}$ is unrestricted when we have a correlator $\langle {\cal J}_{s}{\cal J}_{s}{\cal O}_{0} \rangle$ containing identical supercurrents, which is not the case when the first and second operators are distinct (e.g. $\langle {\cal J}_{2}{\cal J}_{1}{\cal O}_{0} \rangle$).

~

All the subsequent examples contain correlators with all non-zero spins. We present here the final result with all the operators conserved. The intermediate results for correlators comprising both non-conserved and conserved operators are presented in Appendix \ref{appsec:morescft3ptcorr}.

\subsubsection*{$\langle {\cal J}_{1}{\cal J}_{1}{\cal J}_{1} \rangle$}

We consider the correlator that contains three identical spin-1 supercurrents. The non-conserved correlator is given in Eq. (\ref{eq:scfto1o1o1}). This is the first example where all the operators have non-zero spins, and thus the full space of superconformal invariants listed in Table \ref{eq:scftinvariantsfinal} is accessible. The conserved correlator is obtained by implementing the shortening condition on all three operators,
\begin{align}
b_{3}=-2i(7b_{1}-3b_{2})\,,\ 
b_{4}=-2i(5b_{1}-2b_{2})\,,\ 
b_{5}=2b_{1}-b_{2}\,,\ 
b_{6}=-4b_{1}+2b_{2}\,,
\end{align}
which fixes the correlator $\langle {\cal J}_{1}{\cal J}_{1}{\cal J}_{1} \rangle$ in terms of 2 parity-odd coefficients. The fact that there are two structures for this correlator was noted by Osborn in \cite{Osborn4dN1SCFT1998}.

\subsubsection*{$\langle {\cal J}_{2}{\cal J}_{1}{\cal J}_{1} \rangle$}

The structure of the non-conserved correlator can be found in Eq. (\ref{eq:scfto2o1o1}). Applying conservation on all the operators gives us,
\begin{align}
\begin{gathered}
b_{1}=b_{2}=b_{3}=b_{4}=0\,,\quad
a_{3}=\tfrac12(-6a_{1}+a_{2})\,,\quad
a_{4}=\tfrac16(8a_{1}+a_{2})\,,\quad
a_{5}=\tfrac12(8a_{1}-a_{2})\,,\\
a_{6}=30a_{1}-4a_{2}\,,\quad
a_{7}=\tfrac14(-20a_{1}+3a_{2})\,,\quad
a_{8}=\tfrac13(-32a_{1}+5a_{2})\,,\quad
a_{9}=\tfrac13(-7a_{1}+a_{2})\,,\\
a_{10}=\tfrac16(4a_{1}-a_{2})\,,\quad
a_{11}=\tfrac14(36a_{1}-5a_{2})\,,\quad
a_{12}=-\tfrac23(7a_{1}-a_{2})\,.
\end{gathered}
\end{align}
The parity-odd contribution drops out, and the conserved correlator is fixed in terms of 2 parity-even coefficients.

\subsubsection*{$\langle {\cal J}_{3}{\cal J}_{1}{\cal J}_{1} \rangle$}

The non-conserved correlator is given in Eq. (\ref{eq:scfto3o1o1}). The conserved correlator is obtained through the relations
\begin{align}
\begin{gathered}
a_{1}=a_{2}=a_{3}=a_{4}=0\,,\quad
b_{3}=-14ib_{1}+3ib_{2}\,,\quad
b_{4}=-\tfrac23(11ib_{1}-2ib_{2})\,,\quad
b_{5}=-6ib_{1}+ib_{2}\,,\\
b_{6}=3(4ib_{1}-ib_{2})\,,\quad
b_{7}=\tfrac12(-4b_{1}+b_{2})\,,\quad
b_{8}=6b_{1}-b_{2}\,,\quad
b_{9}=\tfrac12(16b_{1}-3b_{2})\,,\\
b_{10}=\tfrac23(-4b_{1}+b_{2})\,,\quad
b_{11}=4b_{1}-b_{2}\,,\quad
b_{12}=\tfrac12b_{2}\,.
\end{gathered}
\end{align}
Here, the parity-even terms vanish, and the conserved correlator has 2 unknown parity-odd coefficients.

\subsubsection*{$\langle {\cal J}_{2}{\cal J}_{2}{\cal J}_{1} \rangle$}

The form of non-conserved correlator is given in Eq. (\ref{eq:scfto2o2o1}). Imposing conservation on all the operators, we get
\begin{align}
\begin{gathered}
a_{1}=a_{2}=a_{3}=a_{4}=a_{5}=a_{6}=a_{7}=0\,,\quad
b_{3}=-\tfrac12b_{1}\,,\quad
b_{4}=\tfrac{1}{13}(23b_{1}-9b_{2})\,,\quad
b_{5}=\tfrac{7}{13}(4b_{1}-b_{2})\,,\\
b_{6}=\tfrac{i}{13}(29b_{1}-4b_{2})\,,\quad
b_{7}=-\tfrac{i}{13}(19b_{1}-8b_{2})\,,\quad
b_{8}=\tfrac{10i}{13}(4b_{1}-b_{2})\,,\quad
b_{9}=i(11b_{1}-2b_{2})\,,\\
b_{10}=6b_{1}-b_{2}\,,\quad
b_{11}=\tfrac{1}{13}(61b_{1}-12b_{2})\,,\quad
b_{12}=-\tfrac{3}{13}(4b_{1}-b_{2})\,,\quad
b_{13}=\tfrac{1}{13}(-45b_{1}+8b_{2})\,,\\
b_{14}=\tfrac{1}{13}(-73b_{1}+15b_{2})\,,\quad
b_{15}=\tfrac{1}{13}(46b_{1}-5b_{2})\,,\quad
b_{16}=6b_{1}-b_{2}\,,\quad
b_{17}=\tfrac{1}{13}(70b_{1}-11b_{2})\,.
\label{eq:scftj2j2j1}
\end{gathered}
\end{align}
The parity-even terms need to vanish, and the conserved correlator is fixed upto 2 parity-odd structures.

\subsubsection*{$\langle {\cal J}_{2}{\cal J}_{2}{\cal J}_{2} \rangle$}

The non-conserved correlator has a form given in Eq. (\ref{eq:scfto2o2o2}). It has no parity-odd contributions. After imposing conservation on all the operators, we get
\begin{align}
\begin{gathered}
a_{2}=11a_{1}\,,\quad
a_{5}=\tfrac{1}{34}(-18a_{1}+4a_{3}+3a_{4})\,,\quad
a_{6}=\tfrac{1}{102}(-178a_{1}+474a_{3}-27a_{4})\,,\\
a_{7}=\tfrac{1}{34}(-20a_{1}-22a_{3}+9a_{4})\,,\quad
a_{8}=\tfrac{1}{51}(-112a_{1}-300a_{3}+81a_{4})\,,\quad
a_{9}=\tfrac{1}{102}(-84a_{1}+166a_{3}-3a_{4})\,,\\
a_{10}=\tfrac{1}{17}(-40a_{1}+330a_{3}-33a_{4})\,,\quad 
a_{11}=\tfrac{1}{17}(2a_{1}+43a_{3}-6a_{4})\,,\quad
a_{12}=\tfrac{1}{34}(-4a_{1}+50a_{3}-5a_{4})\,,\\
a_{13}=\tfrac{1}{68}(-80a_{1}-88a_{3}+19a_{4})\,,\quad
a_{14}=\tfrac{1}{51}(-20a_{1}-39a_{3}+9a_{4})\,,\quad
a_{15}=\tfrac{1}{204}(-88a_{1}+12a_{3}+9a_{4})\,,\\
a_{16}=\tfrac{1}{204}(-112a_{1}-504a_{3}+81a_{4})\,,\quad 
a_{17}=\tfrac{1}{204}(-112a_{1}-504a_{3}+81a_{4})\,,
\end{gathered}
\label{eq:scftj2j2j2}
\end{align}
and the correlator is fixed upto 3 undetermined parity-even coefficients.

We note that our results for conserved correlators coincide with the results of \cite{Buch4dSCFTGen2024}. For a 3-point correlator $\langle{\cal J}_{s_1}{\cal J}_{s_2}{\cal J}_{s_2}\rangle$ with conserved supercurrents of arbitrary spin, the number of independent structures is $2s+2$, where $s={\rm min}(s_1,s_2,s_3)$ \cite{Buch4dSCFTGen2024}.

\section{Discussion}
\label{sec:discuss}

In this paper we studied the constraints of $4d$ $\mathcal{N}=1$ superconformal invariance on correlators of spinning superfield operators which may or may not be conserved supercurrents. Our analysis is general, relying only on the constraints imposed by symmetry, and holds for any $4d$ SCFT with $\mathcal{N}=1$ supersymmetry. The salient results obtained are enumerated below.

\begin{itemize}
\item 
The construction of 3-point invariants and correlators in $4d$ CFTs along the lines of \cite{GPY2011}. Invariants in $4d$ CFTs have been constructed earlier in \cite{Todorov2011,Stanev2012} but the focus was exclusively on conserved current correlators while our analysis in Section \ref{sec:cftreview} holds also for non-conserved operators with spin.
\item 
The construction (in Section \ref{sec:scftinv}) of a complete set of parity-even and parity-odd superconformal 3-point invariants in $4d$ $\mathcal{N}=1$ superspace. One of the novel features of this construction is the use of extended $4d$ CFT relations to define new grassmanian invariants (\ref{ssec:scftinv-grass}). This does not have a counterpart in the $3d$ SCFT analysis \cite{Nizami2013, JainNizami2022}.
\item 
The determination (in Section \ref{sec:scftcorrstr}) of the structure of 3-point spinning superfield correlators in $4d$ SCFTs in terms of the constructed invariants. Unlike most results in the literature, the analysis here is for \textit{non}-conserved operators.
\item 
The constraints (Section \ref{sec:scftconscorr}) on the 3-point correlator resulting from one or more of its operators being a conserved supercurrent. 
    
\end{itemize}

This work can be taken forward in the following ways:
\begin{itemize}
   
\item 
Our analysis was restricted to symmetric traceless operators. In 3-dimensions these are all the bosonic operators of the theory but in 4-dimensions one can have mixed symmetry bosonic operators. One could use grassmanian polarization spinors for such operators. Besides this one can also have fermionic (half-integer spin) operators/currents. The superconformal covariants/invariants we construct in our work should still be useful for this extended analysis.

\item 
It would be natural to extend the analysis for higher supersymmetry. For $\mathcal{N}>1$, the $R$-symmetry group would become non-abelian, but such theories should still be amenable to a similar analysis. Similar work for $3d$ SCFTs \cite{JainNizami2022} also remains restricted to abelian R-symmetry.

\end{itemize}

\section*{Acknowledgments}
\label{sec:ack}

We acknowledge the use of the publicly available Mathematica package \textit{grassmann} by Matthew Headrick, which facilitated computations involving grassmanian variables. AJ is also grateful to Umang (\href{https://github.com/umg-kmr/Brahmagupta}{Brahmagupta}) and Souradeep for access to computational facilities required for various intensive calculations.

\section*{Appendices}
\appendix

\section{Notation and conventions}
\label{appsec:4dconv}

\subsection*{$4d$ Lorentz group}
\label{appssec:4dconv-cft}

We use the `mostly plus' metric $\eta_{\mu\nu}={\rm diag}(-1,+1,+1,+1)$, and denote the spacetime position coordinates as $x^{\mu}$. The 4-vector indices are $\mu,\nu,\hdots$, while the (un)dotted spinor indices are $\alpha,\beta,\tau,\hdots,$ $\dot\alpha,\dot\beta,\dot\tau\hdots$.

We follow the conventions of \cite{BuchKuzSUGRA1998}. The $4d$ Lorentz group $SO(3,1)$ has a universal covering group $Spin(3,1)$, which is isomorphic to $SL(2,\mathbb{C})$. The objects that transform in the fundamental and the complex conjugate representations of $SL(2,\mathbb{C})$ are, respectively, the two-component \textit{left-handed} Weyl spinor $\psi_{\alpha}$, and the \textit{right-handed} Weyl spinor $\bar\psi_{\dot\alpha}$. They are denoted as $(\frac12,0)$ and $(0,\frac12)$, respectively.

The raising and lowering of spinor indices is done by $SL(2,\mathbb{C})$ invariant tensors
\begin{align}
\epsilon^{\alpha\beta}=
\begin{pmatrix}
0&1\\
-1&0
\end{pmatrix}=\epsilon^{\dot\alpha\dot\beta},\qquad
\epsilon_{\alpha\beta}=
\begin{pmatrix}
0&-1\\
1&0
\end{pmatrix}=\epsilon_{\dot\alpha\dot\beta}\,,
\end{align}
where $\epsilon^{\alpha\beta}\epsilon_{\beta\gamma}=\tau^{\alpha}_{\gamma}$, and $\epsilon^{\dot\alpha\dot\beta}\epsilon_{\dot\beta\dot\gamma}=\tau^{\dot\alpha}_{\dot\gamma}$. The raising/lowering conventions are
\begin{align}
\begin{gathered}
\psi^{\beta}=\varepsilon^{\beta\alpha}\psi_{\alpha}\,,\qquad \psi_{\alpha}=\epsilon_{\alpha\beta}\psi^{\beta}\,,\\
\bar\psi^{\dot\beta}=\varepsilon^{\dot\beta\dot\alpha}\bar\psi_{\dot\alpha}\,,\qquad \bar\psi_{\dot\alpha}=\epsilon_{\dot\alpha\dot\beta}\bar\psi^{\dot\beta}\,.
\end{gathered}
\end{align}
Note that the contractions for spinor indices are upper-left to lower-right for undotted spinors, and lower-right to upper-left for dotted spinors,
\begin{align}
\psi\chi=\psi^{\alpha}\chi_{\alpha}\,,\qquad
\bar\psi\bar\chi=\bar\psi_{\dot\alpha}\bar\chi^{\dot\alpha}\,.
\end{align}

The $\sigma,\tilde\sigma$ matrices are also invariant tensors of the Lorentz group, and help transform spacetime 4-vectors to $2\times 2$ spinor matrices $(\frac12,\frac12)$,
\begin{gather}
X_{\alpha\dot\alpha}=x^{\mu}(\sigma_{\mu})_{\alpha\dot\alpha}\,,\quad
\tilde X^{\dot\alpha\alpha}=x^{\mu}(\tilde \sigma_{\mu})^{\dot\alpha\alpha}\,,\quad
X\cdot \tilde X=-x^{2}\mathds{1}=\tilde X\cdot X\,,\\
x^{\mu}=-\frac12(\sigma^{\mu})_{\alpha\dot\alpha}\tilde X^{\dot\alpha\alpha}=-\frac12(\tilde\sigma^{\mu})^{\dot\alpha\alpha} X_{\alpha\dot\alpha}\,.
\end{gather}
where $x^{2}=x^{\mu}x_{\mu}$, and we have employed the implicit index notation for spinor matrices
\begin{align}
X\sim X_{\alpha\dot\alpha}\,,\qquad 
\tilde X\sim \tilde X^{\dot\alpha\alpha}\,.
\end{align}
For 4-vectors, we use lowercase alphabets, while for spinor matrices, we use uppercase, as evident above. Reiterating, we have chosen
\begin{align}
(\sigma_{\mu})_{\alpha\dot\alpha}=(\mathds{1},\sigma^{i}),\quad (\tilde\sigma_{\mu})^{\dot\alpha\alpha}=(\mathds{1},-\sigma^{i})\,,
\end{align}
where $(\tilde\sigma_{\mu})^{\dot\alpha\alpha}=\epsilon^{\dot\alpha\dot\beta}\epsilon^{\alpha\beta}(\sigma_{\mu})_{\beta\dot\beta}$, and $\sigma^{i}$ are the standard $3d$ Pauli matrices. 

Following are some of the properties of the $\sigma,\tilde\sigma$-matrices,
\begin{align}
(\sigma_{\mu}\tilde\sigma_{\nu}+\sigma_{\nu}\tilde\sigma_{\mu})_{\alpha}^{\ \beta}&=-2\eta_{\mu\nu}\tau_{\alpha}^{\  \beta}\,,\\
(\tilde\sigma_{\mu}\sigma_{\nu}+\tilde\sigma_{\nu}\sigma_{\mu})^{\dot\alpha}_{\ \dot\beta}&=-2\eta_{\mu\nu}\tau^{\dot\alpha}_{\  \beta}\,,\\
(\sigma^{\mu})_{\alpha\dot\alpha}(\tilde\sigma_{\mu})^{\dot\beta\beta}&=-2\tau_{\alpha}^{\ \beta}\tau_{\dot\alpha}^{\ \dot\beta}\,,\\
(\sigma^{\mu})_{\alpha\dot\alpha}(\sigma_{\mu})_{\beta\dot\beta}&=-2\epsilon_{\alpha\beta}\epsilon_{\dot\alpha\dot\beta}\,,\\
{\rm Tr}(\sigma_{\mu}\tilde\sigma_{\nu})&=-2\eta_{\mu\nu}\,,\\
\theta^{\alpha}(\sigma^{\mu})_{\alpha\dot\alpha}\bar\theta^{\dot\alpha}&=-\bar\theta_{\dot\alpha}(\tilde \sigma^{\mu})^{\dot\alpha\alpha}\theta_{\alpha}\,,
\end{align}
where $\theta,\bar\theta$ in the last equation are grassmanian.


\section{More 3-point CFT correlators}
\label{appsec:cftconservation}

We present the analysis for various 3-point correlators where the conservation condition
\begin{align}
\left[\frac{\partial}{\partial \bar\lambda^{\dot\alpha}}(\tilde \sigma^{\mu})^{\dot\alpha\alpha}\frac{\partial}{\partial x^{\mu}}\frac{\partial}{\partial \lambda^{\alpha}}
\right]
J_{s}(x,\lambda,\bar\lambda)=0\,
\label{eq:conslaw1}
\end{align}
is imposed on one or more of the spinning operators in the correlator. Note that along with the above equation, the conformal dimension of the conserved spin-$s_{i}$ current is restricted to be canonical, i.e. $\Delta_{i}=s_{i}+2$, or $\tau_{i}=2$. 

\subsubsection*{$\langle O_{s}O_{0}O_{0}\rangle$}

The homogeneity of this correlator is $\lambda_{1}^{s}\bar\lambda_{1}^{s}$. The only allowed parity-even conformally invariant structure is $Q_{1}^{s}$, and it preserves the point-switch symmetry of the correlator under a $2\leftrightarrow 3$ swap only when $s$ is even.

Employing the conservation equation on $x_{1}$, we get the constraint
\begin{align}
\Delta_{3}=\Delta_{2}\,.
\end{align}
Hence, the correlator with a conserved spin-$s$ current and two identical scalar operators with conformal dimension $\Delta$ is
\begin{align}
\langle
J_{s}O_{0}O_{0}
\rangle=\frac{1}{x_{12}^{2}|x_{23}|^{2\Delta-2}\,x_{31}^{2}}\,
Q_{1}^{s}\,.
\end{align}

For odd values of $s$, no structure preserves the $2\leftrightarrow3$ point-switch symmetry, and the correlator vanishes.

\subsubsection*{$\langle O_{s}O_{1}O_{0}\rangle$}

For $s>1$ the correlator possesses no point-switch symmetry, and the linearly independent parity-even invariant structures are just $Q_{1}^{s-1}$ times the structures for $\langle O_{1}O_{1}O_{0}\rangle$: 
\begin{align*}
Q_{1}^{s}Q_{2}\,,\ Q_{1}^{s-1}\hat P_{12}\,.
\end{align*}

The non-conserved correlator has the form
\begin{align}
\langle O_{s}O_{1}O_{0}\rangle=
\frac{1}{|x_{12}|^{\tau_{12,3}}|x_{23}|
^{\tau_{23,1}}|x_{31}|^{\tau_{31,2}}}Q_{1}^{s-1}\left(a_{1} Q_{1}Q_{2}+a_{2}\hat P_{12}\right)\,.
\end{align}

Imposing conservation on the spin-1 current at $x_{2}$ gives the following constraint in terms of $s$,
\begin{align}
a_{2}=-\tfrac{2 (\Delta_{1}-\Delta_{3})}{2+s-\Delta_{1}+\Delta_{3}}a_{1}\,,
\end{align}
and fixes the correlator in terms of a single parity-even structure.

Further imposing conservation on the spin-$s$ current at $x_{1}$ restricts the value $\Delta_{3}=2$, and the correlator takes the form
\begin{align}
\langle J_{s}J_{1}O_{0}\rangle\big|_{\Delta_{3}=2}=\frac{1}{x_{12}^{2}x_{23}^{2}x_{31}^{2}}
\left(
Q_{1}^{s}Q_{2}-s\,Q_{1}^{s-1}\hat P_{12}
\right)\,.
\end{align}

\subsubsection*{$\langle O_{s}O_{2}O_{0}\rangle$}

For $s>2$, there is no point-switch symmetry, and the linearly independent invariant structures are
\begin{align*}
Q_{1}^{s}Q_{2}^{2}\,,\ 
Q_{1}^{s-1}Q_{2}\hat P_{12}\,,\ 
Q_{1}^{s-2}\hat P_{12}^{2}\,,
\end{align*}
and the correlator is obtained by multiplying $Q_{1}^{s-2}$ to $\langle O_{2}O_{2}O_{0}\rangle$ (see Section \ref{ssec:cftreview-confcorr}). 

After conservation on only the spin-2 operator at $x_{2}$, the correlator is fixed in upto a single structure, in terms of the $\Delta_{1},\Delta_{3}$ and spin $s$,
\begin{align}
a_{2}=-\tfrac{2(s+3\Delta_{1}-3\Delta_{3})}{s+4-\Delta_{1}+\Delta_{3}}a_{1}\,,\ 
a_{3}=-\tfrac{2(2s+s^{2}-3(\Delta_{1}-\Delta_{3})^{2})}{(s+4-\Delta_{1}+\Delta_{3})(s+2-\Delta_{1}+\Delta_{3})}a_{1}\,.
\end{align}

On imposing conservation on both the spin-$s$ and spin-2 operators, the conformal dimension of the scalar operator is constrained to be $\Delta_{3}=2$, and the conserved correlator has the form
\begin{align}
\langle J_{s}J_{2}O_{0}\rangle\big|_{\Delta_{3}=2}=
\frac{1}{x_{12}^{2}x_{23}^{2}x_{31}^{2}}
&\left(
Q_{1}^{s}Q_{2}^{2}-2s\, Q_{1}Q_{2}\hat P_{12}+\tfrac{s(s-1)}{2}\,\hat P_{12}^{2}
\right)\,.
\end{align}

\subsubsection*{$\langle O_{2}O_{1}O_{1}\rangle$}

The independent invariant structures with appropriate point-switch symmetry applied are all parity-even,
\begin{align*}
Q_{1}^{2}Q_{2}Q_{3}\,,\ 
Q_{1}^{2}\hat P_{23}\,,\ 
Q_{1}\left(\hat P_{31}Q_{2}+\hat P_{23}Q_{3}\right),\ 
\hat P_{12}\hat P_{31}\,.
\end{align*}
There is one odd structure possible: $Q_{1}P_{123+}$, but it does not preserve the symmetry under $2\leftrightarrow3$ swap.

The non-conserved correlator looks like
\begin{align}
\langle O_{2}O_{1}O_{1}\rangle=\frac{1}{|x_{12}|^{\tau_{12,3}}|x_{23}|
^{\tau_{23,1}}|x_{31}|^{\tau_{31,2}}}
\left(
a_{1} Q_{1}^{2}Q_{2}Q_{3}+ a_{2}Q_{1}^{2}\hat P_{23}+a_{3}
Q_{1}\left(\hat P_{31}Q_{2}+\hat P_{23}Q_{3}\right)+
a_{4}\hat P_{12}\hat P_{31}
\right)\,.
\end{align}
Carrying forward the pattern for identical operators, imposing conservation on \textit{any one} of the two spin-1 operators fixes the correlator in terms of the conformal dimensions of the spin-2 and the other spin-1 operator, with two undermined coefficients. 
\begin{align}
\begin{aligned}
&\langle O_{2}J_{1}O_{1}\rangle:\quad a_{2}=\tfrac{2a_{1}(\Delta_{1}-\Delta_{3})+a_{3}(5-\Delta_{1}-\Delta_{3})}{5+\Delta_{1}-\Delta_{3}}\,,\quad
a_{4}=\tfrac{8 a_{1}(-\Delta_{1}+\Delta_{3})+2a_{3}(-5+\Delta_{1}^{2}-2\Delta_{1}(-4+\Delta_{3})-8\Delta_{3}+\Delta_{3}^{2})}{(-3+\Delta_{1}-\Delta_{3})(5+\Delta_{1}-\Delta_{3})}\,.\\
&\langle O_{2}O_{1}J_{1}\rangle:\quad a_{2}=\tfrac{2a_{1}(\Delta_{1}-\Delta_{2})+a_{3}(5-\Delta_{1}-\Delta_{2})}{5+\Delta_{1}-\Delta_{2}}\,,\quad 
a_{4}=\tfrac{8 a_{1}(-\Delta_{1}+\Delta_{2})+2a_{3}(-5+\Delta_{1}^{2}-2\Delta_{1}(-4+\Delta_{2})-8\Delta_{2}+\Delta_{2}^{2})}{(-3+\Delta_{1}-\Delta_{2})(5+\Delta_{1}-\Delta_{2})}\,.\\
\end{aligned}
\end{align}
If we impose conservation on \textit{both} the spin-1 operators, we get the following relations in terms of conformal dimension of the non-conserved spin-2 operator,
\begin{align}
\langle O_{2}J_{1}J_{1}\rangle\ :\ a_{2}=\tfrac{2a_{1}(\Delta_{1}-3)-a_{3}(\Delta_{1}-8)}{2+\Delta_{1}}\,,\quad
a_{4}=\tfrac{-8 a_{1}(\Delta_{1}-3)+2a_{3}(\Delta_{1}(\Delta_{1}+2)-20)}{(\Delta_{1}-6)(\Delta_{1}+2)}\,.
\end{align}
The conservation condition on the spin-2 operator $O_{2}$ is trivially satisfied and gives no further constraints, keeping the correlator still fixed upto two coefficients, and the conserved correlator looks like
\begin{align}
\langle J_{2}J_{1}J_{1}\rangle=\frac{1}{x_{12}^{2}x_{23}^{2}x_{31}^{2}}
\left(
a_{1} Q_{1}^{2}Q_{2}Q_{3}+
\tfrac{(a_{1}+2a_{3})}{3}Q_{1}^{2}\hat P_{23}+
a_{3}Q_{1}\left(\hat P_{31}Q_{2}+\hat P_{23}Q_{3}\right)+
\tfrac{2(a_{1}-a_{3})}{3}\hat P_{12}\hat P_{31}
\right)\,.
\end{align}

\subsubsection*{$\langle O_{3}O_{1}O_{1}\rangle$}

The allowed structures are severely restricted due to the point-switch symmetry, and we get one parity-even and one parity-odd term,
\begin{align*}
\begin{aligned}
{\rm even}:\quad & 
Q_{1}^{2}\left(Q_{2}\hat P_{31}-Q_{3}\hat P_{12} \right)\,,\\
{\rm odd}:\quad & 
Q_{1}^{2}P_{123+}\,.
\end{aligned}
\end{align*}
If any of the three operators are conserved, then the parity-even structure does not survive. The parity-odd part trivially satisfies conservation on $x_{2}$ and $x_{3}$, but for $x_{1}$ gives the constraint $\Delta_{2}=\Delta_{3}$. Hence, the correlator containing only conserved currents is expressible in terms of a single parity-odd structure
\begin{align}
\langle
J_{3}J_{1}J_{1}\rangle
=\frac{1}{x_{12}^{2}x_{23}^{2}x_{31}^{2}}\,Q_{1}^{2}P_{123+}\,.
\end{align}

\subsubsection*{$\langle O_{s}O_{1}O_{1}\rangle$}

For even values of $s>1$, the allowed linearly independent structures that preserve the point-switch symmetry are all parity-even and easily derivable since they are $Q_{1}^{s-2}$ times the structures for $\langle O_{2}O_{1}O_{1}\rangle$. The correlator with all the operators conserved follows the same prescription as $\langle O_{2}O_{1}O_{1}\rangle$. The form of the fully conserved correlator is fixed upto two unknown coefficients $a_{1},a_{3}$,
\begin{align}
\begin{aligned}
\langle J_{s={\rm even}}J_{1}J_{1}\rangle=\frac{1}{x_{12}^{2}x_{23}^{2}x_{31}^{2}}
\left(
a_{1} Q_{1}^{s}Q_{2}Q_{3}+
\frac{a_{1}(s-1)+2a_{3}}{s+1}Q_{1}^{s}\hat P_{23}+
a_{3}Q_{1}^{s-1}\left(\hat P_{31}Q_{2}+\hat P_{23}Q_{3}\right)\right.+&\\
\left.
\frac{s(s-1)(a_{1}-a_{3})}{s+1}Q_{1}^{s-2}\hat P_{12}\hat P_{31}
\right)&\,.
\end{aligned}
\end{align}
This expression works for $s=2$ as well.

For odd values of $s>1$, the analysis matches that with $\langle O_{3}O_{1}O_{1}\rangle$, and the allowed independent structures are $Q_{1}^{s-2}$ times those for $\langle O_{3}O_{1}O_{1}\rangle$. The parity-odd part trivially satisfies conservation, while the parity-even structure does not and needs to be removed. The conserved correlator only has a parity-odd contribution
\begin{align}
\langle J_{s={\rm odd}}J_{1}J_{1}\rangle=\frac{1}{x_{12}^{2}x_{23}^{2}x_{31}^{2}}
Q_{1}^{s-1}P_{123+}
\,.
\end{align}

\subsubsection*{$\langle O_{2}O_{2}O_{1} \rangle$}

The allowed structures are
\begin{align*}
\begin{aligned}
{\rm even}:\quad & 
Q_{1}Q_{2}\left(Q_{1}\hat P_{23}-Q_{2}\hat P_{31}\right)\,,\ 
\hat P_{12}\left(Q_{1}\hat P_{23}-Q_{2}\hat P_{31}\right)\,,\\
{\rm odd}:\quad & 
Q_{1}Q_{2}P_{123+}\,,\ 
\hat P_{12}P_{123+}\,,
\end{aligned}
\end{align*}
and the non-conserved correlator looks like
\begin{align}
\begin{aligned}
\langle
O_{2}O_{2}O_{1}
\rangle=
\frac{1}{|x_{12}|^{\tau_{12,3}}|x_{23}|
^{\tau_{23,1}}|x_{31}|^{\tau_{31,2}}}
\left[
a_{1}Q_{1}Q_{2}\left(Q_{1}\hat P_{23}-Q_{2}\hat P_{31}\right)+
a_{2}\hat P_{12}\left(Q_{1}\hat P_{23}-Q_{2}\hat P_{31}\right)\right.+&\\
\left.
b_{1}Q_{1}Q_{2}P_{123+}+b_{2}\hat P_{12}P_{123+}
\right]&\,.
\end{aligned}
\end{align}
Imposing conservation of $O_{1}$ at $x_{3}$ gives the constraints
\begin{align}
a_{1}=a_{2}=0\,,
\end{align}
i.e. the parity-odd part trivially satisfies conservation on $O_{1}$ at $x_{3}$, while the parity-even part does not and has to be removed.

When conservation on the spin-2 operators is imposed, we get the following constraints
\begin{align}
\Delta_{1}=\Delta_{2}=4\,,\quad  a_{1}=a_{2}=0\,,\quad b_{2}=\frac{2(\Delta_{3}-4)}{\Delta_{3}+1}b_{1}\,.
\end{align}
Hence, the completely conserved correlator consists of a single parity-odd structure
\begin{align}
\langle
J_{2}J_{2}J_{1}
\rangle=
\frac{1}{x_{12}^{2}x_{23}^{2}x_{31}^{2}}
\left(
Q_{1}Q_{2}-\frac12 \hat P_{12}
\right)P_{123+}\,.
\end{align}

\subsubsection*{$\langle O_{3}O_{2}O_{1} \rangle$}

There is no point-switch symmetry, and the non-conserved correlator contains 8 parity-even and 2 parity-odd linearly independent structures,
\begin{align}
\begin{aligned}
\langle
O_{3}O_{2}O_{1}
\rangle=
\frac{1}{|x_{12}|^{\tau_{12,3}}|x_{23}|
^{\tau_{23,1}}|x_{31}|^{\tau_{31,2}}}
\left[
a_{1}Q_{1}^{3}Q_{2}^{2}Q_{3}+a_{2}Q_{1}^{3}Q_{2}\hat P_{23}+a_{3}Q_{1}^{2}Q_{2}^{2}\hat P_{31}+a_{4}Q_{1}^{2}Q_{2}Q_{3}\hat P_{12}\,+\right.&\\
\left.
a_{5}Q_{1}^{2}\hat P_{12}\hat P_{23}+a_{6}Q_{1}Q_{2}\hat P_{12}\hat P_{31}+a_{7}Q_{1}Q_{3}\hat P_{12}^{2}+a_{8}\hat P_{12}^{2}\hat P_{31}+\left(b_{1}Q_{1}^{2}Q_{2}+b_{2}Q_{1}\hat P_{12}\right)P_{123+}
\right]&
\end{aligned}
\label{eq:j3j2j1cft}
\end{align}

Imposing conservation on all the three operators fixes the form of the correlator in terms of 2 parity-even ($a_{1},a_{2}$) and 1 parity-odd ($b_{1}$) coefficients. The relations are
\begin{align}
\begin{gathered}
a_{3}=\frac13(-a_{1}+4a_{2})\,,\ 
a_{4}=-2(a_{1}-a_{2})\,,\ 
a_{5}=\frac13(4a_{1}-7a_{2})\,,\ 
a_{6}=\frac13(10a_{1}-13a_{2})\,,\\
a_{7}=\frac13(a_{1}-4a_{2})\,,\ 
a_{8}=-a_{1}+a_{2}\,,\qquad
b_{2}=-b_{1}\,.
\end{gathered}
\end{align}
The form of $\langle J_{3}J_{2}J_{1}\rangle$ is easily obtained by applying $\tau_{ij,k}=2$ and the above relations in Eq. (\ref{eq:j3j2j1cft}).

\subsubsection*{$\langle O_{s}O_{2}O_{1} \rangle$}

For $s>3$, the independent structures are $Q_{1}^{s-3}$ times all the structures listed above for $\langle O_{3}O_{2}O_{1}\rangle$,
\begin{align}
\begin{aligned}
\langle
O_{s}O_{2}O_{1}
\rangle=
\frac{1}{|x_{12}|^{\tau_{12,3}}|x_{23}|
^{\tau_{23,1}}|x_{31}|^{\tau_{31,2}}}
\left[
a_{1}Q_{1}^{s}Q_{2}^{2}Q_{3}+a_{2}Q_{1}^{s}Q_{2}\hat P_{23}+a_{3}Q_{1}^{s-1}Q_{2}^{2}\hat P_{31}\right.+&\\
\left.
a_{4}Q_{1}^{s-1}Q_{2}Q_{3}\hat P_{12}+
a_{5}Q_{1}^{s-1}\hat P_{12}\hat P_{23}+a_{6}Q_{1}^{s-2}Q_{2}\hat P_{12}\hat P_{31}+a_{7}Q_{1}^{s-2}Q_{3}\hat P_{12}^{2}+a_{8}Q_{1}^{s-3}\hat P_{12}^{2}\hat P_{31}\right.+&\\
\left.
\left(b_{1}Q_{1}^{s-1}Q_{2}+
b_{2}Q_{1}^{s-2}\hat P_{12}\right)P_{123+}
\right]&\,.
\end{aligned}
\end{align}

The conservation condition reduces the independent structures to 3 (two parity-even and one parity-odd), and we have the following relations for $\langle J_{s}J_{2}J_{1}\rangle$
\begin{align}
\begin{gathered}
b_{2}=\tfrac{(1-s)}{2}b_{1}\,,\qquad a_{3}=\tfrac{(2-s)}{3}a_{1} +\tfrac{(1+s)}{3}a_{2}\,,\ 
a_{4}=(1-s)a_{1}+\tfrac{(1+s)}{2}a_{2}\,,\ 
a_{5}=\tfrac{(1+s)}{3}a_{1}+\tfrac{(1-5s)}{6}a_{2}\,,\\[.5em]
a_{6}=\tfrac{(s-1)(2s-1)}{3}a_{1}+\tfrac{s(3-4s)+1}{6}a_{2}\,,\ 
a_{7}=\tfrac{(s-1)(s-2)}{6}a_{1}+\tfrac{(1-s^{2})}{6}a_{2}\,,\ 
a_{8}=\tfrac{s(s-1)(s-2)}{6}(-a_{1}+a_{2})\,.
\end{gathered}
\end{align}

\subsubsection*{$\langle O_{3}O_{2}O_{2} \rangle$}

The correlator has point-switch symmetry under $2\leftrightarrow3$ swap. The correlator with the allowed conformally invariant structures are
\begin{align}
\begin{aligned}
\langle O_{3}O_{2}O_{2} \rangle=\frac{1}{|x_{12}|^{\tau_{12,3}}|x_{23}|
^{\tau_{23,1}}|x_{31}|^{\tau_{31,2}}}
\left\{
a_{1}Q_{1}^{2}Q_{2}Q_{3}\left(Q_{2}\hat P_{31}-Q_{3}\hat P_{12}\right)+
a_{2}Q_{1}^{2}\hat P_{23}\left(Q_{2}\hat P_{31}-Q_{3}\hat P_{12}\right)\right.+&\\
\left.
a_{3}Q_{1}\left(Q_{2}^{2}\hat P_{31}^{2}-Q_{3}^{2}\hat P_{12}^{2}\right)+
a_{4}\hat P_{12}\hat P_{31}\left(Q_{2}\hat P_{31}-Q_{3}\hat P_{12}\right)\right.+&\\
\left.
\left[
b_{1}Q_{1}^{2}Q_{2}Q_{3}+
b_{2}Q_{1}^{2}\hat P_{23}+
b_{3}Q_{1}\left(Q_{2}\hat P_{31}+ Q_{3}\hat P_{12}\right)+
b_{4}\hat P_{12}\hat P_{31}
\right]P_{123+}
\right\}&\,,
\end{aligned}
\end{align}
i.e. there are 4 parity-even and 4 parity-odd contributions. 

On imposing conservation on any or all of the operators, we get the following constraints
\begin{align}
a_{1}=a_{2}=a_{3}=a_{4}=0\,,\quad
b_{3}=\frac13(-b_{1}+4b_{2})\,,\ b_{4}=\frac13(b_{1}-b_{2})\,.
\end{align}
Hence, the conserved correlator $\langle J_{3}J_{2}J_{2}\rangle$ is completely parity-odd and determined upto 2 unknown coefficients.

\subsubsection*{$\langle O_{4}O_{2}O_{2} \rangle$}

There is a proliferation of allowed conformally invariant structures as we move up the spin, and we find 10 parity-even and 4 parity-odd linearly independent structures which preserve the $2\leftrightarrow3$ symmetry of this correlator,\footnote{The multiplicative factor containing $x_{ij}$'s has been omitted for convenience.}
\begin{align}
\begin{aligned}
\langle O_{4}O_{2}O_{2} \rangle\sim
\left\{
a_{1}Q_{1}^{4}Q_{2}^{2}Q_{3}^{2}+
a_{2}Q_{1}^{4}Q_{2}Q_{3}\hat P_{23}+
a_{3}Q_{1}^{4}\hat P_{23}^{2}+
a_{4}Q_{1}^{3}Q_{2}Q_{3}\left(Q_{2}\hat P_{31}+Q_{3}\hat P_{12}\right)+
a_{10}\hat P_{12}^{2}\hat P_{31}^{2}
\right.+&\\
\left.
a_{5}Q_{1}^{3}\hat P_{23}\left(Q_{2}\hat P_{31}+Q_{3}\hat P_{12}\right)+
a_{6}Q_{1}^{2}\left(Q_{2}^{2}\hat P_{31}^{2}+Q_{3}^{2}\hat P_{12}^{2}\right)+
a_{7}Q_{1}^{2}Q_{2}Q_{3}\hat P_{12}\hat P_{31}+
a_{8}Q_{1}^{2}\hat P_{12}\hat P_{23}\hat P_{31}\right.+&\\
\left.
a_{9}Q_{1}\hat P_{12}\hat P_{31}\left(Q_{2}\hat P_{31}+Q_{3}\hat P_{12}\right)+
b_{1}Q_{1}^{2}\left(Q_{2}\hat P_{31}-Q_{3}\hat P_{12}\right)P_{123+}
\right\}&\,.
\end{aligned}
\end{align}

After applying the conservation condition on the correlator, we get the following relations
\begin{align}
\begin{gathered}
a_{4}=\frac13(-4a_{1}+5a_{2})\,,\ 
a_{5}=\frac53a_{1}-\frac73a_{2}+5a_{3}\,,\ 
a_{6}=\frac16(a_{1}-5a_{2}+15a_{3})\,,\ 
a_{7}=\frac13(22a_{1}-23a_{2}+30a_{3})\\
a_{8}=-\frac{20}3a_{1}+\frac{25}{3}a_{2}-8a_{3}\,,\ 
a_{9}=-3a_{1}+4a_{2}-5a_{3}\,,\ 
a_{10}=a_{1}-a_{2}+a_{3}\,,\qquad
b_{1}=0\,.
\end{gathered}
\end{align}
Hence, only parity-even structures survive the conservation condition, and the $\langle J_{4}J_{2}J_{2}\rangle$ correlator is fixed upto 3 undetermined parity-even coefficients.

\subsubsection*{$\langle O_{s}O_{2}O_{2} \rangle$}

For odd values of $s\geq 3$, the structures are easily derivable from $\langle O_{3}O_{2}O_{2} \rangle$,
\begin{align}
\begin{aligned}
\langle O_{s={\rm odd}}O_{2}O_{2} \rangle\sim
\left\{
a_{1}Q_{1}^{s-1}Q_{2}Q_{3}\left(Q_{2}\hat P_{31}-Q_{3}\hat P_{12}\right)+
a_{2}Q_{1}^{s-1}\hat P_{23}\left(Q_{2}\hat P_{31}-Q_{3}\hat P_{12}\right)\right.+&\\
\left.
a_{3}Q_{1}^{s-2}\left(Q_{2}^{2}\hat P_{31}^{2}-Q_{3}^{2}\hat P_{12}^{2}\right)+
a_{4}Q_{1}^{s-3}\hat P_{12}\hat P_{31}\left(Q_{2}\hat P_{31}-Q_{3}\hat P_{12}\right)\right.+&\\
\left.
Q_{1}^{s-3}\left[
b_{1}Q_{1}^{2}Q_{2}Q_{3}+
b_{2}Q_{1}^{2}\hat P_{23}+
b_{3}Q_{1}\left(Q_{2}\hat P_{31}+Q_{3}\hat P_{12}\right)+
b_{4}\hat P_{12}\hat P_{31}
\right]P_{123+}
\right\}&\,.
\end{aligned}
\end{align}
The number of independent structures stay the same, 4 parity-even and 4 parity-odd structures.

After conservation, the parity-even part drops out, as it does for $s=3$, and we get the $\langle J_{s={\rm odd}}J_{2}J_{2}\rangle$ correlator determined upto 2 parity-odd coefficients
\begin{align}
b_{3}=\tfrac{(2-s)}{3}b_{1}+\tfrac{(1+s)}{3}b_{2}\,,\ 
b_{4}=\tfrac{(s-2)(s-1)}{6}b_{1}-\tfrac{(s-2)(s-1)}{6}b_{2}\,.
\end{align}

For even values of $s\geq4$, the analysis follows that of $\langle O_{4}O_{2}O_{2}\rangle$. The independent structures are $Q_{1}^{s-4}$ times the structures for $\langle O_{4}O_{2}O_{2}\rangle$, 
\begin{align}
\begin{aligned}
\langle O_{s={\rm even}}O_{2}O_{2} \rangle\sim
\left\{
a_{1}Q_{1}^{s}Q_{2}^{2}Q_{3}^{2}+
a_{2}Q_{1}^{s}Q_{2}Q_{3}\hat P_{23}+
a_{3}Q_{1}^{s}\hat P_{23}^{2}+
a_{4}Q_{1}^{s-1}Q_{2}Q_{3}\left(Q_{2}\hat P_{31}+Q_{3}\hat P_{12}\right)
\right.+&\\
\left.
a_{5}Q_{1}^{s-1}\hat P_{23}\left(Q_{2}\hat P_{31}+Q_{3}\hat P_{12}\right)+
a_{6}Q_{1}^{s-2}\left(Q_{2}^{2}\hat P_{31}^{2}+Q_{3}^{2}\hat P_{12}^{2}\right)+
a_{7}Q_{1}^{s-2}Q_{2}Q_{3}\hat P_{12}\hat P_{31}+
a_{8}Q_{1}^{s-2}\hat P_{12}\hat P_{23}\hat P_{31}\right.+&\\
\left.
a_{9}Q_{1}^{s-3}\hat P_{12}\hat P_{31}\left(Q_{2}\hat P_{31}+Q_{3}\hat P_{12}\right)+
a_{10}Q_{1}^{s-4}\hat P_{12}^{2}\hat P_{31}^{2}+
b_{1}Q_{1}^{s-2}\left(Q_{2}\hat P_{31}-Q_{3}\hat P_{12}\right)P_{123+}
\right\}&\,.
\end{aligned}
\end{align}
The conserved correlator $\langle J_{s={\rm even}}J_{2}J_{2}\rangle$ is fixed upto 3 parity-even structures, and there is no parity-odd contribution. The relations are
\begin{align}
\begin{gathered}
a_{4}=\tfrac{2(s-2)}{3}a_{1}+\tfrac{(1+s)}{3}a_{2}\,,\ 
a_{5}=\tfrac{(1+s)}{3}a_{1}+\tfrac{(1-2s)}{3}a_{2}+(1+s)a_{3}\,,\\
a_{6}=\tfrac{(s-3)(s-2)}{12}a_{1}+\tfrac{(2+s-s^{2})}{12}a_{2}+\tfrac{(s+1)(s+2)}{12}a_{3}\,,\ 
a_{7}=\tfrac{(8+s(5s-11))}{6}a_{1}+\tfrac{(1-2s(s-1))}{3}a_{2}+\tfrac{s(s+1)}{2}a_{3}\,,\\ 
a_{8}=\tfrac{(4+s-3s^{2})}{6}a_{1}+\tfrac{(1+2s(s-1))}{3}a_{2}+\tfrac{(4+s(7-5s))}{6}a_{3}\,,\ 
a_{9}=\tfrac{(s-2)(s-1)^{2}}{6}a_{1}+\tfrac{s(s-2)(s-1)}{6}a_{2}-\tfrac{(s-2)(s-1)(s+1)}{6}a_{3}\,,\\
a_{10}=\tfrac{s(s-3)(s-2)(-1+s)}{24}(a_{1}-a_{2}+a_{3})\,,\qquad b_{1}=0\,.
\end{gathered}
\end{align}

\section{Relations between superconformal invariants}
\label{appsec:relations}

We enumerate the various relations that hold between the superconformal invariants in Section \ref{ssec:scftinv-relation}.

For \ref{r2}, the relations are:
\begin{gather}
\hat R_{ij}^{2}+2\hat P_{ij}\left(\hat P_{ij}+Q_{ij+}\right)R'=0\,,\\
\hat R_{12}\hat R_{23}-
\left(
	2\hat P_{12}\hat P_{23}+\hat P_{12}Q_{23+}+\hat P_{23}Q_{12+}+Q_{12+}Q_{23+}+\hat P_{31}Q_{2+}^{2}
\right)R'=0\quad\text{and perm}\,,\\
\hat R_{12}R_{123}-2
\left[
	(Q_{12+}+\hat P_{12})(Q_{1+}\hat P_{23}+Q_{2+}\hat P_{31})-Q_{3+}\hat P_{12}^{2}+Q_{12+}^{2}Q_{3+}
\right]R'=0\quad\text{and perm}\,,\\
R_{123}^{2}-8
\left[
3 Q_{123+}^{2}+5Q_{123+}\sum_{\rm cyc}Q_{1+}\hat P_{23}+5\sum_{\rm cyc}Q_{12+}\hat P_{23}\hat P_{31}+2\sum_{\rm cyc}Q_{1+}^{2}\hat P_{23}^{2}	
\right]R'=0\,,\\
\hat R_{ij}T'=\hat R_{ij}Q_{i-}=\hat R_{ij}Q_{j-}=0\,,\\
\hat R_{12}Q_{3-}+P_{123+}R'=0\quad\text{and perm}\,,\\
R_{123}T'-3i\,P_{123+}R'=0\,,\\
R_{123}Q_{i-}-2Q_{i+}P_{123+}R'=0\,.
\end{gather}

For \ref{r3}, we have:
\begin{gather}
2T'^{2}+ R'=0\,,\\
Q_{i-}^{2}=0\,,\\
Q_{i-}Q_{j-}-\left(
\hat P_{ij}+Q_{ij+}
\right)R'=0\,,\\
2Q_{i-}T'-i Q_{i+}R'=0\,,\\
6 P_{123+}T'-i R_{123}+i\sum_{\rm cyc}Q_{1+}\hat R_{23}-i\left(9 Q_{123+}+\sum_{\rm cyc}Q_{1+}\hat P_{23}\right)R'=0\,,\\
\begin{aligned}
6 P_{123+}Q_{1-}-Q_{1+} R_{123}+2(3\hat P_{31}+2Q_{31+})\hat R_{12}-2Q_{1+}^{2}\hat R_{23}+2(3\hat P_{12}+2 Q_{12+})\hat R_{31}&+\\
2\left[
Q_{1+}\left(-3Q_{123+}+\sum_{\rm cyc}Q_{1+}\hat P_{23}\right)+6 \hat P_{12}\hat P_{31}
\right]R'=0\quad\text{and perm}\,.&
\end{aligned}
\end{gather}
We also find the superconformal version of Eq. (\ref{eq:p123square}),
\begin{align}
\begin{aligned}
P_{123+}^{2}&=Q_{123+}^{2}+2Q_{123+}\sum_{\rm cyc}Q_{1+}\hat P_{12}+\left(\sum_{\rm cyc}Q_{1+}\hat P_{12}\right)^{2}+4\hat P_{12}\hat P_{23}\hat P_{31}+\\
&\frac13 Q_{123+}\left(R_{123}+\sum_{\rm cyc}Q_{1+}\hat R_{12}\right)+
\frac13 \left(\sum_{\rm cyc}Q_{1+}\hat P_{23}\right)\left(R_{123}+\sum_{\rm cyc}Q_{1+}\hat R_{12}\right)+\\
&\left[
\frac{18}3 Q_{123+}^{2}+\frac{22}3Q_{123+}\sum_{\rm cyc}Q_{1+}\hat P_{23}+\frac43\sum_{\rm cyc}Q_{1+}^{2}\hat P_{23}^{2}+\frac{14}3\sum_{\rm cyc}Q_{12+}\hat P_{23}\hat P_{31}+2\hat P_{12}\hat P_{23}\hat P_{31}
\right]R'\,.
\end{aligned}
\label{eq:p123squarescft}
\end{align}

\section{More 3-point SCFT correlators with conservation constraints}
\label{appsec:morescft3ptcorr}

We present more examples of 3-point correlators with one or more operators subjected to the shortening condition Eq. (\ref{eq:shortening}). The fully conserved correlators can be found in Section \ref{sec:scftconscorr}.

\subsubsection*{$\langle {\cal J}_{3}{\cal O}_{1}{\cal O}_{0} \rangle$}

The non-conserved correlator is given in Eq. (\ref{eq:scftoso1o0}) with $s=3$. When the spin-3 operator is conserved, we get the relations:
\begin{align}
\Delta_{3}=\Delta_{2}-1\,,\quad
\begin{gathered}
a_{2}=8a_{1}\,,\quad
a_{3}=-3a_{1}\,,\quad
a_{4}=-12a_{1}\,,\quad
a_{5}=-3a_{1}\,,\\
b_{2}=3b_{1}\,,\quad
b_{3}=\tfrac{3i}{2}b_{1}\,,\quad
b_{4}=-\tfrac{3i}{2}b_{1}\,,\quad
b_{5}=-i\tfrac{3i}{2}b_{1}\,.
\end{gathered}
\end{align}
The conformal dimension of the scalar operator is restricted and the correlator is fixed in terms of 1 parity-even and 1 parity-odd coefficients.

\subsubsection*{$\langle {\cal O}_{3}{\cal J}_{1}{\cal O}_{0} \rangle$}

The non-conserved correlator is again given in Eq. (\ref{eq:scftoso1o0}) with $s=3$. When we apply the shortening condition on the spin-1 operator, we get
\begin{align}
\begin{gathered}
a_{2}=\tfrac{(7-\Delta_{1}+\Delta_{3})(5+\Delta_{1}-\Delta_{3})}{4}a_{1}\,,\ 
a_{3}=\tfrac{2(\Delta_{1}-\Delta_{3})}{-5+\Delta_{1}-\Delta_{3}}a_{1}\,,\ 
a_{4}=-\tfrac{(\Delta_{1}-\Delta_{3})(5+\Delta_{1}-\Delta_{3})}{2}a_{1}\,,\ 
a_{5}=-\tfrac{3+\Delta_{1}-\Delta_{3}}{2}a_{1}\,,\\
b_{2}=\tfrac{3(1+\Delta_{1}-\Delta_{3})}{2(-1+\Delta_{1}-\Delta_{3})}b_{1}\,,\ 
b_{3}=\tfrac{3i(5-(\Delta_{1}-\Delta_{3})^2)}{2(5-\Delta_{1}+\Delta_{3})(1-\Delta_{1}+\Delta_{3})}b_{1}\,,\ 
b_{4}=-\tfrac{3i(1+\Delta_{1}-\Delta_{3})}{4(1-\Delta_{1}+\Delta_{3})}b_{1}\,,\ 
b_{5}=-\tfrac{i(5+\Delta_{1}-\Delta_{3})}{4(1-\Delta_{1}+\Delta_{3})}b_{1}\,.
\end{gathered}
\end{align}
Note that here we get no constraints on the dimensions of ${\cal O}_{3}$ and ${\cal O}_{0}$. The correlator is still fixed in terms of 1 parity-even and 1 parity-odd coefficients.

\subsubsection*{$\langle {\cal J}_{2}{\cal O}_{2}{\cal O}_{0} \rangle$}

The non-conserved correlator is given in Eq. (\ref{eq:scfto2o2o0}). As alluded to in the main text, imposing conservation on any of the spin-2 operators gives identical relations. For instance, the relations for $\langle {\cal J}_{2}{\cal O}_{2}{\cal O}_{0} \rangle$ are,
\begin{align}
\begin{gathered}
b_{1}=b_{2}=0\,,\quad
a_{2}=-\tfrac{(-8+\Delta_{2}-\Delta_{3})(3+\Delta_{2}-\Delta_{3})}{4}a_{1}\,,\quad
a_{3}=\tfrac{2(-8+3\Delta_{2}^{2}-6\Delta_{2}\Delta_{3}+3\Delta_{3}^{2})}{24+\Delta_{2}^{2}+10\Delta_{3}+\Delta_{3}^{2}-2\Delta_{1}(5+\Delta_{3})}a_{1}\,,\\
a_{4}=\tfrac{(24-3\Delta_{2}^{3}-12\Delta_{3}-8\Delta_{3}^{2}+3\Delta_{3}^{3}+\Delta_{2}^{2}(-8+9\Delta_{3})+\Delta_{2}(12+16\Delta_{3}-9\Delta_{3}^{2}))}{2(-6+\Delta_{2}-\Delta_{3})}a_{1}\,,\quad
a_{5}=\tfrac{2(2+3\Delta_{2}-3\Delta_{3})}{-6+\Delta_{2}-\Delta_{3}}a_{1}\,,\\
a_{6}=-\tfrac{(4+3\Delta_{2}^{2}-6\Delta_{2}(-2+\Delta_{3})-12\Delta_{3}+3\Delta_{3}^{2})}{2}a_{1}\,,\quad
a_{7}=-\tfrac{2(-2+\Delta_{2}+\Delta_{2}^{2}-\Delta_{3}-2\Delta_{2}\Delta_{3}+\Delta_{3}^{2})}{-6+\Delta_{2}+\Delta_{3}}a_{1}\,,\quad
a_{8}={\scriptstyle (-2-\Delta_{2}+\Delta_{3})}a_{1}\,.
\end{gathered}
\end{align}
We get identical relations for $\langle {\cal O}_{2}{\cal J}_{2}{\cal O}_{0} \rangle$ when the second spin-2 operator is conserved.

\subsubsection*{$\langle {\cal J}_{1}{\cal O}_{1}{\cal O}_{1} \rangle$}

The non-conserved correlator is given in Eq. (\ref{eq:scfto1o1o1}). Note that we expect equivalent relations when the shortening condition is imposed on any one of the spin-1 operators. For $\langle {\cal J}_{1}{\cal O}_{1}{\cal O}_{1} \rangle$, we get,
\begin{align}
\Delta_{3}=\Delta_{2}\,,\qquad
b_{3}=-2i(7b_{1}-3b_{2})\,,\ 
b_{4}=-2i(5b_{1}-2b_{2})\,,\ 
b_{5}=2b_{1}-b_{2}\,,\ 
b_{6}=-4b_{1}+2b_{2}\,.
\end{align}
Similarly, imposing conservation on $\check z_2,\check z_3$ gives us equivalent relations for $\langle {\cal O}_{1}{\cal J}_{1}{\cal O}_{1} \rangle,\,\langle {\cal O}_{1}{\cal O}_{1}{\cal J}_{1} \rangle$ respectively.

\subsubsection*{$\langle {\cal J}_{2}{\cal O}_{1}{\cal O}_{1} \rangle$}

The non-conserved correlator is given in Eq. (\ref{eq:scfto2o1o1}). When conservation is imposed on only the spin-2 operator, one obtains the relations:
\begin{align}
\Delta_{2}=\Delta_{3}\,,\quad 
\begin{gathered}
a_{5}=\tfrac1{48}(37a_{1}-4a_{2}-15(3a_{3}-a_{4}))\,,\quad
a_{6}=\tfrac14(27a_{1}-4a_{2}-9(3a_{3}-a_{4}))\,,\\
a_{7}=\tfrac{1}{32}(-5a_{1}+4a_{2}+45a_{3}-15a_{4})\,,\quad 
a_{8}=\tfrac14(9a_{1}+15a_{3}-5a_{4})\,,\\
a_{9}=\tfrac{1}{16}(-27a_{1}+4a_{2}+3a_{3}-a_{4})\,,\quad
a_{10}=\tfrac{1}{16}(21a_{1}-4a_{2}+3a_{3}-a_{4})\,,\\
a_{11}=\tfrac{1}{32}(71a_{1}-12a_{2}-63a_{3}+21a_{4})\,,\quad 
a_{12}=\tfrac{1}{16}(-23a_{1}+4a_{2}+15a_{3}-5a_{4})\,,\\
b_{1}=b_{2}=b_{3}=b_{4}=0\,.
\end{gathered}
\end{align}
Thus, the correlator is fixed in terms of 4 parity-even coefficients, and the parity-odd part drops out. The dimensions of the non-conserved spin-1 operators are also restricted to be equal.

\subsubsection*{$\langle {\cal O}_{2}{\cal J}_{1}{\cal J}_{1} \rangle$}

If we instead consider the correlator with two spin-1 supercurrents and a general spin-2 operator, we get the relations:
\begin{align}
\begin{gathered}
a_{3}=\tfrac{(4a_{2}+a_{1}(-24-4\Delta_{1}+\Delta_{1}^{2}))}{8}\,,\ 
a_{4}=\tfrac{-4a_{2}(288+76\Delta_{1}-48\Delta_{1}^{2}+3\Delta_{1}^{3})+a_{1}(7168+3072\Delta_{1}-1232\Delta_{1}^{2}-196\Delta_{1}^{3}+60\Delta_{1}^{4}-3\Delta_{1}^{5})}{96(-8+\Delta_{1})}\,,\\
a_{5}=-\tfrac{4a_{2}(-20+2\Delta_{1}+\Delta_{1}^{2})+a_{1}(384+112\Delta_{1}-68\Delta_{1}^{2}-2\Delta_{1}^{3}+\Delta_{1}^{4})}{4(-8+\Delta_{1})(-6+\Delta_{1})}\,,\ 
a_{6}=\tfrac{4a_{2}(-112-4\Delta_{1}+8\Delta_{1}^{2}+\Delta_{1}^{3})+a_{1}(2304+864\Delta_{1}-304\Delta_{1}^{2}-76\Delta_{1}^{3}+4\Delta_{1}^{4}+\Delta_{1}^{5})}{16(-8+\Delta_{1})}\,,\\
a_{7}=\tfrac{-4a_{2}(2+\Delta_{1})+a_{1}\Delta_{1}(48+2\Delta_{1}-\Delta_{1}^{2})}{8(-8+\Delta_{1})}\,,\ 
a_{8}=\tfrac{4a_{2}\Delta_{1}(-64-6\Delta_{1}+3\Delta_{1}^{2})+a_{1}(-2048+1536\Delta_{1}+736\Delta_{1}^{2}-160\Delta_{1}^{3}-18\Delta_{1}^{4}+3\Delta_{1}^{5})}{96(-8+\Delta_{1})}\,,\\
a_{9}=\tfrac{-8a_{2}(-3+\Delta_{1})+a_{1}(-112+6\Delta_{1}+17\Delta_{1}^{2}-2\Delta_{1}^{3})}{6(-8+\Delta_{1})}\,,\ 
a_{10}=\tfrac{4a_{2}(-48+2\Delta_{1}+3\Delta_{1}^{2})+a_{1}(1024+384\Delta_{1}-176\Delta_{1}^{2}-10\Delta_{1}^{3}+3\Delta_{1}^{4})}{48(-8+\Delta_{1})}\,,\\
a_{11}=\tfrac{(-3+\Delta_{1})(4a_{2}(6+\Delta_{1})+a_{1}(-128-64\Delta_{1}+2\Delta_{1}^{2}+\Delta_{1}^{3}))}{8(-8+\Delta_{1})}\,,\ 
a_{12}=-\tfrac{4a_{2}(12+\Delta_{1})+a_{1}(-256-96\Delta_{1}+8\Delta_{1}^{2}+\Delta_{1}^{3})}{24(-8+\Delta_{1})}\,,\\
b_{1}=b_{2}=b_{3}=b_{4}=0\,.
\end{gathered}
\end{align}
Thus, $\langle {\cal O}_{2}{\cal J}_{1}{\cal J}_{1} \rangle$ is fixed in terms of 2 parity-even coefficients, and the parity-odd contribution drops out. The correlator with two conserved spin-1 supercurrents and a general spinning superfield operator was exclusively studied in \cite{Manenti4d3pt2018}.

\subsubsection*{$\langle {\cal O}_{3}{\cal J}_{1}{\cal J}_{1} \rangle$}

The non-conserved correlator is given in Eq. (\ref{eq:scfto3o1o1}). When both the spin-1 operators are conserved, we get the constraints,
\begin{align}
\begin{gathered}
b_{3}=-\tfrac{i(4b_{2}(9-19\Delta_{1}+2\Delta_{1}^{2})+b_{1}(-273+389\Delta_{1}+55\Delta_{1}^{2}-29\Delta_{1}^{3}+2\Delta_{1}^{4}))}{2(-6+\Delta_{1}+\Delta_{1}^{2})}\,,\ 
b_{4}=-\tfrac{i(4b_{2}(-9+\Delta_{1})+b_{1}(133+51\Delta_{1}-17\Delta_{1}^{2}+\Delta_{1}^{3}))}{4(-2+\Delta_{1})}\,,\\
b_{5}=-\tfrac{i(12b_{2}(1-6\Delta_{1}+\Delta_{1}^{2})+b_{1}(-147+392\Delta_{1}+24\Delta_{1}^{2}-32\Delta_{1}^{3}+3\Delta_{1}^{4}))}{2(-6+\Delta_{1}+\Delta_{1}^{2})}\,,\ 
b_{6}=-\tfrac{i(4b_{2}(-9-16\Delta_{1}+5\Delta_{1}^{2})+b_{1}(69+416\Delta_{1}-34\Delta_{1}^{2}-40\Delta_{1}^{3}+5\Delta_{1}^{4}))}{2(-6+\Delta_{1}+\Delta_{1}^{2})}\,,\\
b_{7}=\tfrac{4b_{2}(21+20\Delta_{1}-5\Delta_{1}^{2})+b_{1}(-321-548\Delta_{1}+30\Delta_{1}^{2}+44\Delta_{1}^{3}-5\Delta_{1}^{4})}{2(-21-4\Delta_{1}+\Delta_{1}^{2})}\,,\ 
b_{8}=-\tfrac{4b_{2}-b_{1}(-9-8\Delta_{1}+\Delta_{1}^{2})}{4}\,,\\
b_{9}=\tfrac{4b_{2}(9+16\Delta_{1}-5\Delta_{1}^{2})+b_{1}(-117-408\Delta_{1}+42\Delta_{1}^{2}+40\Delta_{1}^{3}-5\Delta_{1}^{4})}{4(-6+\Delta_{1}+\Delta_{1}^{2})}\,,\ 
b_{10}=-\tfrac{(-9+\Delta_{1})(4b_{2}+b_{1}(-21-4\Delta_{1}+\Delta_{1}^{2}))}{8(-2+\Delta_{1})}\,,\\
b_{11}=\tfrac{-4b_{2}+b_{1}(21+4\Delta_{1}-\Delta_{1}^{2})}{4}\,,\ 
b_{12}=\tfrac{-12b_{2}(1-6\Delta_{1}+\Delta_{1}^{2})+b_{1}(135-402\Delta_{1}-20\Delta_{1}^{2}+34\Delta_{1}^{3}-3\Delta_{1}^{4})}{4(-6+\Delta_{1}+\Delta_{1}^{2})}\,,\\
a_{1}=a_{2}=a_{3}=a_{4}=0\,.
\end{gathered}
\end{align}
For this (partially) conserved correlator, the parity-even part vanishes, and we are left with 2 unknown parity-odd coefficients.

\subsubsection*{$\langle {\cal O}_{2}{\cal O}_{2}{\cal J}_{1} \rangle$}

The non-conserved correlator has the form given in Eq. (\ref{eq:scfto2o2o1}). When the spin-1 operator is constrained to be conserved, we get the relations:
\begin{align}
\Delta_{1}=\Delta_{2}\,,\qquad
\begin{gathered}
b_{9}=\tfrac{2i}{6}((14b_{1}-2b_{2}-9b_{3}+3b_{4}-3b_{5})+3b_{6}+8b_{8})\,,\ 
b_{10}=6b_{1}-b_{2}\,,\\
b_{11}=b_{1}+2b_{3}-b_{4}+3b_{5}\,,\ 
b_{12}=-4b_{1}+b_{2}-ib_{8}\,,\\
b_{13}=\tfrac13(-7b_{1}+b_{2}-12b_{3}+6b_{4}-15b_{5}-4ib_{8})\,,\ 
b_{14}=-b_{1}-3b_{3}+b_{4}-4b_{5}+\tfrac i2b_{7}\,,\\ 
b_{15}=2b_{1}-\tfrac i2b_{8}\,,\ 
b_{16}=\tfrac{1}{12}(40b_{1}-4b_{2}-6b_{3}+6b_{4}- 6b_{5}-3ib_{6}-8ib_{8})\,,\\
b_{17}=\tfrac13(10b_{1}-b_{2}-2ib_{8})\,,\quad
a_{1}=a_{2}=a_{3}=a_{4}=a_{5}=a_{6}=a_{7}=0\,.
\end{gathered}
\end{align}
Similar to other results, the conservation fixes the dimensions of the spin-2 operators to be equal, and the correlator is known upto 8 parity-odd coefficients, while the parity-even structures vanish.

\subsubsection*{$\langle {\cal J}_{2}{\cal J}_{2}{\cal O}_{1} \rangle$}

When we consider the correlator with two conserved spin-2 supercurrents along with a spin-1 operator, we get the relations:
\begin{align}
\begin{gathered}
b_{2}=-\tfrac{(226-145\Delta_{3}-40\Delta_{3}^{2}+7\Delta_{3}^{3})}{28(-1+\Delta_{3})}b_{1}\,,\ 
b_{3}=\tfrac{2(-4+\Delta_{3})}{1+\Delta_{3}}b_{1}\,,\ 
b_{4}=\tfrac{(451-592\Delta_{3}+171\Delta_{3}^{2}-14\Delta_{3}^{3})}{28(-1+\Delta_{3})}b_{1}\,,\ 
b_{5}=\tfrac{(-9+\Delta_{3})}{4}b_{1}\,,\\
b_{6}=\tfrac{i(215-81\Delta_{3}-11\Delta_{3}^{2}+5\Delta_{3}^{3})}{7(-1+\Delta_{3}^{2})}b_{1}\,,\ 
b_{7}=\tfrac{i(-461+513\Delta_{3}-131\Delta_{3}^{2}+15\Delta_{3}^{3})}{14(-1+\Delta_{3}^{2})}b_{1}\,,\
b_{8}=-\tfrac{i(3+10\Delta_{3}+3\Delta_{3}^{2})}{14(-1+\Delta_{3})}b_{1}\,,\\
b_{9}=-\tfrac{i(-303+227\Delta_{3}-25\Delta_{3}^{2}+5\Delta_{3}^{3})}{14(-1+\Delta_{3}^{2})}b_{1}\,,\ 
b_{10}=-\tfrac{(-79+26\Delta_{3}+5\Delta_{3}^{2})}{28(-1+\Delta_{3})}b_{1}\,,\ 
b_{11}=-\tfrac{(5-2\Delta_{3}+5\Delta_{3}^{2})}{14(-1+\Delta_{3})}b_{1}\,,\\
b_{12}=\tfrac{3(-6+\Delta_{3}+\Delta_{3}^{2})}{14(-1+\Delta_{3})}b_{1}\,,\ 
b_{13}=\tfrac{(71-62\Delta_{3}+15\Delta_{3}^{2})}{14(-1+\Delta_{3})}b_{1}\,,\ 
b_{14}=\tfrac{(-307+401\Delta_{3}-117\Delta_{3}^{2}+15\Delta_{3}^{3})}{14(-1+\Delta_{3}^{2})}\,,\\
b_{15}=-\tfrac{(-109+10\Delta_{3}+3\Delta_{3}^{2})}{28(-1+\Delta_{3})}b_{1}\,,\ 
b_{16}=-\tfrac{(89-30\Delta_{3}+5\Delta_{3}^{2})}{28(-1+\Delta_{3})}b_{1}\,,\ 
b_{17}=-\tfrac{(-55+22\Delta_{3}+\Delta_{3}^{2})}{28(-1+\Delta_{3})}b_{1}\,,\\
a_{1}=a_{2}=a_{3}=a_{4}=a_{5}=a_{6}=a_{7}=0\,.
\end{gathered}
\end{align}
The correlator is fixed upto a single parity-odd coefficient, and depends on the dimension of the spin-1 operator. It is worth noting that the fully conserved correlator $\langle {\cal J}_{2}{\cal J}_{2}{\cal J}_{1} \rangle$ in Eq. (\ref{eq:scftj2j2j1}) is fixed in terms of 2 unknown coefficients, while the partially conserved correlator above is fixed in terms of a single coefficient.

\subsubsection*{$\langle {\cal J}_{2}{\cal O}_{2}{\cal O}_{2} \rangle$}

The analysis of this correlator is similar to $\langle {\cal J}_{1}{\cal O}_{1}{\cal O}_{1} \rangle$. Since there are three identical spin-2 operators, the constraints for conservation on any one is identical to the others. The non-conserved correlator is in Eq. (\ref{eq:scfto2o2o2}). On implementing conservation on $\check z_{1}$, the we get exactly the relations as in Eq. (\ref{eq:scftj2j2j2}) along with the constraint $\Delta_{2}=\Delta_{3}$. Thus, the correlator is fixed upto two parity-even coefficients. 

Similarly, if conservation is instead imposed on $\check z_{2}$ (or $\check z_{3}$), we get the same relations and an equivalent restriction $\Delta_{3}=\Delta_{1}$ (or $\Delta_{1}=\Delta_{2}$).

\newpage

\bibliographystyle{JHEP}

\bibliography{4dn1v2}

\providecommand{\href}[2]{#2}\begingroup\raggedright\begin{thebibliography}{10}

\bibitem{GPY2011}
S.~Giombi, S.~Prakash and X.~Yin, \emph{{A Note on CFT Correlators in Three
  Dimensions}}, \href{https://doi.org/10.1007/JHEP07(2013)105}{\emph{JHEP}
  {\bfseries 07} (2013) 105} [\href{https://arxiv.org/abs/1104.4317}{{\ttfamily
  1104.4317}}].

\bibitem{Park4dN1SCFT1997}
J.-H.~Park, \emph{{N=1 superconformal symmetry in four-dimensions}},
  \href{https://doi.org/10.1142/S0217751X98000755}{\emph{Int. J. Mod. Phys. A}
  {\bfseries 13} (1998) 1743}
  [\href{https://arxiv.org/abs/hep-th/9703191}{{\ttfamily hep-th/9703191}}].

\bibitem{Park4dSCFT1999}
J.-H.~Park, \emph{{Superconformal symmetry and correlation functions}},
  \href{https://doi.org/10.1016/S0550-3213(99)00432-0}{\emph{Nucl. Phys. B}
  {\bfseries 559} (1999) 455}
  [\href{https://arxiv.org/abs/hep-th/9903230}{{\ttfamily hep-th/9903230}}].

\bibitem{Park6d4dSCFT1998}
J.-H.~Park, \emph{{Superconformal symmetry in six-dimensions and its reduction
  to four-dimensions}},
  \href{https://doi.org/10.1016/S0550-3213(98)00720-2}{\emph{Nucl. Phys. B}
  {\bfseries 539} (1999) 599}
  [\href{https://arxiv.org/abs/hep-th/9807186}{{\ttfamily hep-th/9807186}}].

\bibitem{Osborn4dN1SCFT1998}
H.~Osborn, \emph{{N=1 superconformal symmetry in four-dimensional quantum field
  theory}}, \href{https://doi.org/10.1006/aphy.1998.5893}{\emph{Annals Phys.}
  {\bfseries 272} (1999) 243}
  [\href{https://arxiv.org/abs/hep-th/9808041}{{\ttfamily hep-th/9808041}}].

\bibitem{Osborn4dSCFTchiral2000}
F.A.~Dolan and H.~Osborn, \emph{{Implications of N=1 superconformal symmetry
  for chiral fields}},
  \href{https://doi.org/10.1016/S0550-3213(00)00553-8}{\emph{Nucl. Phys. B}
  {\bfseries 593} (2001) 599}
  [\href{https://arxiv.org/abs/hep-th/0006098}{{\ttfamily hep-th/0006098}}].

\bibitem{KuzenkoN2SCFT1999}
S.M.~Kuzenko and S.~Theisen, \emph{{Correlation functions of conserved currents
  in N=2 superconformal theory}},
  \href{https://doi.org/10.1088/0264-9381/17/3/307}{\emph{Class. Quant. Grav.}
  {\bfseries 17} (2000) 665}
  [\href{https://arxiv.org/abs/hep-th/9907107}{{\ttfamily hep-th/9907107}}].

\bibitem{OsbornCFTGenD1993}
H.~Osborn and A.C.~Petkou, \emph{{Implications of conformal invariance in field
  theories for general dimensions}},
  \href{https://doi.org/10.1006/aphy.1994.1045}{\emph{Annals Phys.} {\bfseries
  231} (1994) 311} [\href{https://arxiv.org/abs/hep-th/9307010}{{\ttfamily
  hep-th/9307010}}].

\bibitem{OsbornCFTGenD1996}
J.~Erdmenger and H.~Osborn, \emph{{Conserved currents and the energy momentum
  tensor in conformally invariant theories for general dimensions}},
  \href{https://doi.org/10.1016/S0550-3213(96)00545-7}{\emph{Nucl. Phys. B}
  {\bfseries 483} (1997) 431}
  [\href{https://arxiv.org/abs/hep-th/9605009}{{\ttfamily hep-th/9605009}}].

\bibitem{Polyakov1970}
A.M.~Polyakov, \emph{{Conformal symmetry of critical fluctuations}},
  {\emph{JETP Lett.} {\bfseries 12} (1970) 381}.

\bibitem{Schreier1971}
E.J.~Schreier, \emph{{Conformal symmetry and three-point functions}},
  \href{https://doi.org/10.1103/PhysRevD.3.980}{\emph{Phys. Rev. D} {\bfseries
  3} (1971) 980}.

\bibitem{Park3DSCFT1999}
J.-H.~Park, \emph{{Superconformal symmetry in three-dimensions}},
  \href{https://doi.org/10.1063/1.1290056}{\emph{J. Math. Phys.} {\bfseries 41}
  (2000) 7129} [\href{https://arxiv.org/abs/hep-th/9910199}{{\ttfamily
  hep-th/9910199}}].

\bibitem{CPPR2011}
M.S.~Costa, J.~Penedones, D.~Poland and S.~Rychkov, \emph{{Spinning Conformal
  Correlators}}, \href{https://doi.org/10.1007/JHEP11(2011)071}{\emph{JHEP}
  {\bfseries 11} (2011) 071} [\href{https://arxiv.org/abs/1107.3554}{{\ttfamily
  1107.3554}}].

\bibitem{MZ3dCFTHigh2011}
J.~Maldacena and A.~Zhiboedov, \emph{{Constraining Conformal Field Theories
  with A Higher Spin Symmetry}},
  \href{https://doi.org/10.1088/1751-8113/46/21/214011}{\emph{J. Phys. A}
  {\bfseries 46} (2013) 214011}
  [\href{https://arxiv.org/abs/1112.1016}{{\ttfamily 1112.1016}}].

\bibitem{MZ3dCFTBroken2012}
J.~Maldacena and A.~Zhiboedov, \emph{{Constraining conformal field theories
  with a slightly broken higher spin symmetry}},
  \href{https://doi.org/10.1088/0264-9381/30/10/104003}{\emph{Class. Quant.
  Grav.} {\bfseries 30} (2013) 104003}
  [\href{https://arxiv.org/abs/1204.3882}{{\ttfamily 1204.3882}}].

\bibitem{Stanev2012}
Y.S.~Stanev, \emph{{Correlation Functions of Conserved Currents in Four
  Dimensional Conformal Field Theory}},
  \href{https://doi.org/10.1016/j.nuclphysb.2012.07.027}{\emph{Nucl. Phys. B}
  {\bfseries 865} (2012) 200}
  [\href{https://arxiv.org/abs/1206.5639}{{\ttfamily 1206.5639}}].

\bibitem{Todorov2011}
I.~Todorov, \emph{{Conformal field theories with infinitely many conservation
  laws}}, \href{https://doi.org/10.1063/1.4790408}{\emph{J. Math. Phys.}
  {\bfseries 54} (2013) 022303}
  [\href{https://arxiv.org/abs/1207.3661}{{\ttfamily 1207.3661}}].

\bibitem{Karateev4d2014}
E.~Elkhidir, D.~Karateev and M.~Serone, \emph{{General Three-Point Functions in
  4D CFT}}, \href{https://doi.org/10.1007/JHEP01(2015)133}{\emph{JHEP}
  {\bfseries 01} (2015) 133} [\href{https://arxiv.org/abs/1412.1796}{{\ttfamily
  1412.1796}}].

\bibitem{AlbaDiabd42013}
V.~Alba and K.~Diab, \emph{{Constraining conformal field theories with a higher
  spin symmetry in d=4}},  \href{https://arxiv.org/abs/1307.8092}{{\ttfamily
  1307.8092}}.

\bibitem{AlbaDiab2015}
V.~Alba and K.~Diab, \emph{{Constraining conformal field theories with a higher
  spin symmetry in $d > 3$ dimensions}},
  \href{https://doi.org/10.1007/JHEP03(2016)044}{\emph{JHEP} {\bfseries 03}
  (2016) 044} [\href{https://arxiv.org/abs/1510.02535}{{\ttfamily
  1510.02535}}].

\bibitem{Nizami2013}
A.A.~Nizami, T.~Sharma and V.~Umesh, \emph{{Superspace formulation and
  correlation functions of 3d superconformal field theories}},
  \href{https://doi.org/10.1007/JHEP07(2014)022}{\emph{JHEP} {\bfseries 07}
  (2014) 022} [\href{https://arxiv.org/abs/1308.4778}{{\ttfamily 1308.4778}}].

\bibitem{JainNizami2022}
A.~Jain and A.A.~Nizami, \emph{{Superconformal invariants and spinning
  correlators in 3d ${{{\mathcal {N}}}}=2$ SCFTs}},
  \href{https://doi.org/10.1140/epjc/s10052-022-11016-2}{\emph{Eur. Phys. J. C}
  {\bfseries 82} (2022) 1065}
  [\href{https://arxiv.org/abs/2205.11157}{{\ttfamily 2205.11157}}].

\bibitem{BuchKuz3dSCFT2015}
E.I.~Buchbinder, S.M.~Kuzenko and I.B.~Samsonov, \emph{{Superconformal field
  theory in three dimensions: Correlation functions of conserved currents}},
  \href{https://doi.org/10.1007/JHEP06(2015)138}{\emph{JHEP} {\bfseries 06}
  (2015) 138} [\href{https://arxiv.org/abs/1503.04961}{{\ttfamily
  1503.04961}}].

\bibitem{Buch3dSCFTGen2023}
E.I.~Buchbinder and B.J.~Stone, \emph{{Three-point functions of conserved
  supercurrents in 3D N=1 SCFT: General formalism for arbitrary superspins}},
  \href{https://doi.org/10.1103/PhysRevD.107.106001}{\emph{Phys. Rev. D}
  {\bfseries 107} (2023) 106001}
  [\href{https://arxiv.org/abs/2302.00593}{{\ttfamily 2302.00593}}].

\bibitem{Buch3dSCFTMixed2021}
E.I.~Buchbinder and B.J.~Stone, \emph{{Mixed three-point functions of conserved
  currents in three-dimensional superconformal field theory}},
  \href{https://doi.org/10.1103/PhysRevD.103.086023}{\emph{Phys. Rev. D}
  {\bfseries 103} (2021) 086023}
  [\href{https://arxiv.org/abs/2102.04827}{{\ttfamily 2102.04827}}].

\bibitem{Buch3dSCFTN1spin22021}
E.I.~Buchbinder and B.J.~Stone, \emph{{Three-point functions of a superspin-2
  current multiplet in 3D, N=1 superconformal theory}},
  \href{https://doi.org/10.1103/PhysRevD.104.106004}{\emph{Phys. Rev. D}
  {\bfseries 104} (2021) 106004}
  [\href{https://arxiv.org/abs/2108.01865}{{\ttfamily 2108.01865}}].

\bibitem{Buch3dSCFTOdd2023}
E.I.~Buchbinder and B.J.~Stone, \emph{{Grassmann-odd three-point functions of
  conserved supercurrents in 3D N=1 SCFT}},
  \href{https://doi.org/10.1103/PhysRevD.108.046001}{\emph{Phys. Rev. D}
  {\bfseries 108} (2023) 046001}
  [\href{https://arxiv.org/abs/2305.02233}{{\ttfamily 2305.02233}}].

\bibitem{Buch4DSCFTN12022}
E.I.~Buchbinder, J.~Hutomo and G.~Tartaglino-Mazzucchelli, \emph{{Three-Point
  Functions of Higher-Spin Supercurrents in 4D ${\cal N}=1$ Superconformal
  Field Theory}}, \href{https://doi.org/10.1002/prop.202200133}{\emph{Fortsch.
  Phys.} {\bfseries 70} (2022) 2200133}
  [\href{https://arxiv.org/abs/2208.07057}{{\ttfamily 2208.07057}}].

\bibitem{Buch4dSCFTN1higherspin2021}
E.I.~Buchbinder, J.~Hutomo and S.M.~Kuzenko, \emph{{Three-point functions of
  higher-spin spinor current multiplets in $ \mathcal{N} $ = 1 superconformal
  theory}}, \href{https://doi.org/10.1007/JHEP10(2021)058}{\emph{JHEP}
  {\bfseries 10} (2021) 058}
  [\href{https://arxiv.org/abs/2106.14498}{{\ttfamily 2106.14498}}].

\bibitem{Buch4dCFTFerm2022}
E.I.~Buchbinder and B.J.~Stone, \emph{{Three-point functions of a fermionic
  higher-spin current in 4D conformal field theory}},
  \href{https://doi.org/10.1103/PhysRevD.105.125004}{\emph{Phys. Rev. D}
  {\bfseries 105} (2022) 125004}
  [\href{https://arxiv.org/abs/2204.04899}{{\ttfamily 2204.04899}}].

\bibitem{Buch4dSCFTN1spinor2021}
E.I.~Buchbinder, J.~Hutomo and S.M.~Kuzenko, \emph{{Correlation functions of
  spinor current multiplets in $ \mathcal{N} $ = 1 superconformal theory}},
  \href{https://doi.org/10.1007/JHEP07(2021)165}{\emph{JHEP} {\bfseries 07}
  (2021) 165} [\href{https://arxiv.org/abs/2103.09472}{{\ttfamily
  2103.09472}}].

\bibitem{Buch4dCFTGen2023}
E.I.~Buchbinder and B.J.~Stone, \emph{{Three-point functions of conserved
  currents in 4D CFT: General formalism for arbitrary spins}},
  \href{https://doi.org/10.1103/PhysRevD.108.086017}{\emph{Phys. Rev. D}
  {\bfseries 108} (2023) 086017}
  [\href{https://arxiv.org/abs/2307.11435}{{\ttfamily 2307.11435}}].

\bibitem{Buch4dSCFTGen2024}
E.I.~Buchbinder, J.~Hutomo and B.J.~Stone, \emph{{Three-point functions of
  higher-spin supercurrents in 4D N=1 SCFT: General formalism for arbitrary
  superspins}}, \href{https://doi.org/10.1103/PhysRevD.110.085013}{\emph{Phys.
  Rev. D} {\bfseries 110} (2024) 085013}
  [\href{https://arxiv.org/abs/2407.17106}{{\ttfamily 2407.17106}}].

\bibitem{StoneThesis2024}
B.J.~Stone, \emph{{Correlation functions of conserved currents in
  (super)conformal field theory}}, Ph.D. thesis, Western Australia U., 2023.
\newblock \href{https://arxiv.org/abs/2407.17384}{{\ttfamily 2407.17384}}.

\bibitem{Manenti4d3pt2018}
A.~Manenti, A.~Stergiou and A.~Vichi, \emph{{R-current three-point functions in
  4d $\mathcal{N}=1$ superconformal theories}},
  \href{https://doi.org/10.1007/JHEP12(2018)108}{\emph{JHEP} {\bfseries 12}
  (2018) 108} [\href{https://arxiv.org/abs/1804.09717}{{\ttfamily
  1804.09717}}].

\bibitem{SachinJain2023}
S.~Jain, D.K.~S, D.~Mazumdar and S.~Yadav, \emph{{A foray on SCFT$_{3}$ via
  super spinor-helicity and Grassmann twistor variables}},
  \href{https://doi.org/10.1007/JHEP09(2024)027}{\emph{JHEP} {\bfseries 09}
  (2024) 027} [\href{https://arxiv.org/abs/2312.03059}{{\ttfamily
  2312.03059}}].

\bibitem{BuchKuzSUGRA1998}
I.L.~Buchbinder and S.M.~Kuzenko, \emph{{Ideas and methods of supersymmetry and
  supergravity, Or a walk through superspace}}, CRC Press (1998),
  \href{https://doi.org/10.1201/9780367802530}{10.1201/9780367802530}.

\bibitem{GilliozCFT2022}
M.~Gillioz, \emph{{Conformal field theory for particle physicists}},
  SpringerBriefs in Physics, Springer (2023),
  \href{https://doi.org/10.1007/978-3-031-27086-4}{10.1007/978-3-031-27086-4},
  [\href{https://arxiv.org/abs/2207.09474}{{\ttfamily 2207.09474}}].

\bibitem{Stanev1988}
Y.S.~Stanev, \emph{{Stress - Energy Tensor and U(1) Current Operator Product
  Expansions in Conformal {QFT}}}, {\emph{Bulg. J. Phys.} {\bfseries 15} (1988)
  93}.

\bibitem{FZmultiplet1975}
S.~Ferrara and B.~Zumino, \emph{{Transformation Properties of the
  Supercurrent}},
  \href{https://doi.org/10.1016/0550-3213(75)90063-2}{\emph{Nucl. Phys. B}
  {\bfseries 87} (1975) 207}.

\end{thebibliography}\endgroup

\end{document}